\documentclass[onecolumn, twoside]{IEEEtran}

\usepackage[ruled]{algorithm2e}
\renewcommand{\algorithmcfname}{ALGORITHM}
\SetAlFnt{\small}
\SetAlCapFnt{\small}
\SetAlCapNameFnt{\small}
\SetAlCapHSkip{0pt}
\IncMargin{-\parindent}
\usepackage{graphicx}
\usepackage{algorithmic}
\usepackage{subfig}
\usepackage{multirow}
\usepackage{array}
\usepackage{amsmath} 
\usepackage{amssymb}
\usepackage[maxfloats=30]{morefloats}
\usepackage{amsthm}
\theoremstyle{definition}

\begin{document}
	
	\title{\huge Stable Desynchronization for Wireless Sensor Networks: \\(II) Performance Evaluation}
	
	\author{
		\IEEEauthorblockN{Supasate Choochaisri, Kittipat Apicharttrisorn, Chalermek Intanagonwiwat}\\
		\IEEEauthorblockA{Chulalongkorn University, Bangkok, Thailand\\Email:\{supasate.c, kittipat.api, intanago\}@gmail.com}
	}
	
	\date{}
	\maketitle

\begin{abstract}
In this paper, we evaluate M-DWARF's performance by experimentation and simulation. We validate its functionalities on TelosB motes \cite{telosb} and compare its performance with EXT-DESYNC \cite{MK09DESYNC}, and LIGHT-WEIGHT \cite{5062165} on TOSSIM \cite{Levis:2003:TAS:958491.958506}. On simulation, we test the algorithms on several multi-hop topologies and discuss both the average and problematic cases. In addition, we investigate the impacts of period length ($T$) and compare channel utilization fairness among the desynchronization algorithms. Finally, we propose a method to reduce desynchronization overhead. (This paper is the second part of the series Stable Desynchronization for Wireless Sensor Networks - (I) Concepts and Algorithms (II) Performance Evaluation (III) Stability Analysis)

\end{abstract}

\label{sec:performance_evaluation}
\section{Experimental Results on TelosB Motes}

In a sense of TDMA, we assume communication models as follows. First, all nodes have an identical unit circle of communication ranges on the same frequency spectrum. Second, transmission and reception ranges are equal; that is, if node A can receive node B's packets, node A's transmission will interfere with node B's.

We implement M-DWARF into TelosB motes \cite{telosb} using TinyOS \cite{tinyos} and record a set of video to demonstrate its functionalities. All the motes are pre-installed with our M-DWARF implementation code and LED is on when a mote is firing a message and off when receiving a message. A state of desynchrony can be observed with nodes taking turn blinking the LEDs with equal intervals.

In \cite{exp_10_n_join}, we incrementally add new nodes by inserting AA batteries into each TelosB mote until the single-hop network contains 10 nodes. At the beginning, nodes converge to a state of desynchrony within a few time periods ($T$). As the number of nodes increases, the convergence is a little bit slower with a few more periods. We physically arrange all the motes on the table into a circular form for succinct demonstation and a final state of 10-node desynchrony can be observed by LEDs running smoothly and circularly. From the desynchronized 10-node network, we also demonstrate leaving nodes in \cite{exp_10_n_leave}; that is, we take off the nodes' batteries  one-by-one and observe that within a few periods, the network can turn back to a state of desynchrony as each node leaves. 
\begin{figure}
	\centering
	\includegraphics[width=2.5in]{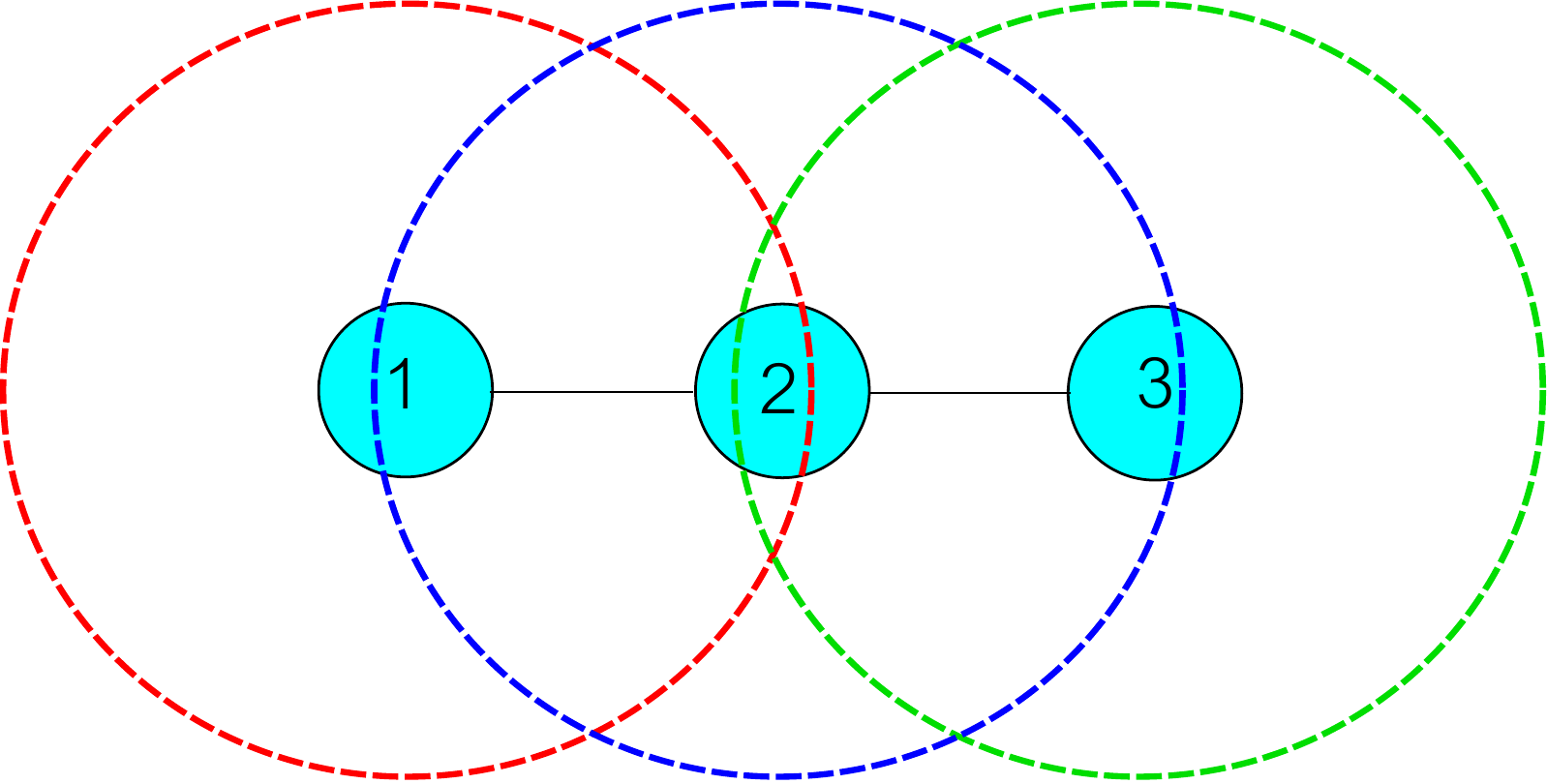}
	\caption{The hidden terminal problem.}
	\label{fig:3nodes-chain-hidden}
\end{figure}
We also demonstrate our M-DWARF for multi-hop desynchronization. In \cite{exp_3_n_multihop}, a simple three-node two-hop network (as in Figure \ref{fig:3nodes-chain-hidden}) is used to show that M-DWARF successfully solves the hidden terminal problem. In the video, the first and last nodes are in two-hop communication via the middle node, all of which must avoid using the same time phase. M-DWARF successfully forces the first and last nodes \emph{not} to blink at the same time; therefore, collision at the middle node is evaded. Moreover, in \cite{exp_4_n_multihop}, we demonstrate, in a four-node-three-hop network, that nodes which are beyond two-hop communication can occupy the same time phase not only without interference but also with increased channel utilization. This can be achieved by employing our \emph{Force Absorption} technique. In the video, the first and third nodes are in two-hop communication via the second one and so are the second and last nodes via the third one. Therefore, the first and last nodes can share the same time phase without interference, so their LEDs can blink at the same time.

These experiments show that M-DWARF can be implemented into TelosB, an off-the-shelf mote platform, and that it fast converges the network to a state of desynchrony. They also validate M-DWARF functionalities. With \emph{Relative Time Relaying}, the desynchronized network can avoid the hidden terminal problem, and with \emph{Force Absorption}, nodes that are beyond two-hop can share the same time phase, allowing the channel utilization to be improved. In the next section, we further evaluate M-DWARF on TOSSIM \cite{Levis:2003:TAS:958491.958506} and compare it with the other state-of-the-art desynchronization algorithms.

\section{Simulation Results on TOSSIM}
We implement M-DWARF, EXT-DESYNC \cite{MK09DESYNC}, and LIGHTWEIGHT \cite{5062165} on TinyOS 2.1.2 and evaluate them on the TOSSIM simulator and have concluded simulation parameters in the Table \ref{tab:simulation_parameter}.
\begin{table}
	\centering
	\renewcommand{\arraystretch}{1.2}
	\caption{Simulation Parameters}
\begin{tabular}{ | c | m{10cm} |} \hline
	Parameter & Value \\ \hline \hline
	Header &  2-byte sender node ID, and 2-byte neighbor ID (for all the algorithms) and additional 2-byte relative phase (for M-DWARF and EXT-DESYNC) \\ \hline
	Radio header & 11-byte CC2420 header \\ \hline
	Time period & 1000 (default), 2000, or 3000 milliseconds  \\ \hline
	Step size & 0.95 (only for EXT-DESYNC and as originally implemented in the paper) \\ \hline
	Topology & Single-hop and Multi-hop (see Figure \ref{fig:multihop-eval}) \\ \hline
	Number of nodes & Varying (see Figure \ref{fig:multihop-eval}) \\ \hline
	Initial phase & Randomized between 0 and the time period \\ \hline
\end{tabular}
\label{tab:simulation_parameter}
\end{table}

\subsection{Simulation on Single-Hop Topologies}
We begin the experiments by evaluating M-DWARF on single-hop networks. In both the M-DWARF and EXT-DESYNC, each firing message contains relative phases of one-hop neighbors. Therefore, to limit the size of a message, we vary the one-hop network size from 4 to 32 nodes.

\begin{figure*}
	\centerline {
		\subfloat[]{\includegraphics[width=3.0in]{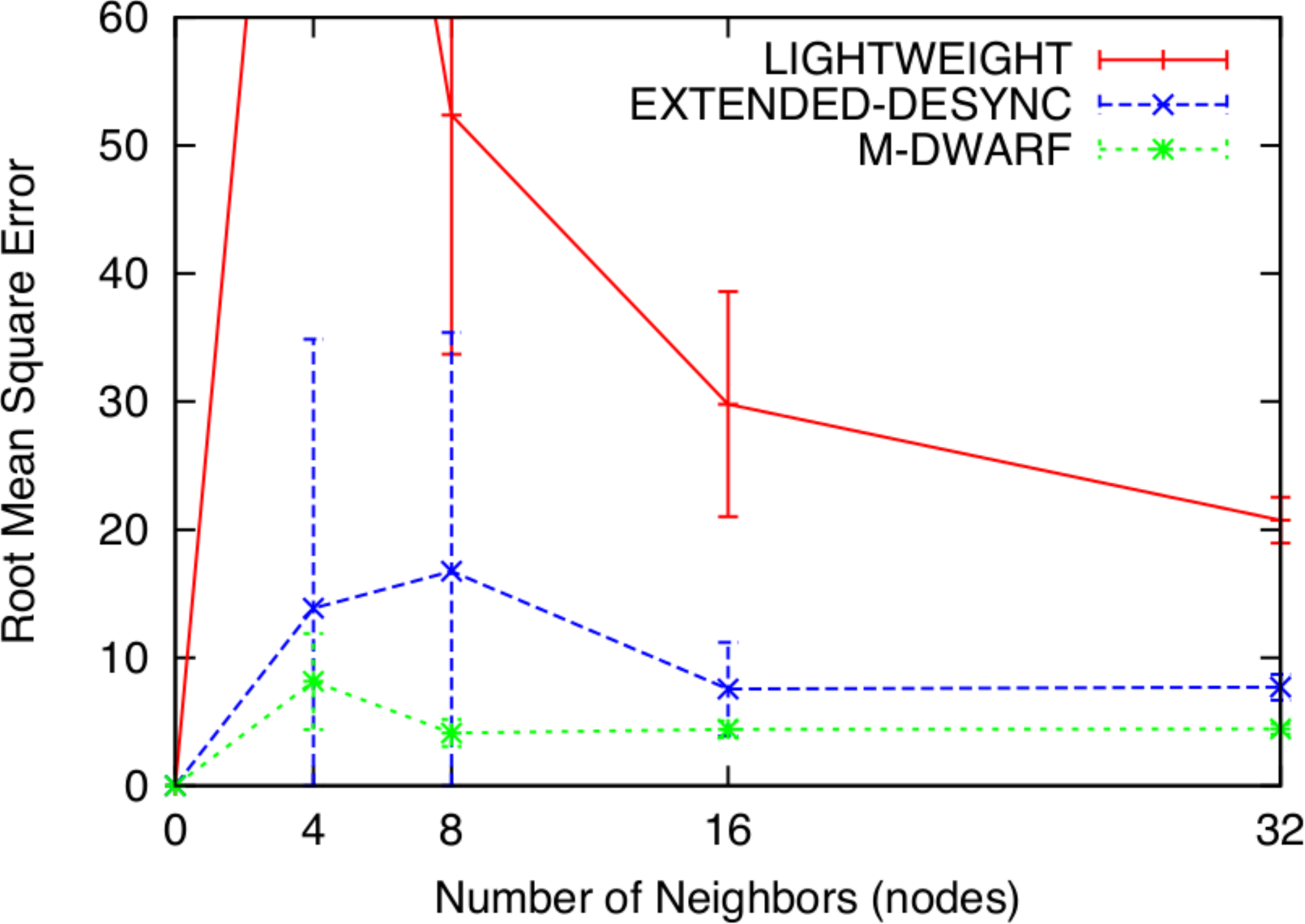}%
			\label{fig:rmse300rounds-mhop}}
		\hfil
		\subfloat[]{\includegraphics[width=3.0in]{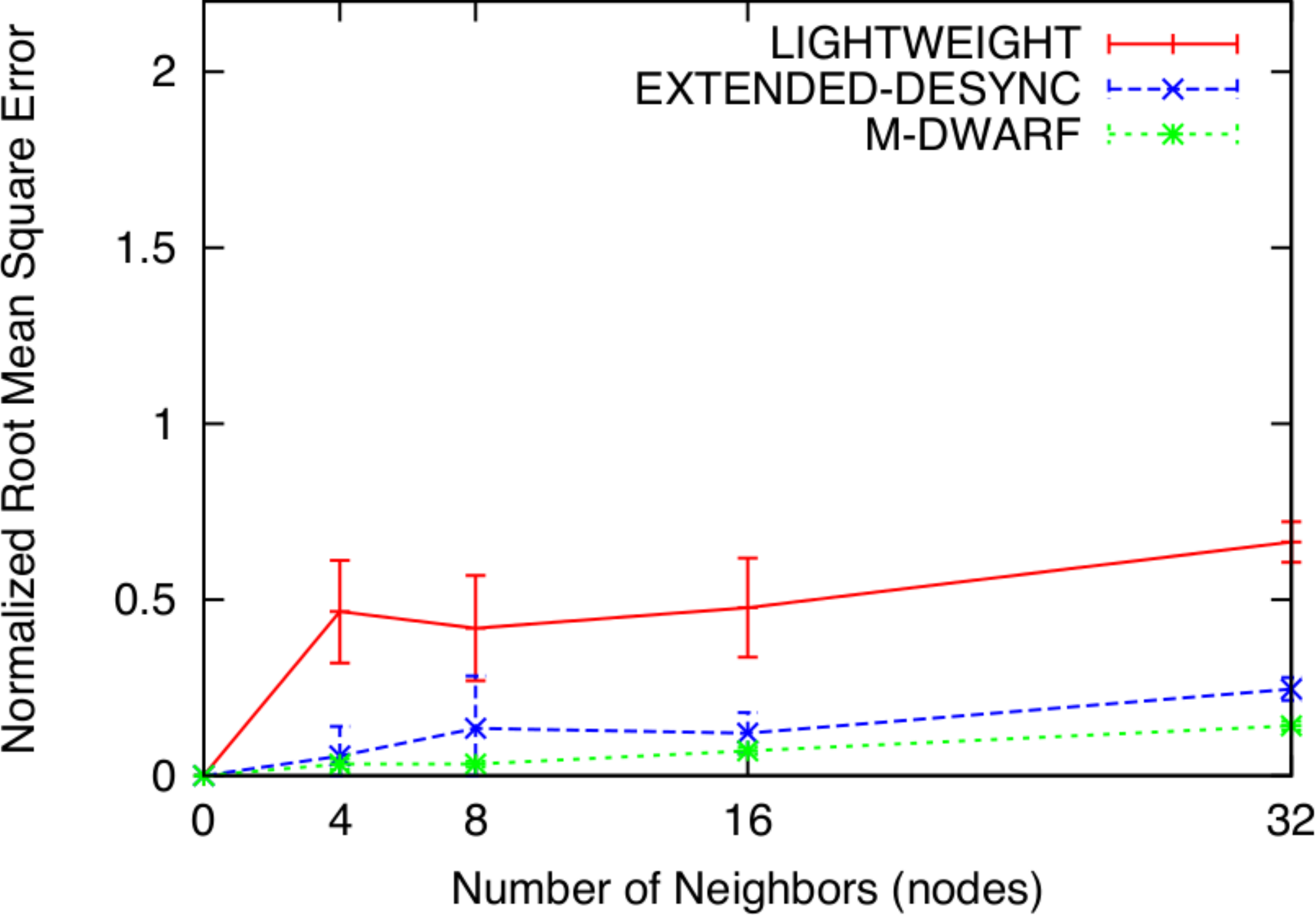}%
			\label{fig:nrmse300rounds-mhop}}
	}
	\caption{(a) Root mean square error after 300 time periods. (b) Root mean square error normalized by perfect phase difference after 300 time periods.}
	
\end{figure*}

\paragraph{Desynchronization Error}
We run the simulation for 300 time periods to measure the desynchronization error. 
In each network size, we run the simulation for 30 times.
Then, we measure the average root mean square error (RMSE) and normalized root mean square error.
\begin{equation}
\label{eq:desync_error}
ERR_i = \Delta \phi_{i,j} - T/n
\end{equation}
\begin{equation}
\label{eq:rmse}
RMSE = \sqrt{\frac{\sum_{i = 1}^{n}{ERR_i^2}}{n}}
\end{equation}
\begin{equation}
\label{eq:nrmse}
NRMSE = \frac{RMSE}{T/n}
\end{equation}
where node $j$ is the next phase neighbor of node $i$.
$\Delta \phi_{i,j}$ is the phase difference between node $i$ and node $j$ on the time period $T$.
Given that $n$ is a total number of nodes, $T/n$ is the perfect phase difference.
Figure  \ref{fig:rmse300rounds-mhop} illustrates the result of absolute desynchronization error and Figure \ref{fig:nrmse300rounds-mhop} illustrates the result of the normalized desynchronization error in each network size after 300 time periods.

The result indicates that, in all network sizes (4 - 32 nodes), M-DWARF achieves significantly better desynchrony states than EXT-DESYNC and LIGHT-WEIGHT do. 
EXT-DESYNC faces the same predicament of direct and continuing error propagation from phase neighbors, as DESYNC does, because they are based on the same mechanism.
This results in a large error even after convergence.
In contrast, both DWARF and M-DWARF are robust to this error propagation because they use the total received forces from all neighbors. An erroneous force from one neighbor does not directly impact the phase adjustment of a node; as a result, M-DWARF maintains the lowest level of desynchronization errrors, compared to those of the other two algorithms.
For LIGHTWEIGHT, the desynchronization errors of small networks are extremely large because the algorithm randomly chooses a free time slot without considering the equitable separation or fairness. However, when the network size becomes larger, the error becomes lower because the perfect phase interval length is reduced.

\begin{figure*}
	\centerline{
		\subfloat[Sparse]{\includegraphics[scale=0.4]{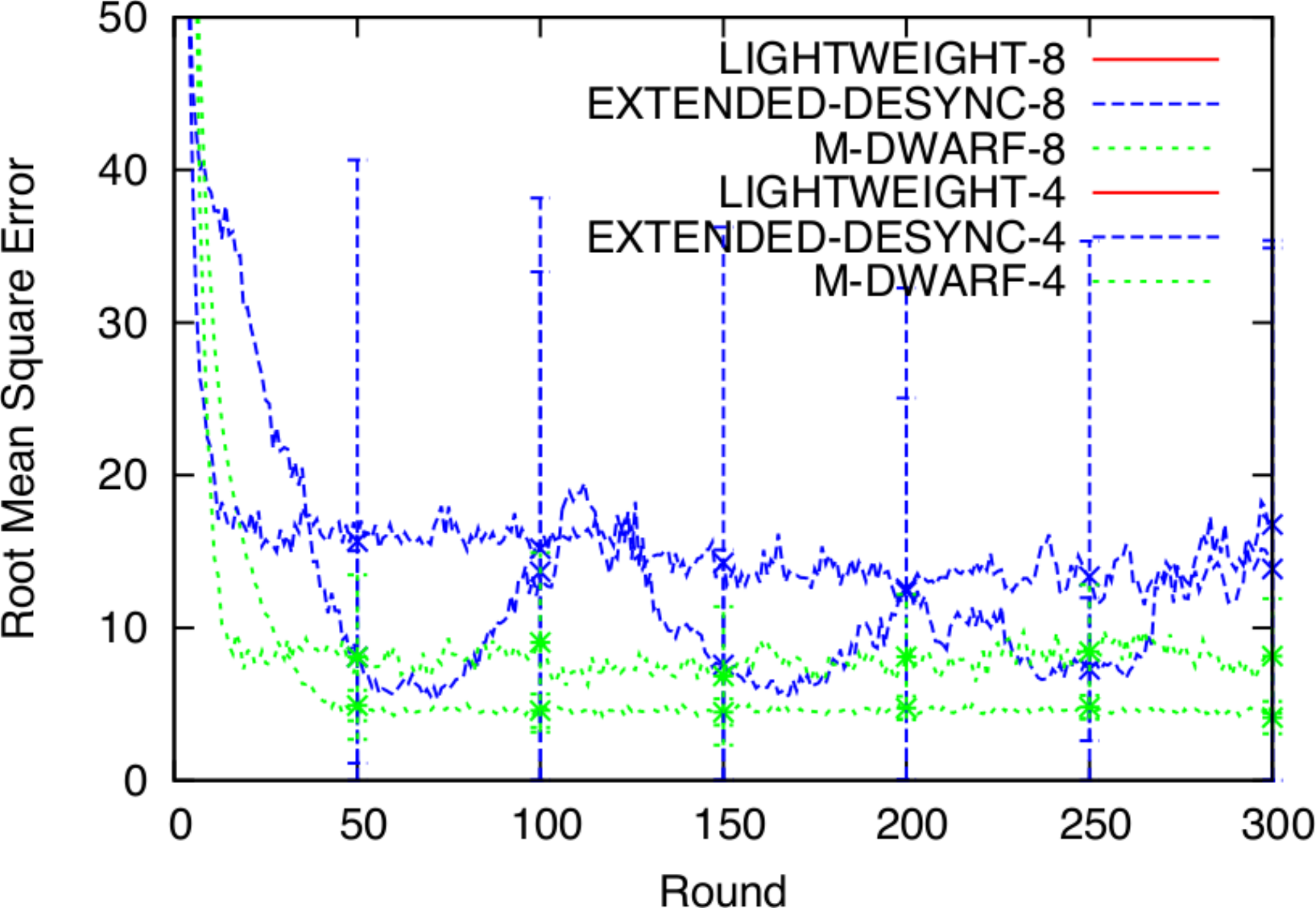}%
			\label{fig:rmse-sparse-mhop}}
		\hfil
		\subfloat[Dense]{\includegraphics[scale=0.4]{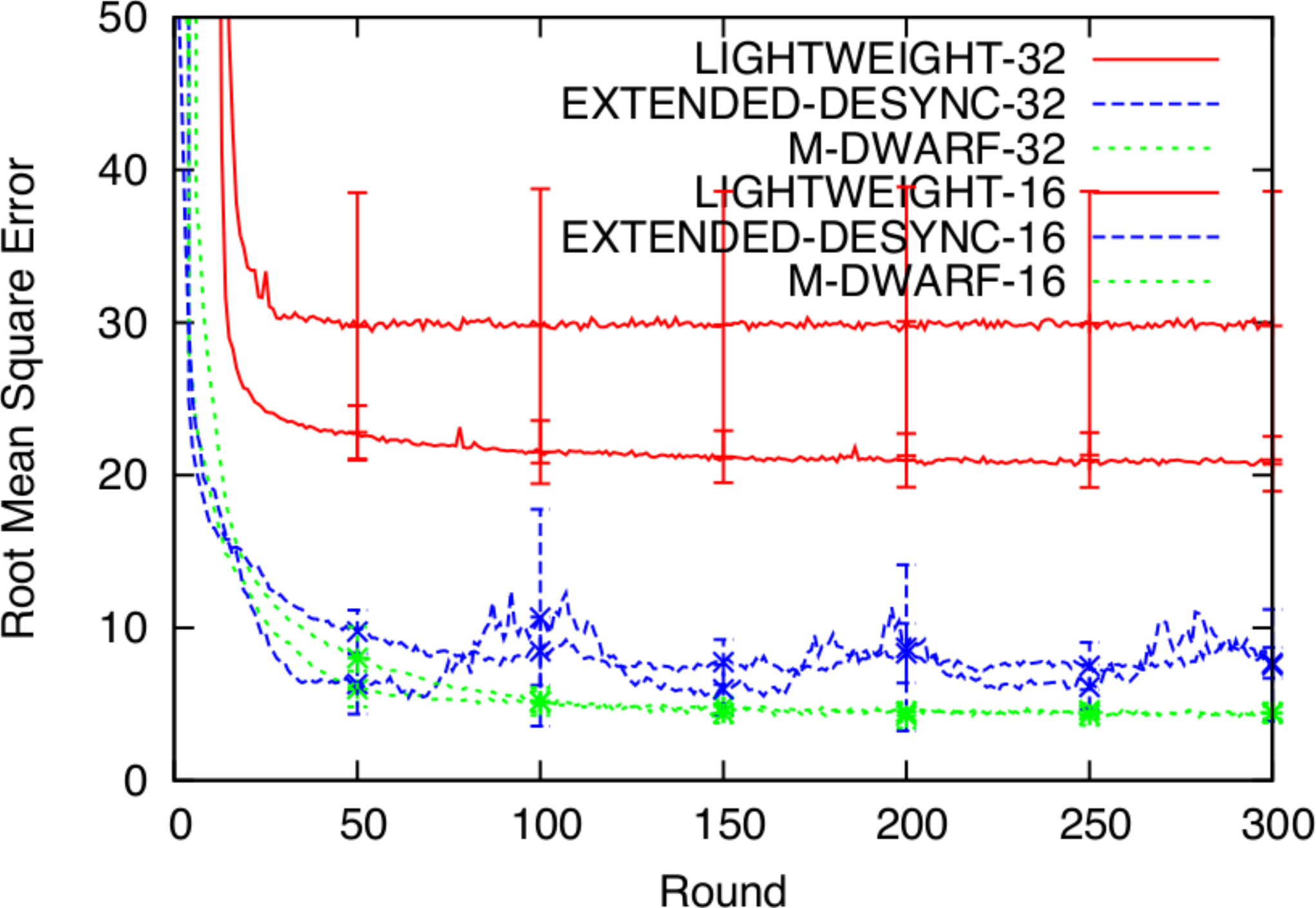}%
			\label{fig:rmse-dense-mhop}}
	}
	\caption{Convergence time and absolute root mean square error}
	\label{fig:rmse-convergence-mhop}
	
\end{figure*}

\begin{figure*}
	\centerline{
		\subfloat[Sparse]{\includegraphics[scale=0.4]{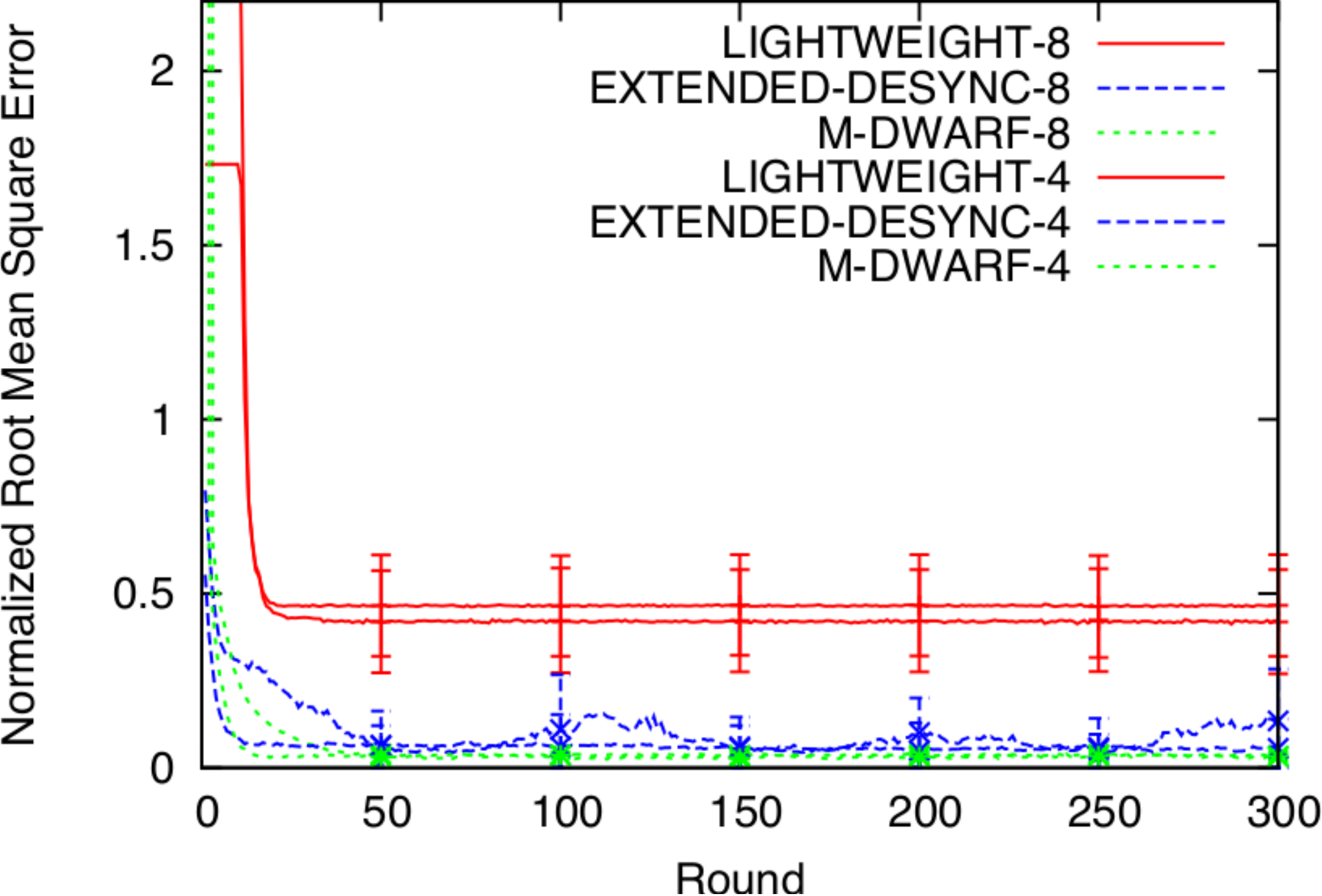}%
			\label{fig:nrmse-sparse-mhop}}
		\hfil
		\subfloat[Dense]{\includegraphics[scale=0.4]{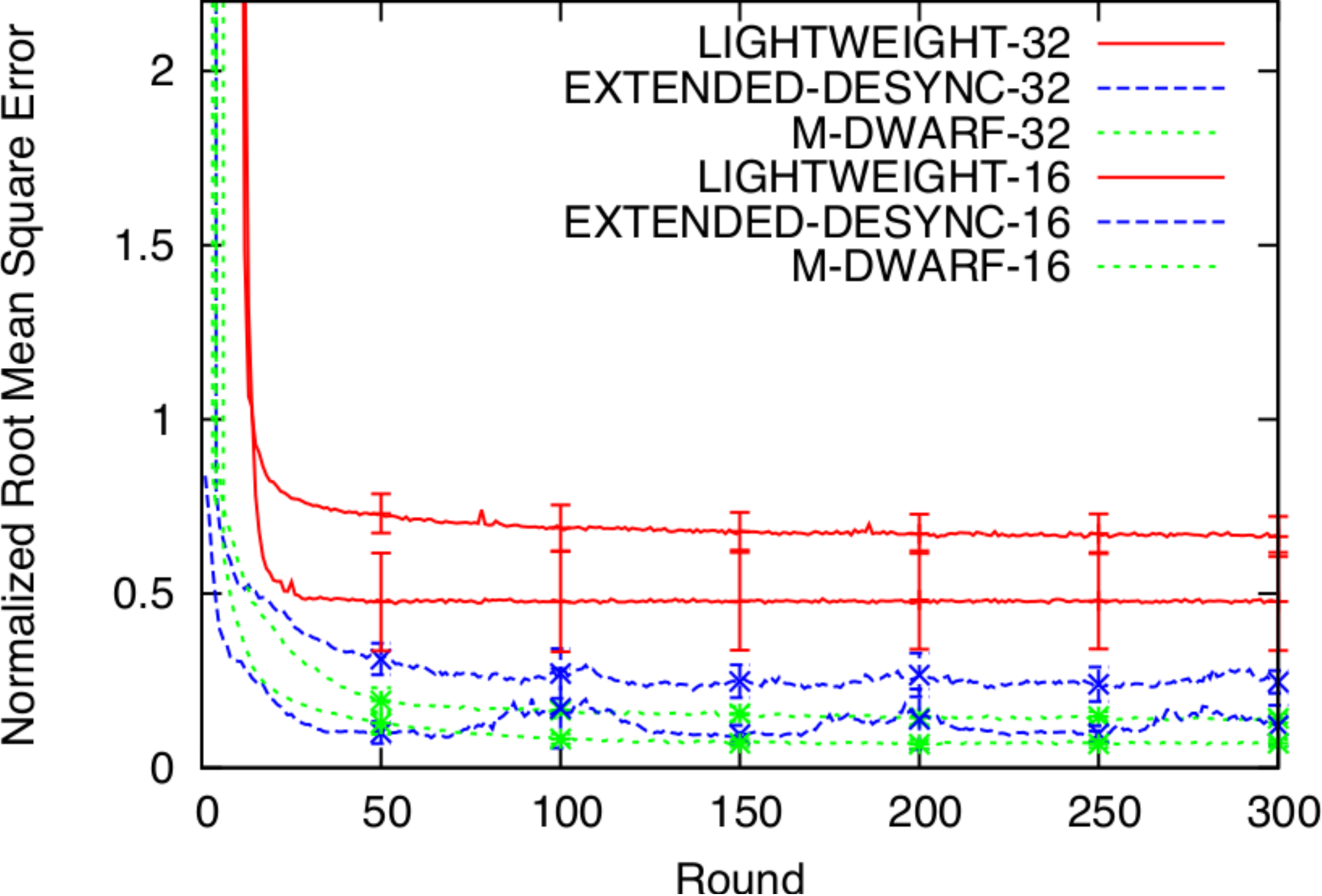}%
			\label{fig:nrmse-dense-mhop}}
	}
	\caption{Convergence time and root mean square error normalized by expected phase difference}
	\label{fig:nrmse-convergence-mhop}
	
\end{figure*}

\paragraph{Convergence Time}
In this section, we measure the absolute root mean square error and normalized root mean square error for each time period to investigate the convergence time.
Figure \ref{fig:rmse-sparse-mhop} and \ref{fig:nrmse-sparse-mhop} show the results of sparse networks, and Figure \ref{fig:rmse-dense-mhop} and \ref{fig:nrmse-dense-mhop} show the results of dense networks.

\begin{figure*}
	\centerline{
		\subfloat[8 nodes]{\includegraphics[scale=0.4]{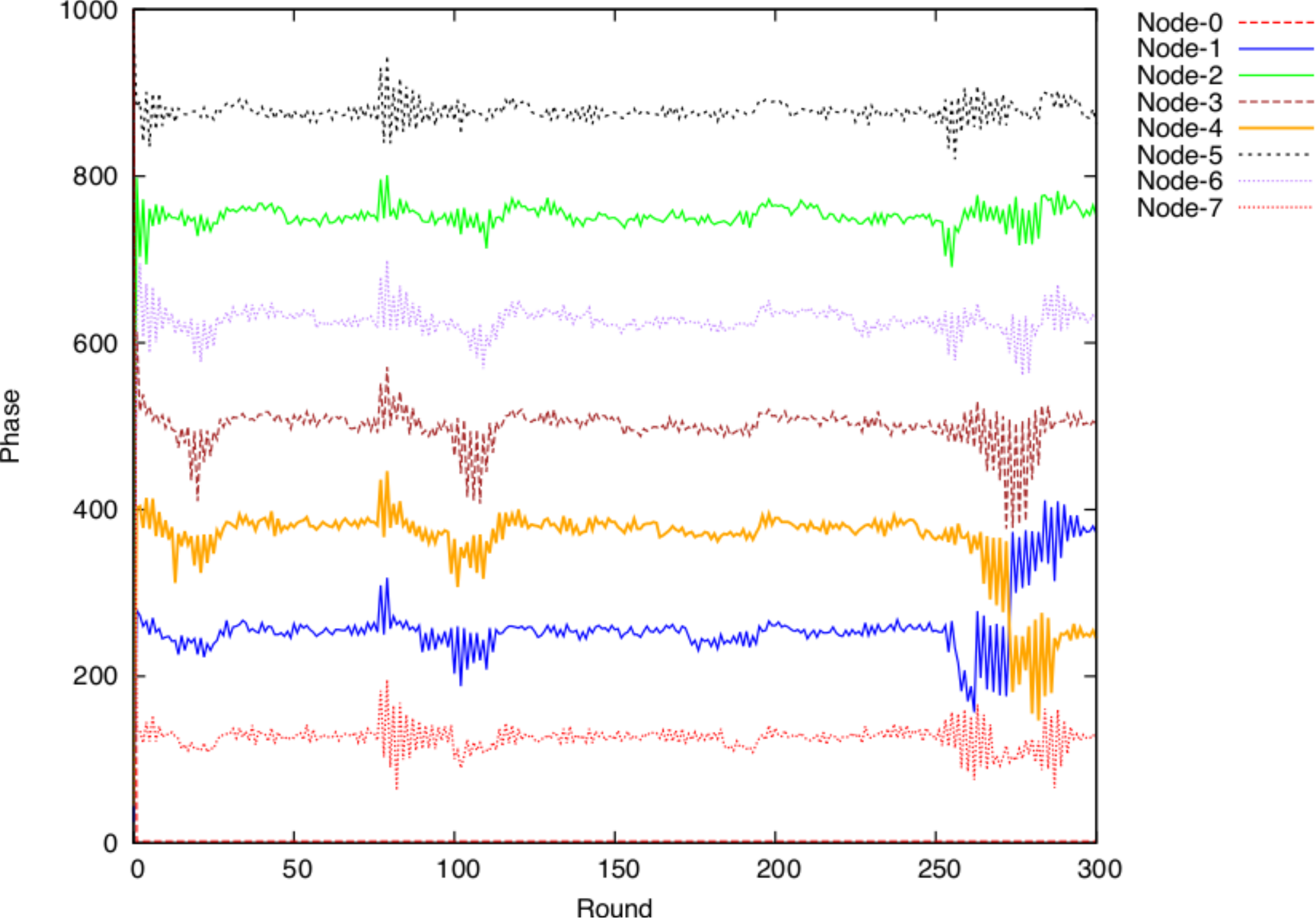}}
		\hfil
		\subfloat[16 nodes]{\includegraphics[scale=0.4]{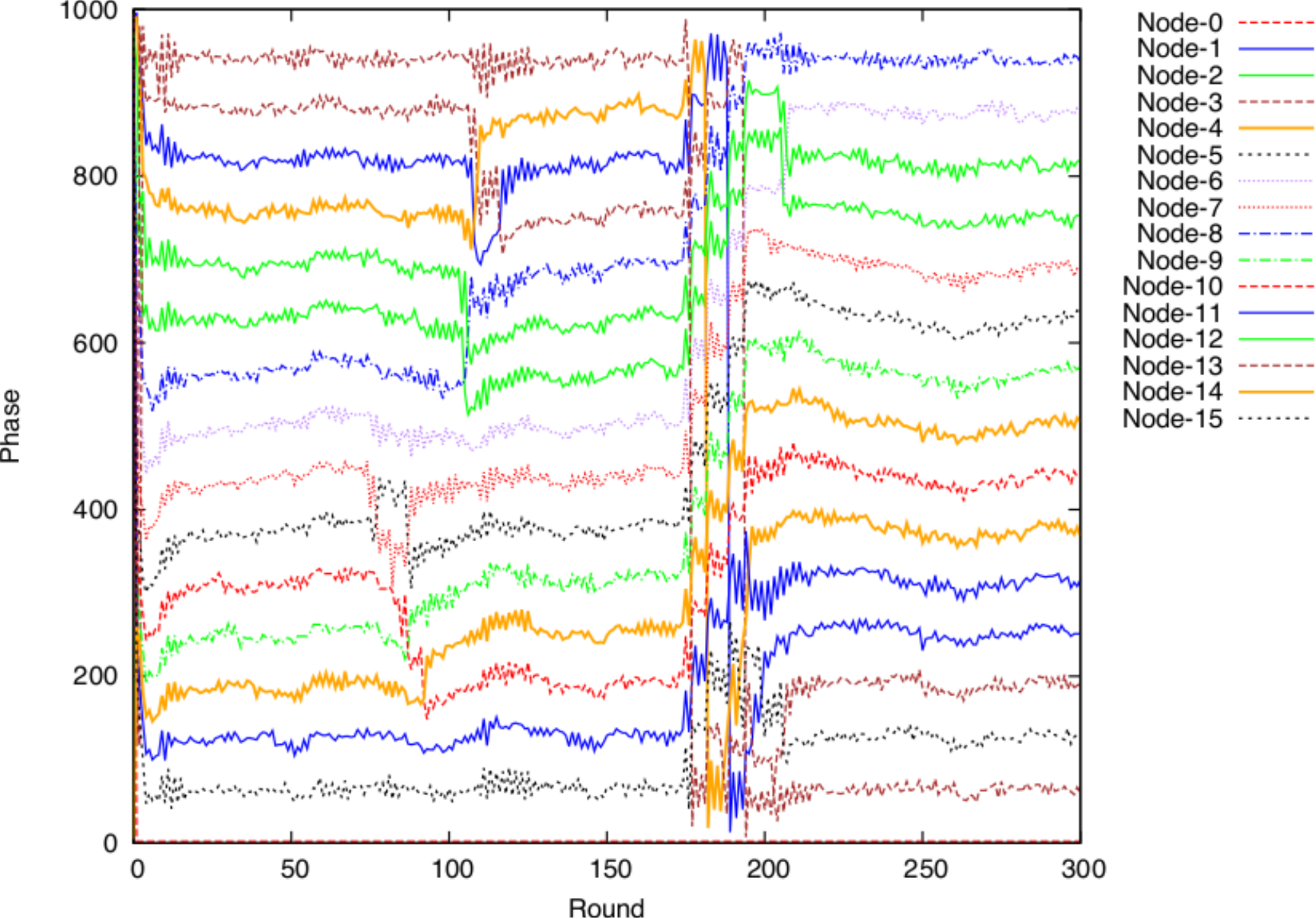}}
	}
	\caption{Fluctuation in some cases of EXT-DESYNC.}
	\label{fig:fluctuate}
\end{figure*}

In all network sizes, the convergence speed of M-DWARF is comparable to that of EXT-DESYNC. However, M-DWARF converges with the lowest desynchronization error. For LIGHTWEIGHT, the algorithm converges fast and stable. However, due to the random slot selection process, the desynchronization errors in all network sizes are large.

We note that the errors of EXT-DESYNC for 8-node and 16-node networks fluctuate. The reason is that, in our 30 EXT-DESYNC simulations, there are some cases in 8-node and 16-node networks that the phases of some nodes highly fluctuate as depicted in Figure \ref{fig:fluctuate}. These cases affect the average value. This fluctuating behaviour is caused by the mechanism of EXT-DESYNC (and DESYNC) that relies on only two phase neighbor information.

The result of single-hop networks indicates that M-DWARF, which augments DWARF with the two mechanisms, still performs very well on single-hop networks without loss of generality. 

\subsection{Simulation on Multi-Hop Topologies}

\begin{figure*}
	\centering{
		\subfloat[6-node Star]{\includegraphics[scale=0.3]{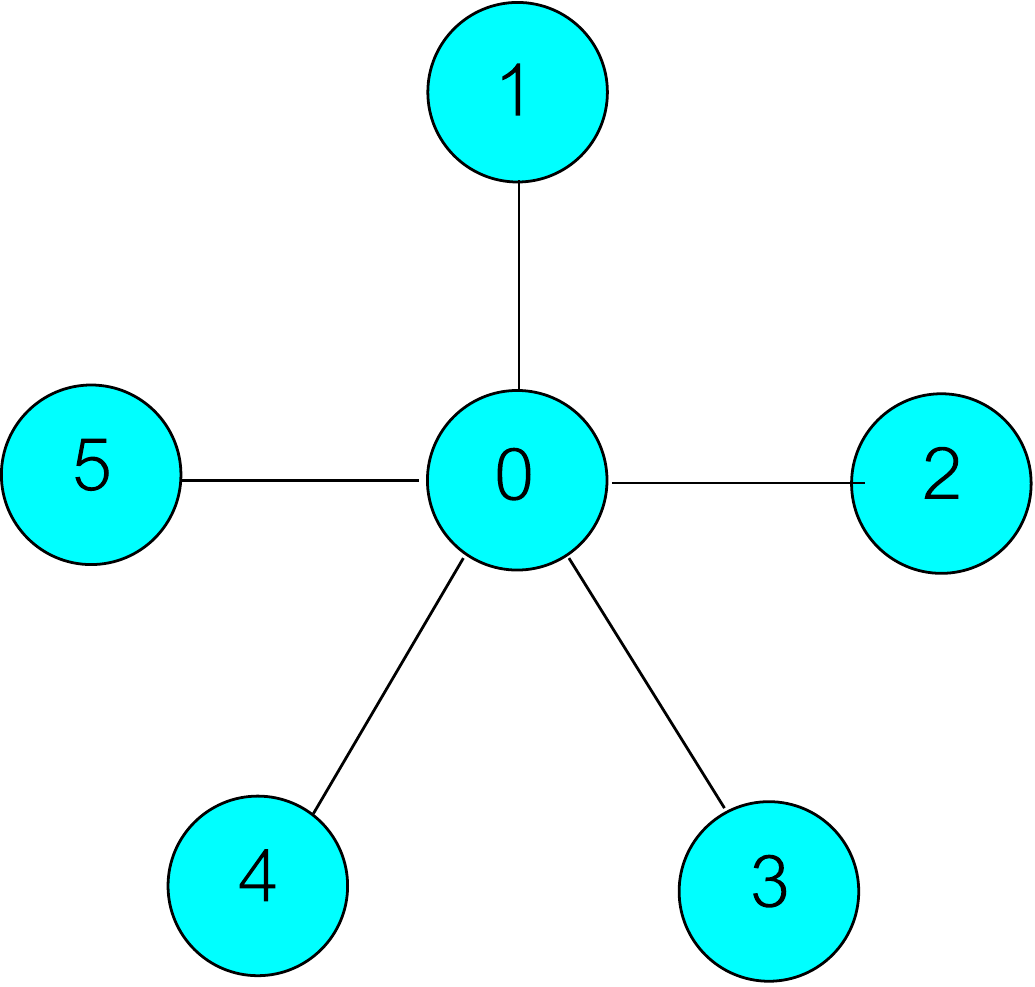}%
			\label{fig:6nodes-star-eval}}
		\hspace{1cm}
		\subfloat[20-node Star]{\includegraphics[scale=0.15]{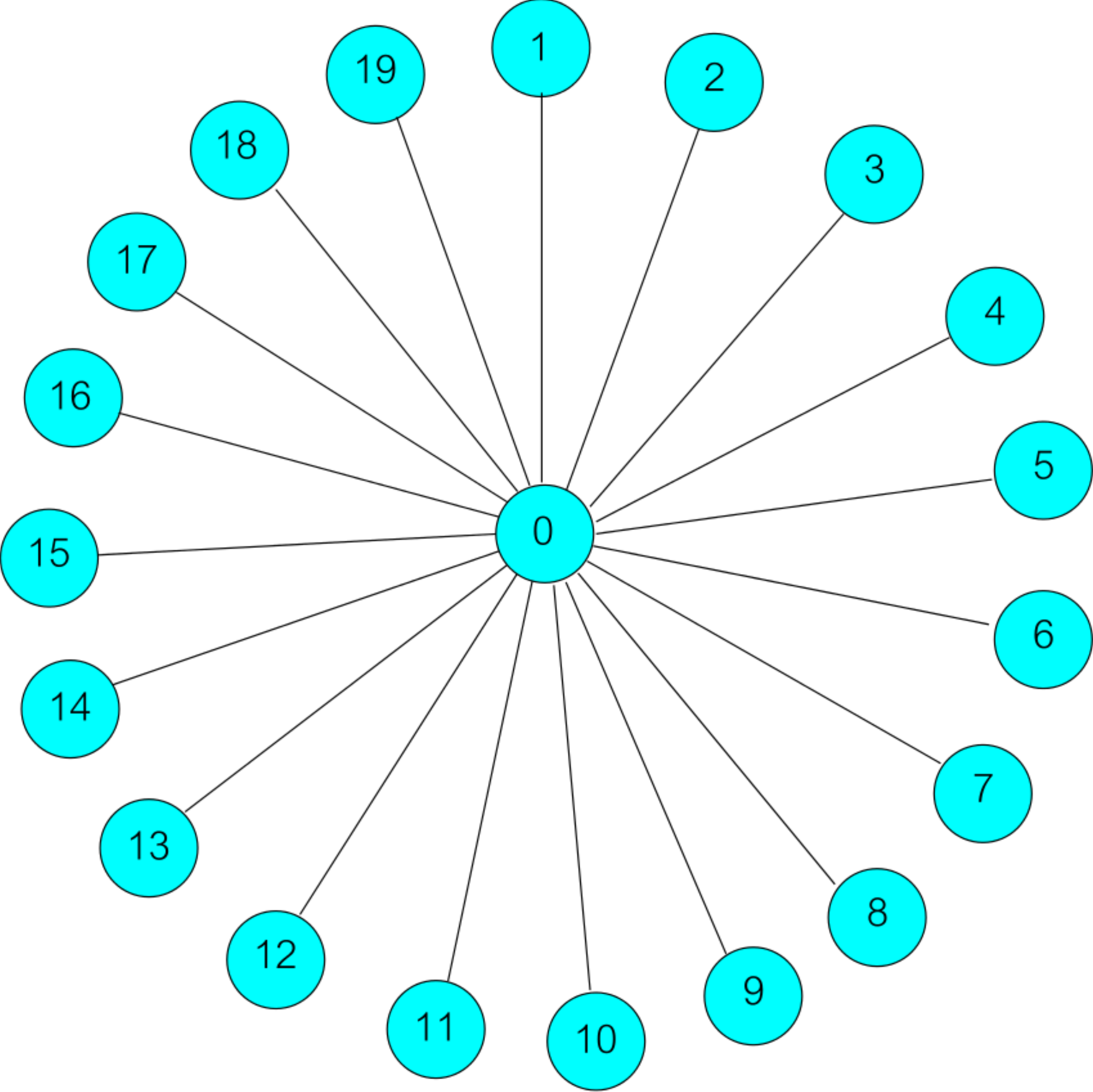}%
			\label{fig:20nodes-star-eval}}
		\hspace{1cm}
		\subfloat[3-node Chain]{\includegraphics[scale=0.15]{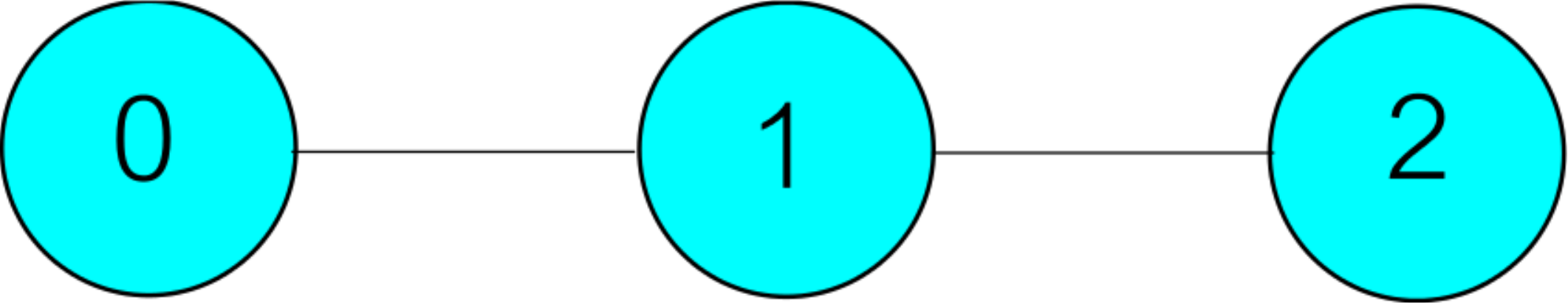}%
			\label{fig:3nodes-chain-eval}}
		
		\subfloat[10-node Chain]{\includegraphics[scale=0.25]{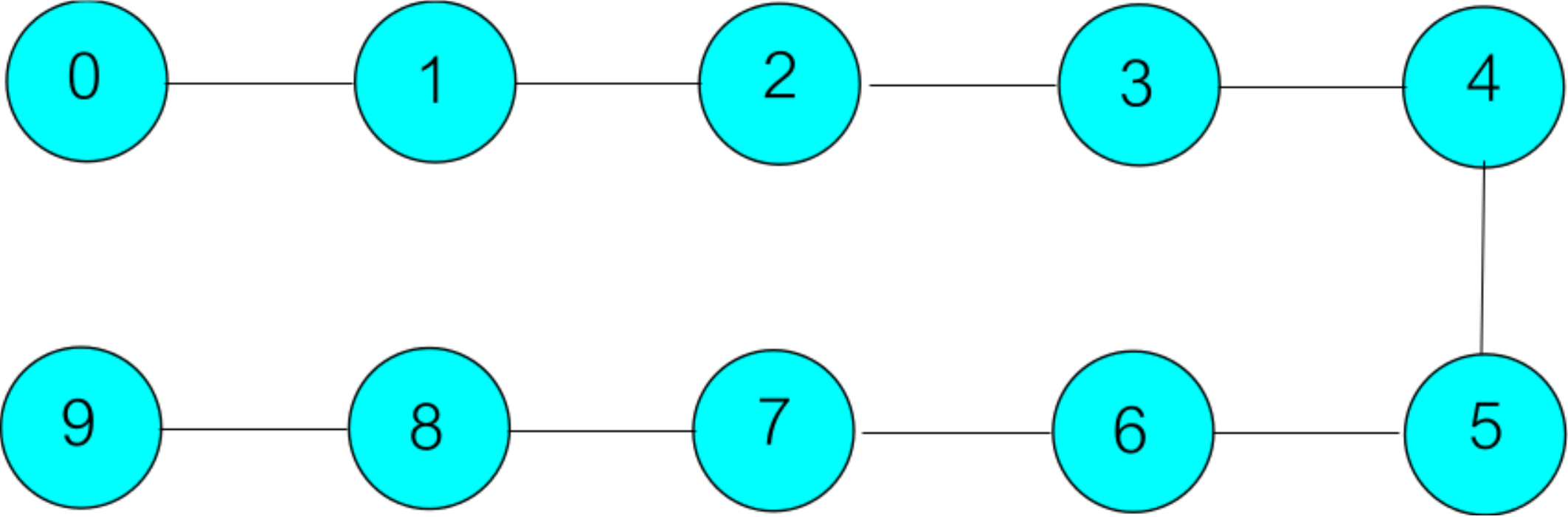}%
			\label{fig:10nodes-chain-eval}}
		\hspace{1cm}
		\subfloat[4-node Cycle]{\includegraphics[scale=0.1]{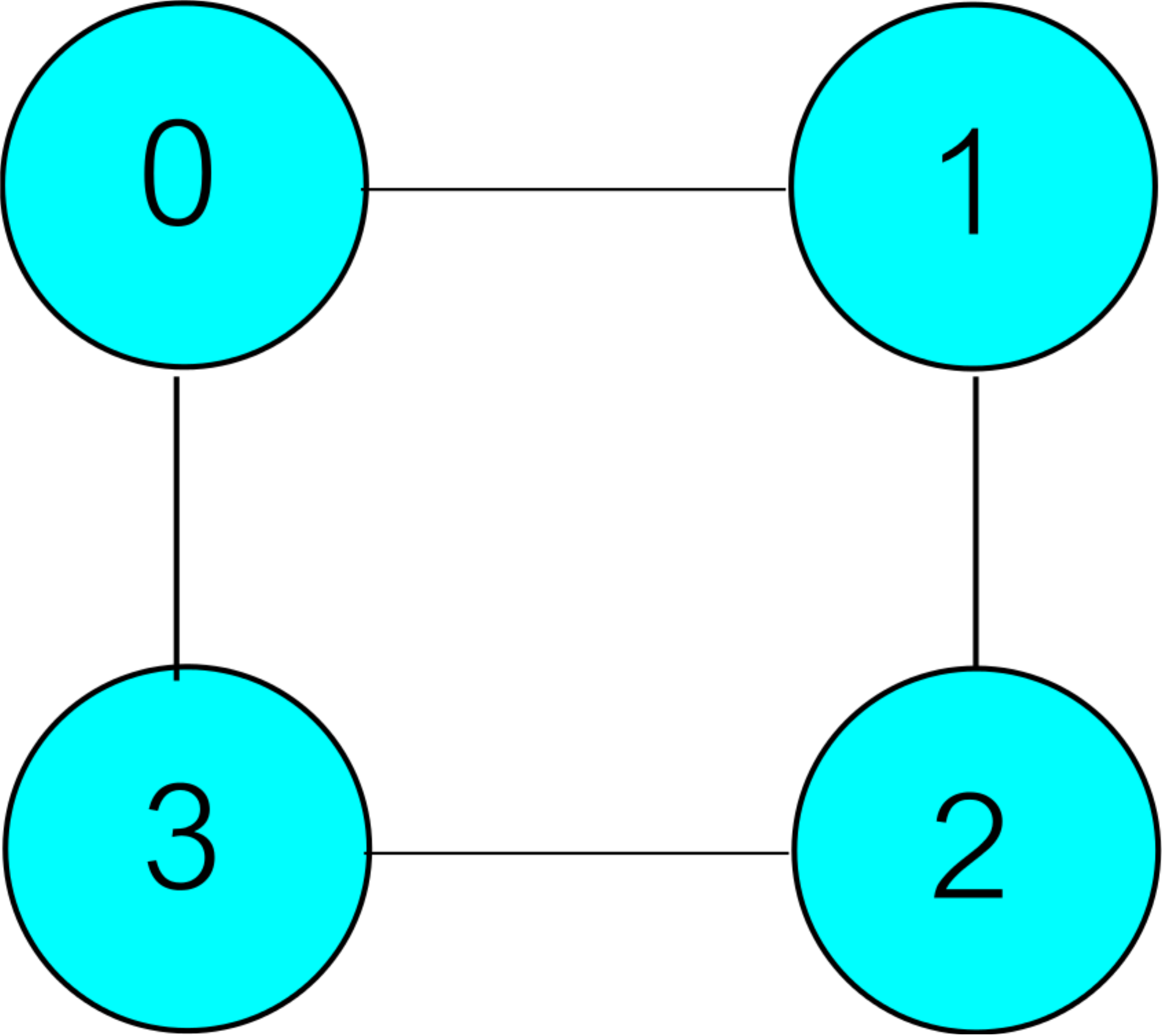}%
			\label{fig:4nodes-cycle-eval}}
		\hspace{1cm}
		\subfloat[10-node Cycle]{\includegraphics[scale=0.25]{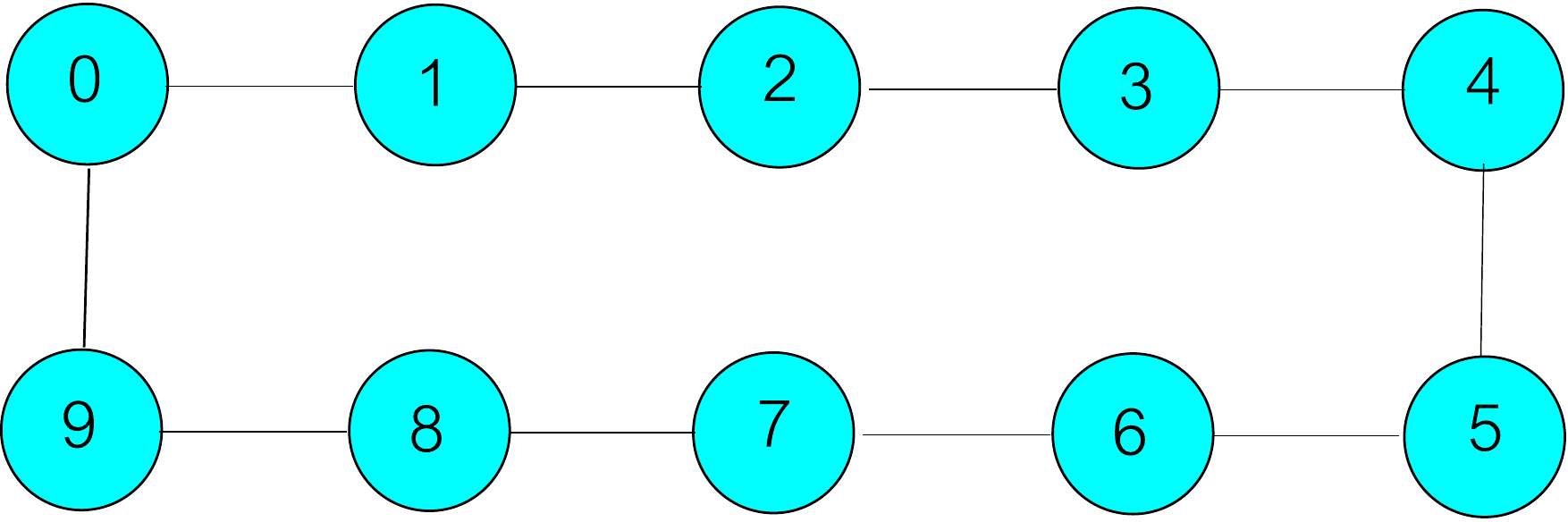}%
			\label{fig:10nodes-cycle-eval}}
		
		\subfloat[6-node Dumbbell]{\includegraphics[scale=0.3]{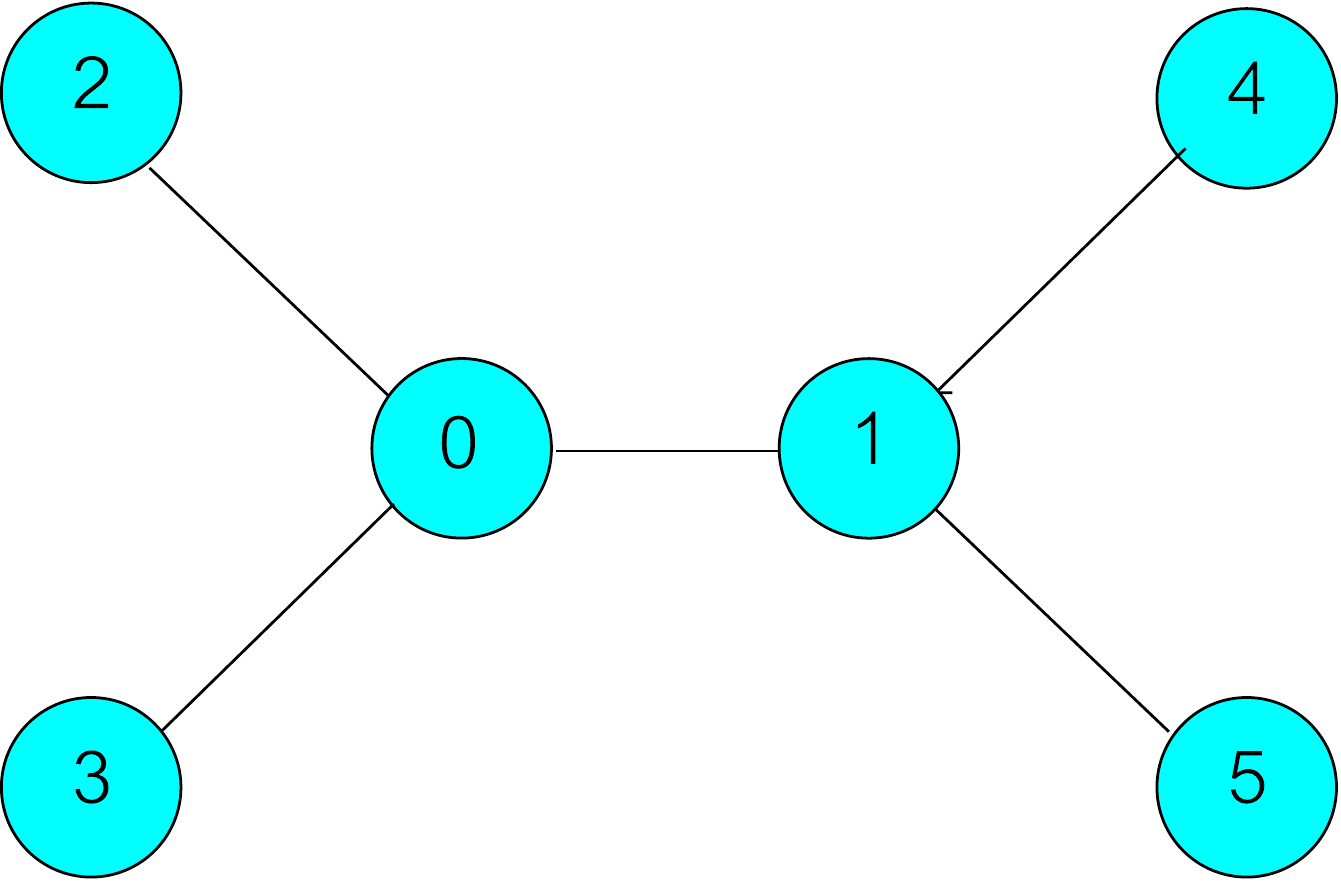}%
			\label{fig:6nodes-butterfly-eval}}
		\hspace{1in}
		\subfloat[20-node Dumbbell]{\includegraphics[scale=0.2]{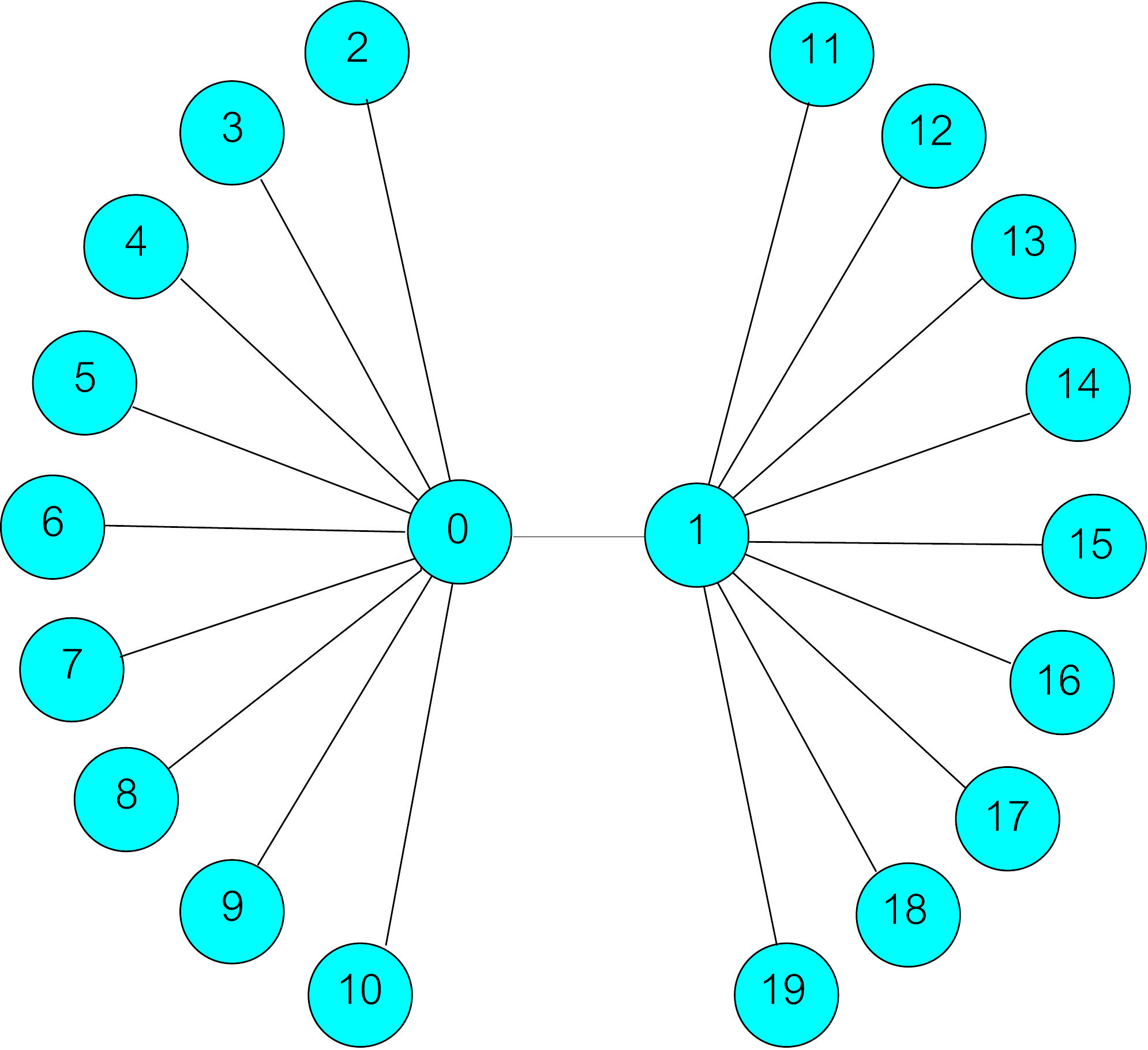}%
			\label{fig:20nodes-butterfly-eval}}
		
		\subfloat[10-node Mesh]{\includegraphics[width=2.5in]{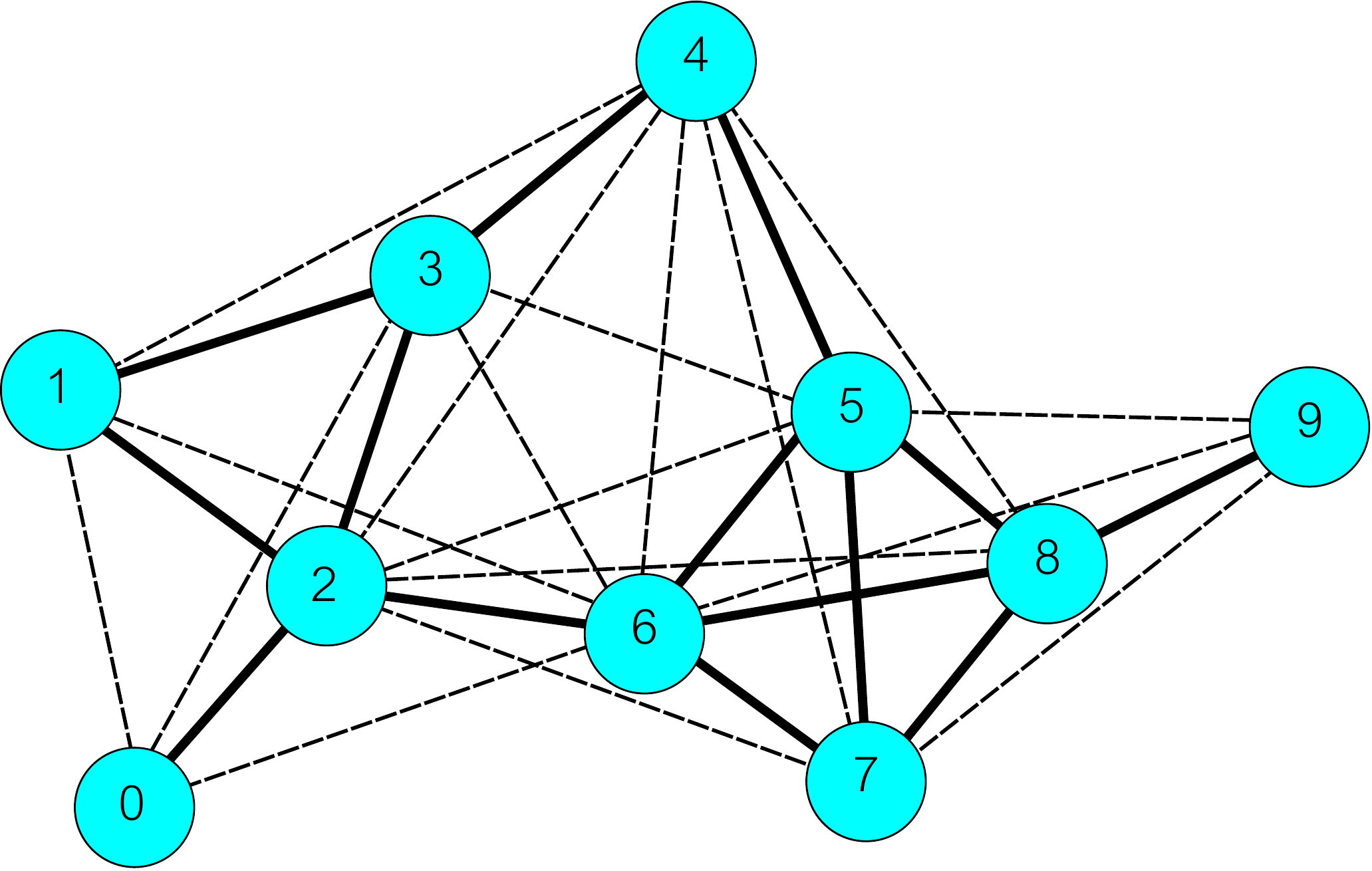}%
			\label{fig:10nodes-topo}}
	}
	\caption{Multihop topologies}
	\label{fig:multihop-eval}
\end{figure*}

For multi-hop networks, we evaluate M-DWARF on several topologies including star, chain, cycle, dumbbell, and mesh topologies (Figure \ref{fig:multihop-eval}). In each topology, we simulate for 30 times and show the relative phase graph results. We show the relative phase graphs that plot relative phases and simulation rounds and represent the average cases and the problematic cases of each algorithm. For each topology, we provide relative phase graphs to visualize the performance of the three desynchronization algorithms. At the end of this section, we qualify the algorithms and compare their performance using a matrix of Convergence, Stability, and Fairness in Table \ref{tab:simresults}.
 
\paragraph{Star Topology}
The star topology is the simplest case for multi-hop desynchronization. There is one center node that can transmit and receive messages with all other nodes in the network. In contrast, other nodes can  transmit and receive only with the center node. Figure \ref{fig:6nodes-star-eval} and \ref{fig:20nodes-star-eval} illustrate 6-node and 20-node star topologies that we use for performance evaluation. In the star topology, every node is connected to each other within two-hop communication; as a result, all nodes must desynchronize to use different time slots. 

\begin{figure*}
	\centering{
		\subfloat[M-DWARF]{\includegraphics[scale=0.27]{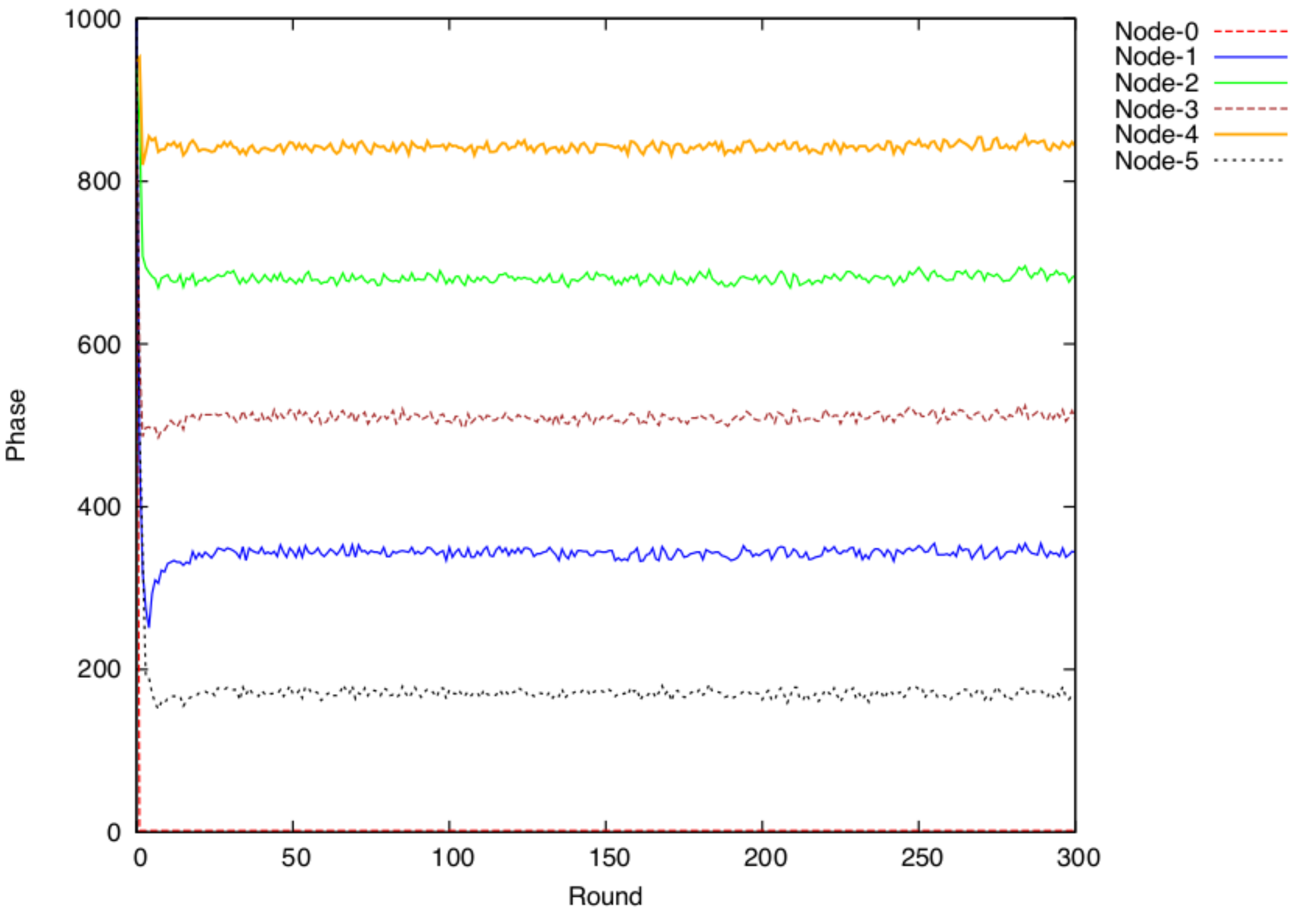}%
			\label{fig:6nodes-star-result-mdwarf-good}}
		\subfloat[EXT-DESYNC]{\includegraphics[scale=0.27]{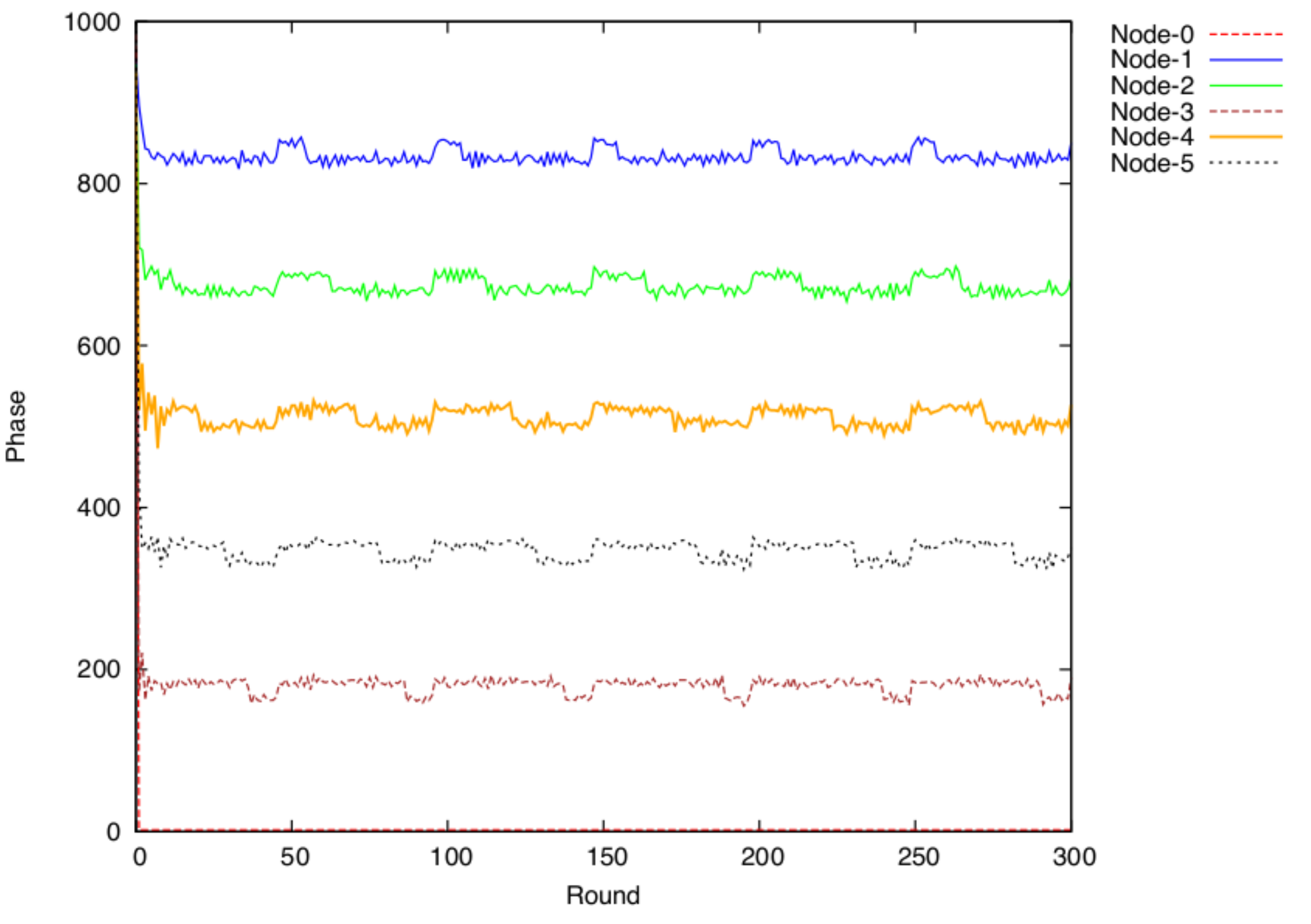}%
			\label{fig:6nodes-star-result-extdesync-good}}
		\subfloat[LIGHTWEIGHT]{\includegraphics[scale=0.27]{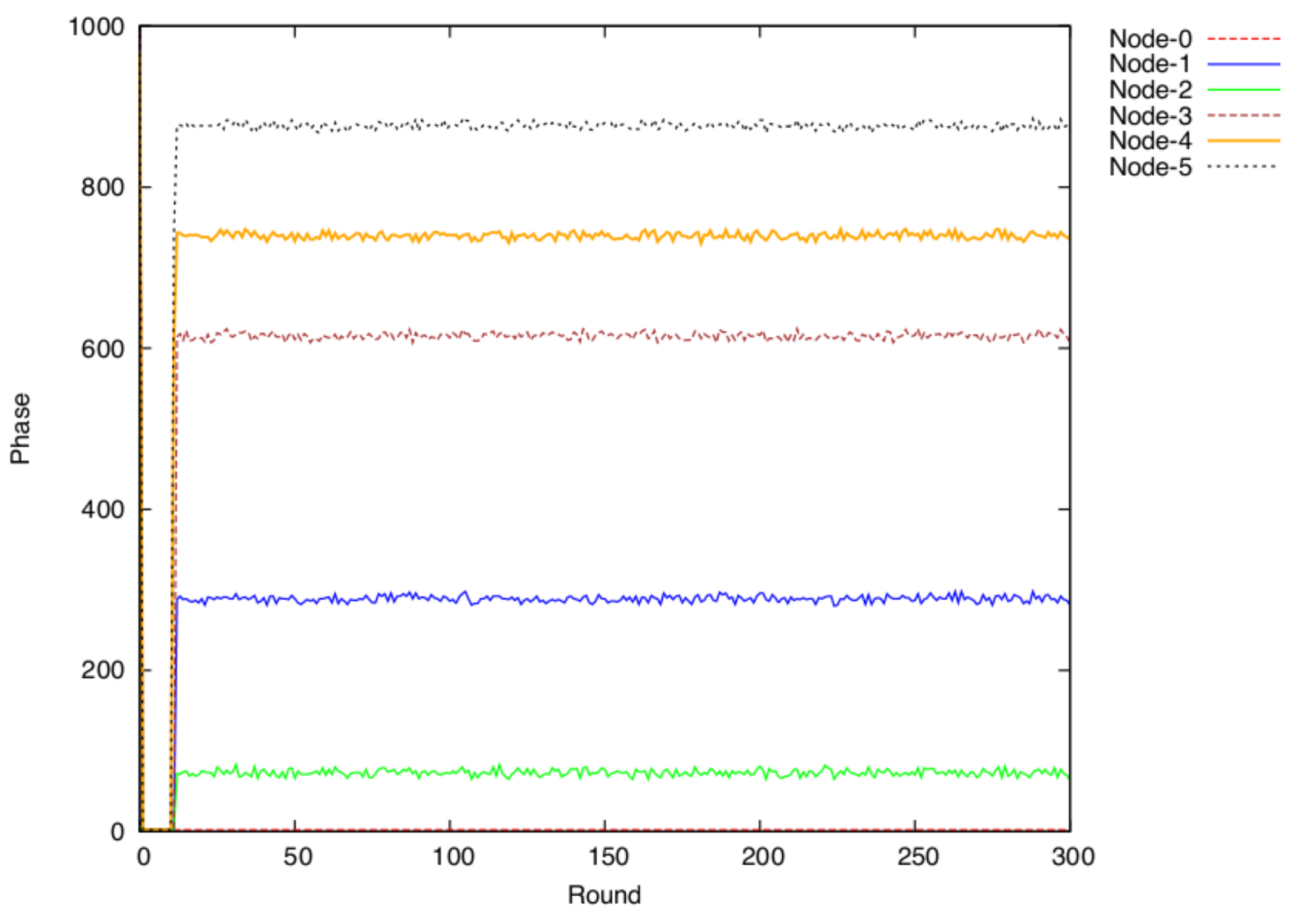}%
			\label{fig:6nodes-star-result-light-good}}
	}
	\caption{6-node star topology evaluation (average case).}
	\label{fig:6nodes-star-result-good}
	
\end{figure*}

\begin{figure*}
	\centering{
		\subfloat[M-DWARF]{\includegraphics[scale=0.27]{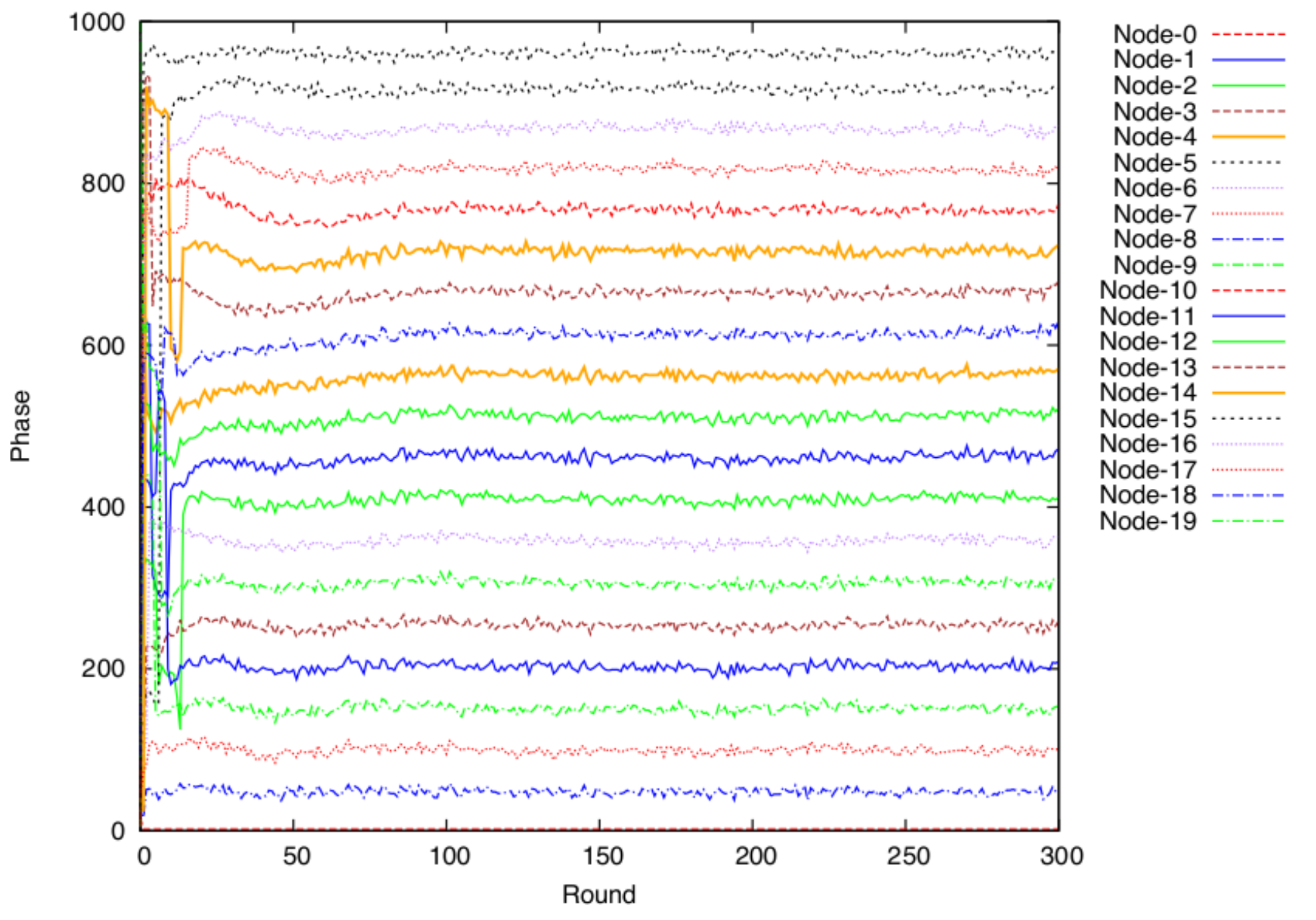}%
			\label{fig:20nodes-star-result-mdwarf-good}}
		\subfloat[EXT-DESYNC]{\includegraphics[scale=0.27]{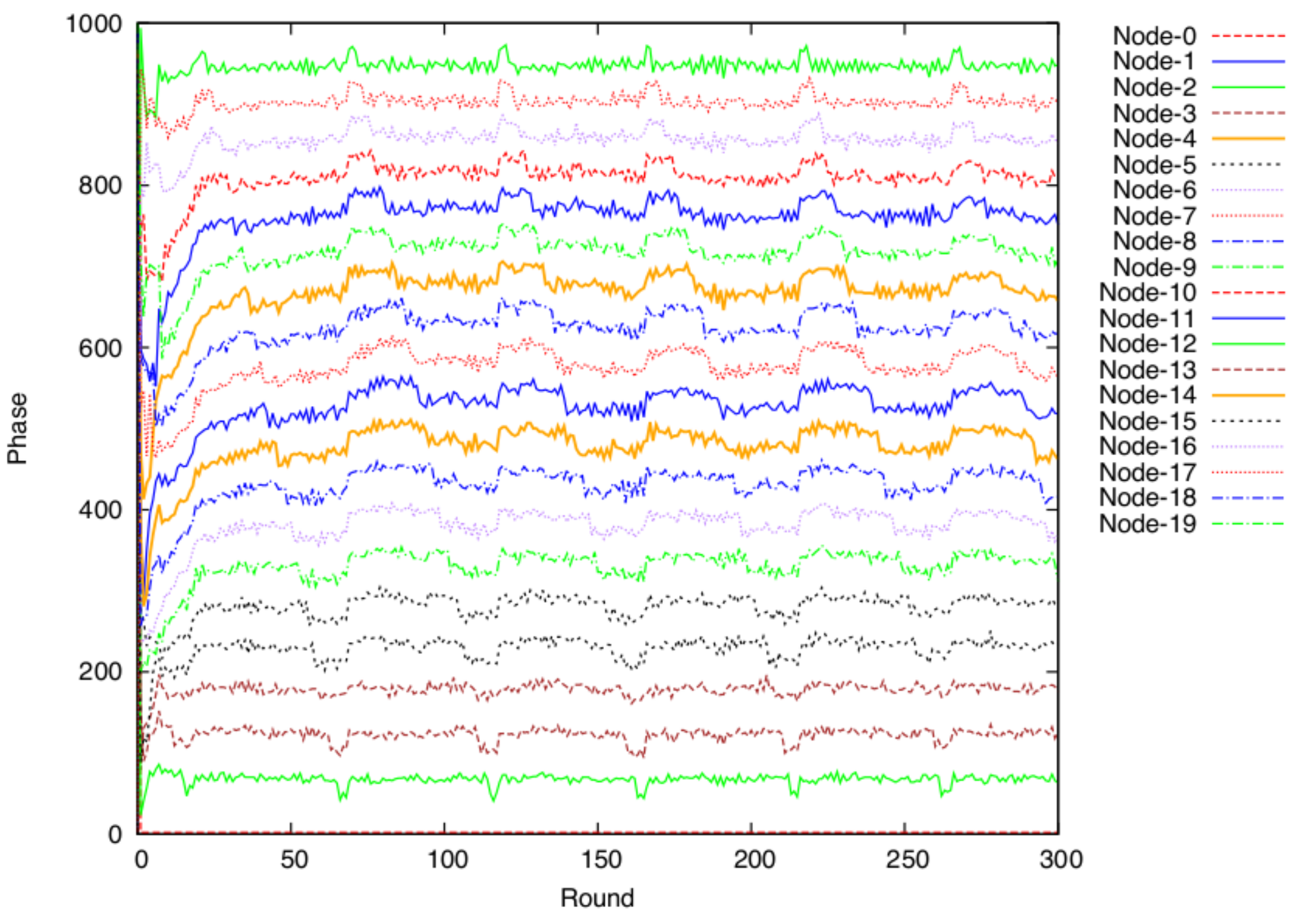}%
			\label{fig:20nodes-star-result-extdesync-good}}
		\subfloat[LIGHTWEIGHT]{\includegraphics[scale=0.27]{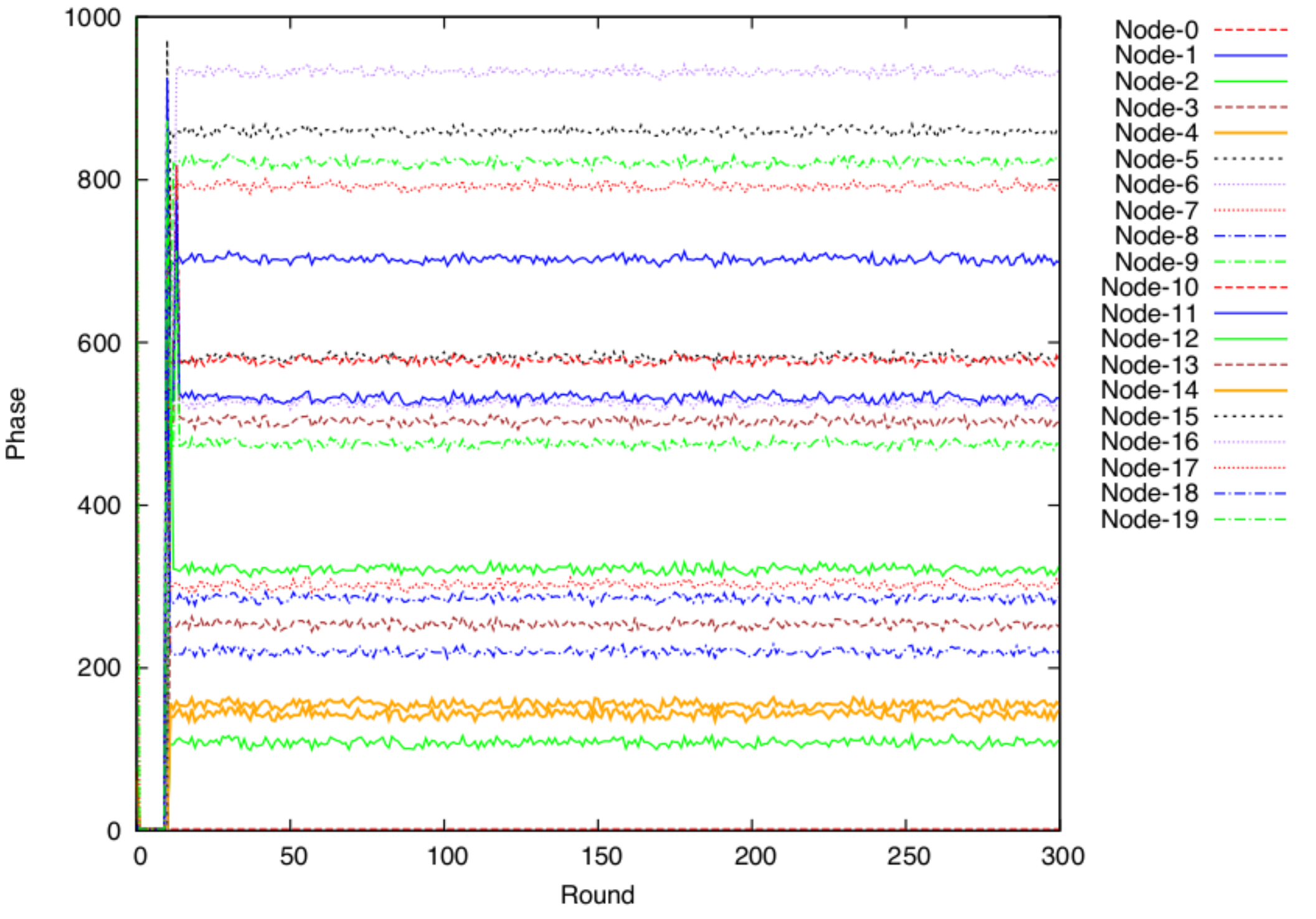}%
			\label{fig:20nodes-star-result-light-good}}
	}
	\caption{20-node star topology evaluation (average case).}
	\label{fig:20nodes-star-result-good}
	
\end{figure*}
Figure \ref{fig:6nodes-star-result-good} and \ref{fig:20nodes-star-result-good} shows the simulation results of three algorithms on 6-node and 20-node star topologies respectively.
Due to the relative phase relaying mechanism, both M-DWARF and EXT-DESYNC perceive relative phases of neighbors within two hops, which are all other nodes in the star topologies. Consequently, both algorithms can separate 6 nodes into almost equivalent 6 time slots. However, M-DWARF achieves almost perfect desynchony whereas EXT-DESYNC slightly fluctuates in the sparse 6-node network and highly fluctuates in the dense 20-node network.
\begin{figure*}
	\centering{
		\subfloat[M-DWARF]{\includegraphics[scale=0.27]{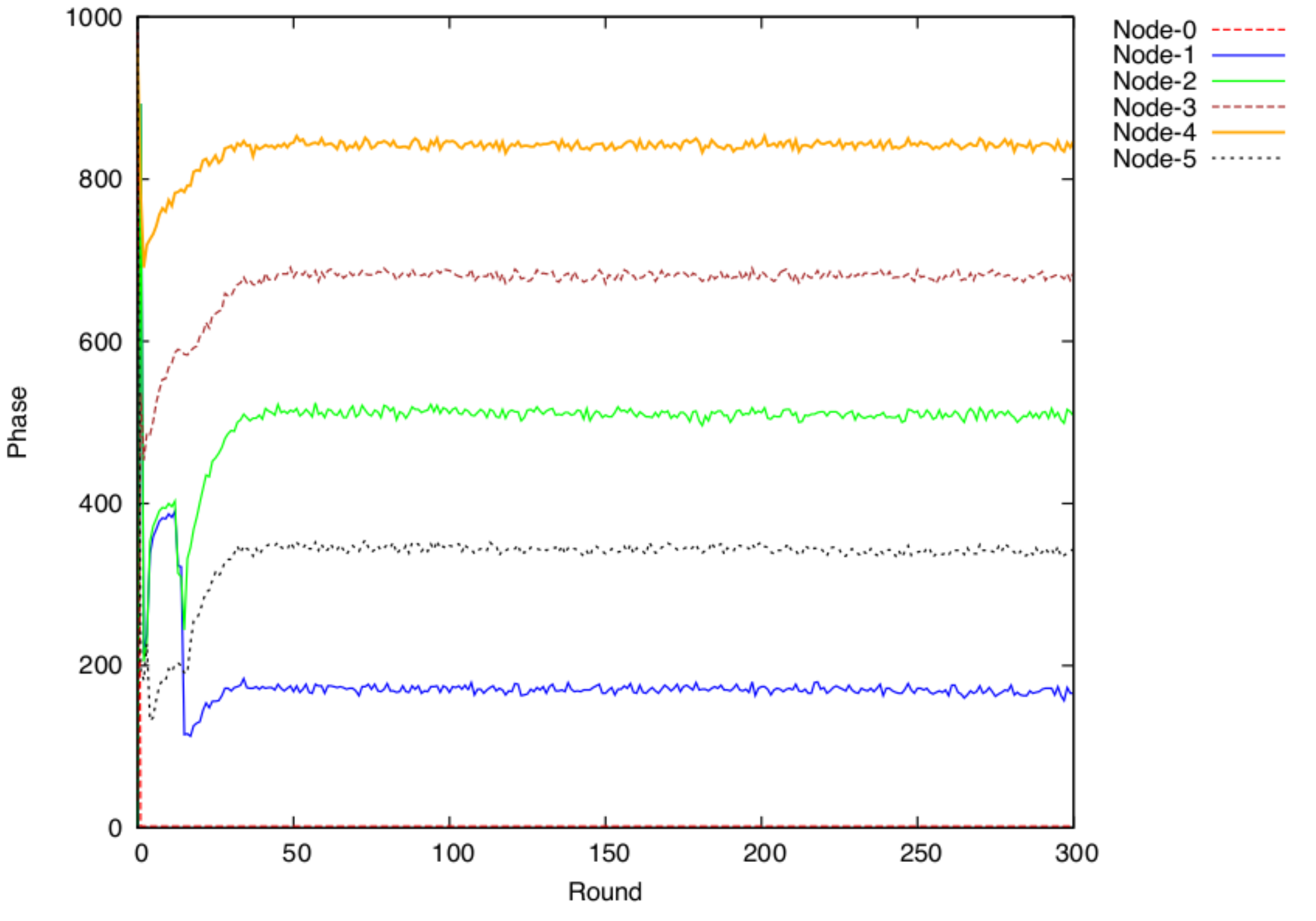}%
			\label{fig:6nodes-star-result-mdwarf-bad}}
		\subfloat[EXT-DESYNC]{\includegraphics[scale=0.27]{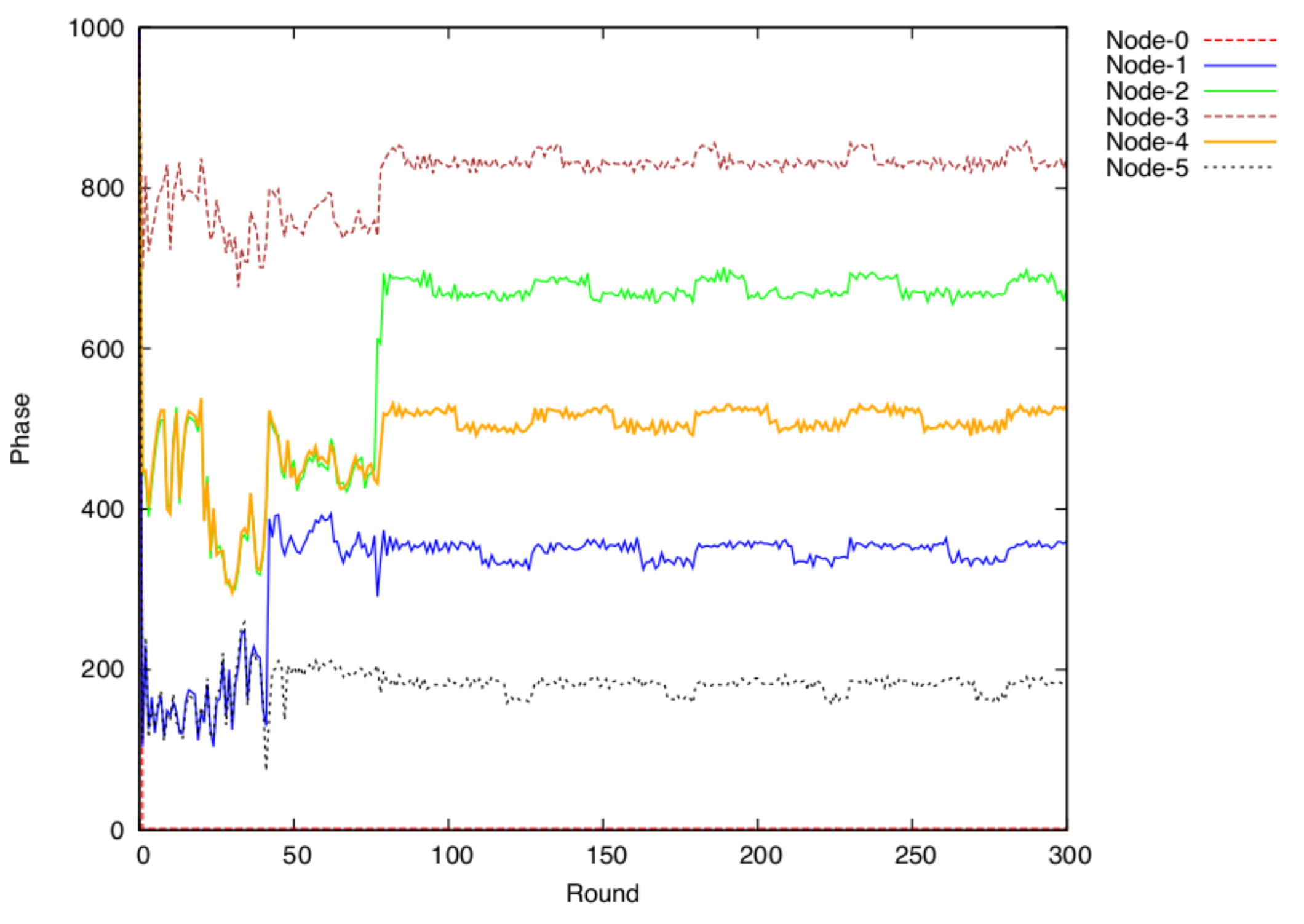}%
			\label{fig:6nodes-star-result-extdesync-bad}}
		\subfloat[LIGHTWEIGHT]{\includegraphics[scale=0.27]{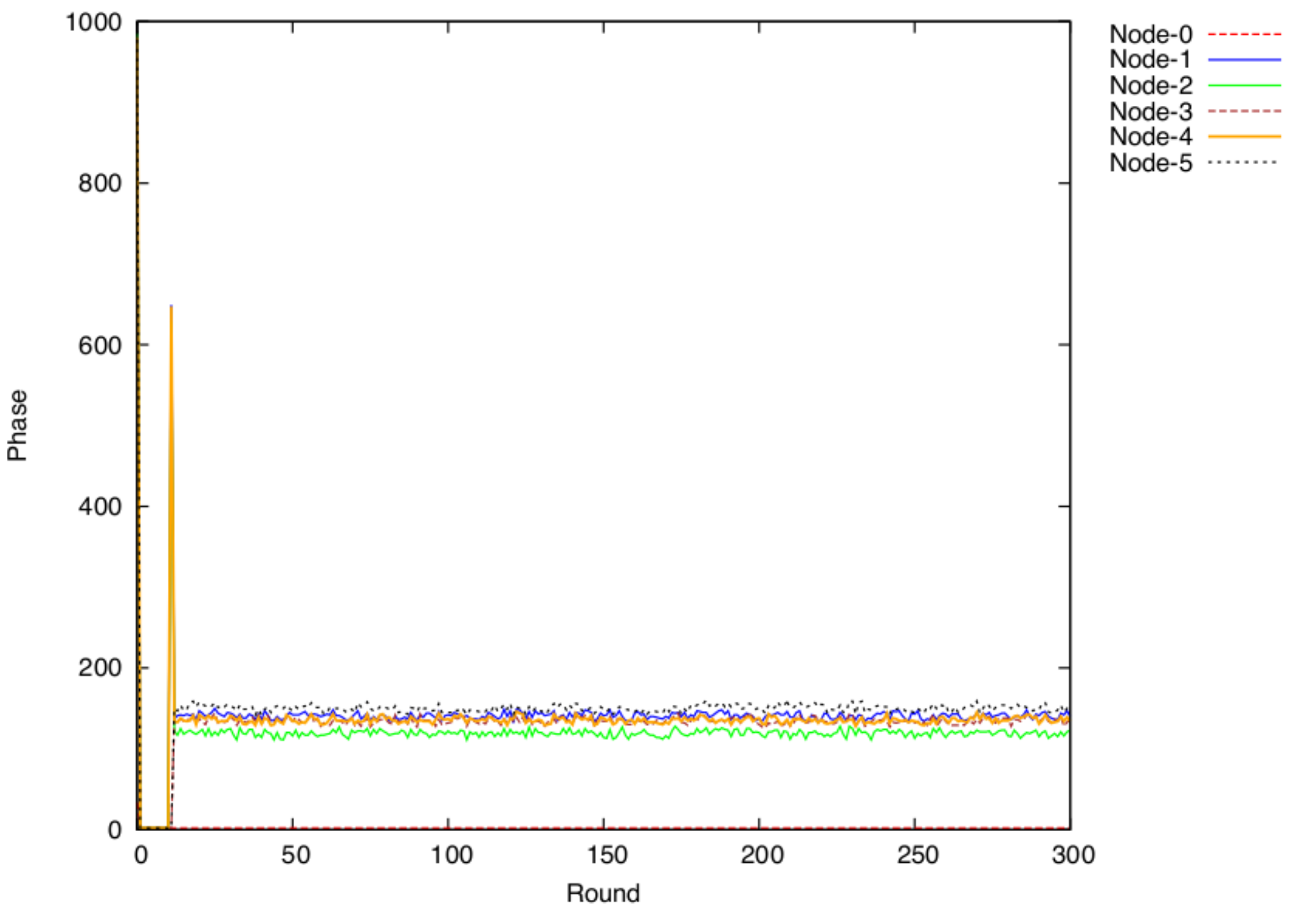}%
			\label{fig:6nodes-star-result-light-bad}}
	}
	\caption{6-node star topology evaluation (problematic case).}
	\label{fig:6nodes-star-result-bad}
	
\end{figure*}
\begin{figure*}
	\centering{
		\subfloat[M-DWARF]{\includegraphics[scale=0.27]{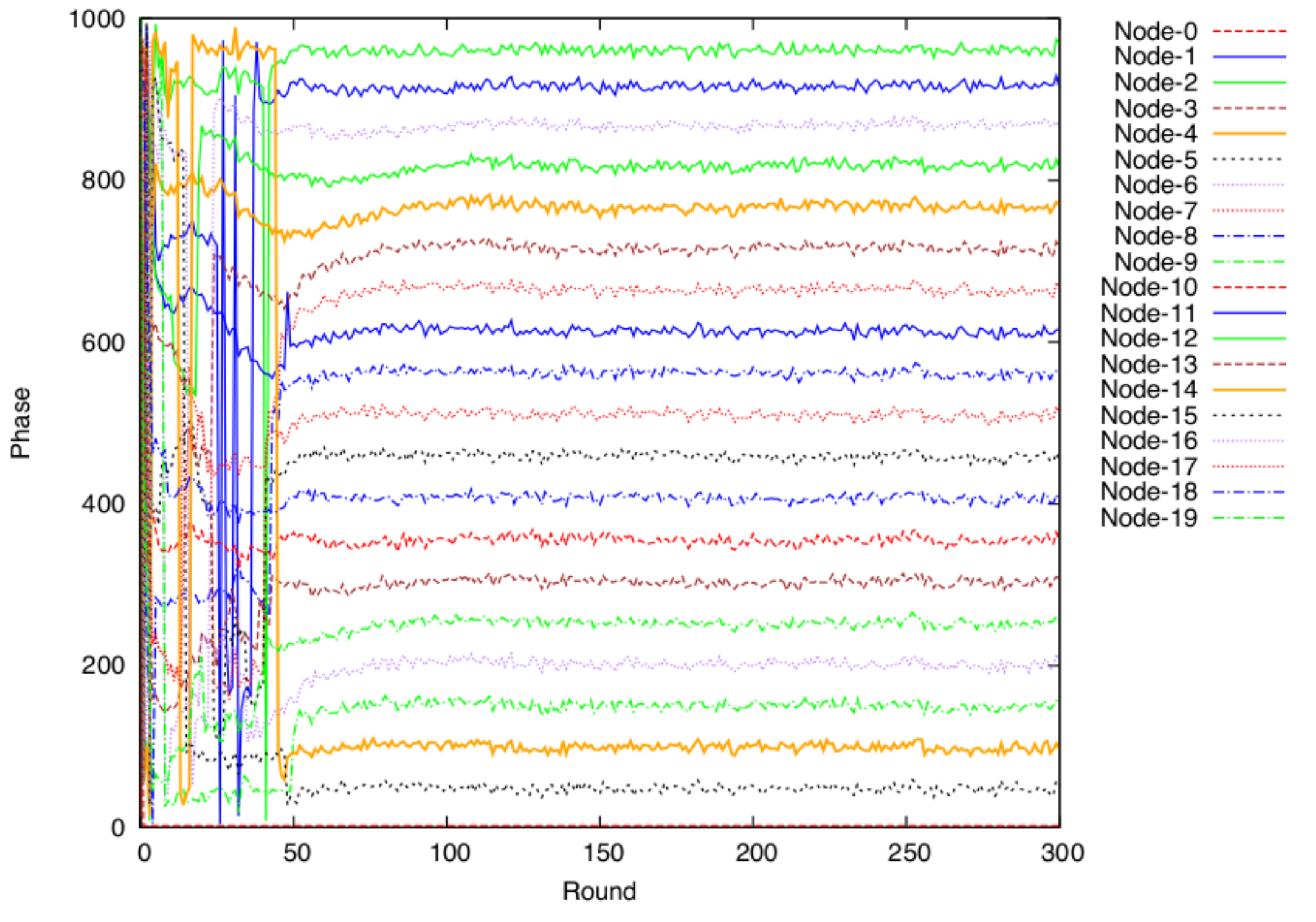}%
			\label{fig:20nodes-star-result-mdwarf-bad}}
		\subfloat[EXT-DESYNC]{\includegraphics[scale=0.27]{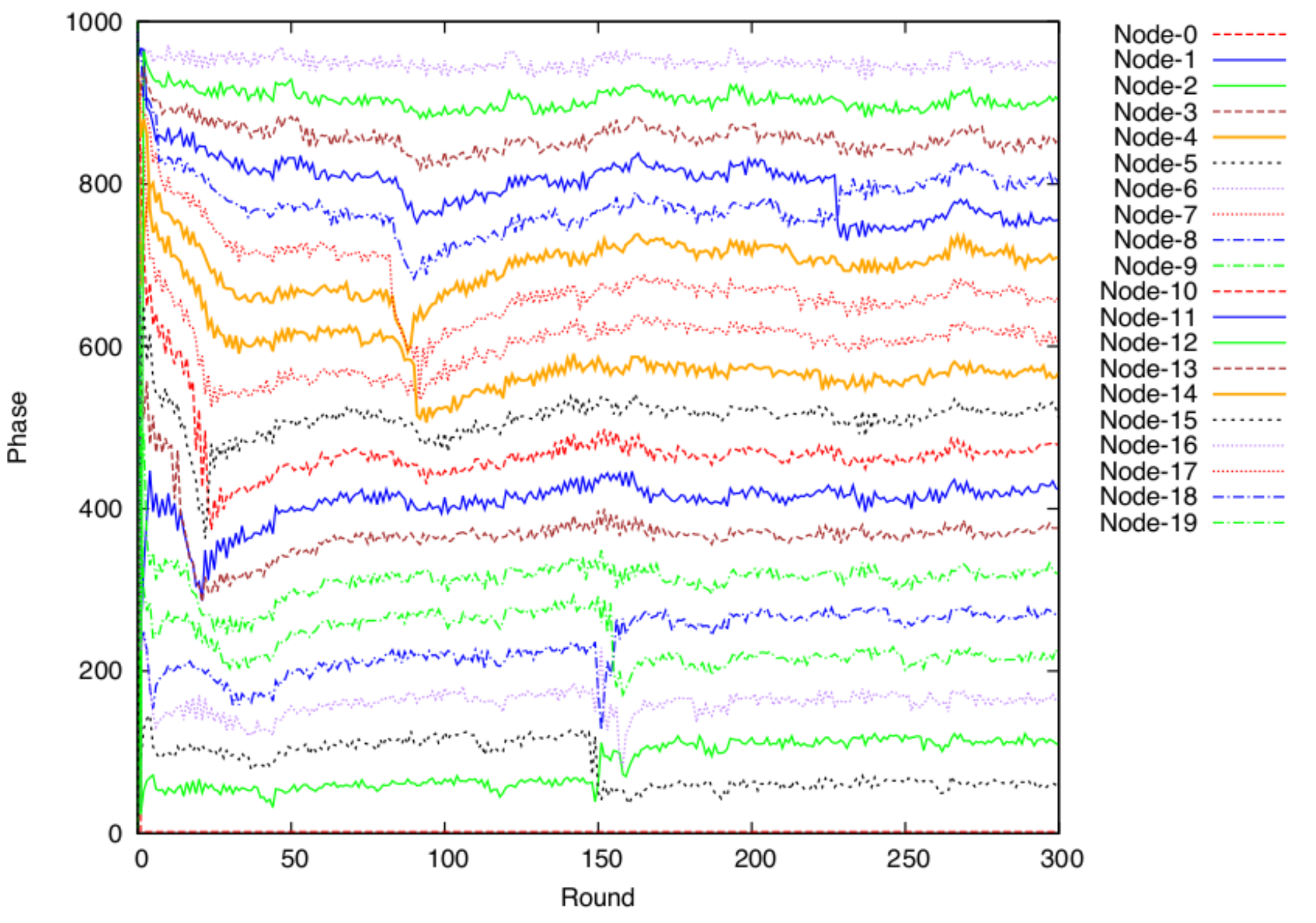}%
			\label{fig:20nodes-star-result-extdesync-bad}}
		\subfloat[LIGHTWEIGHT]{\includegraphics[scale=0.27]{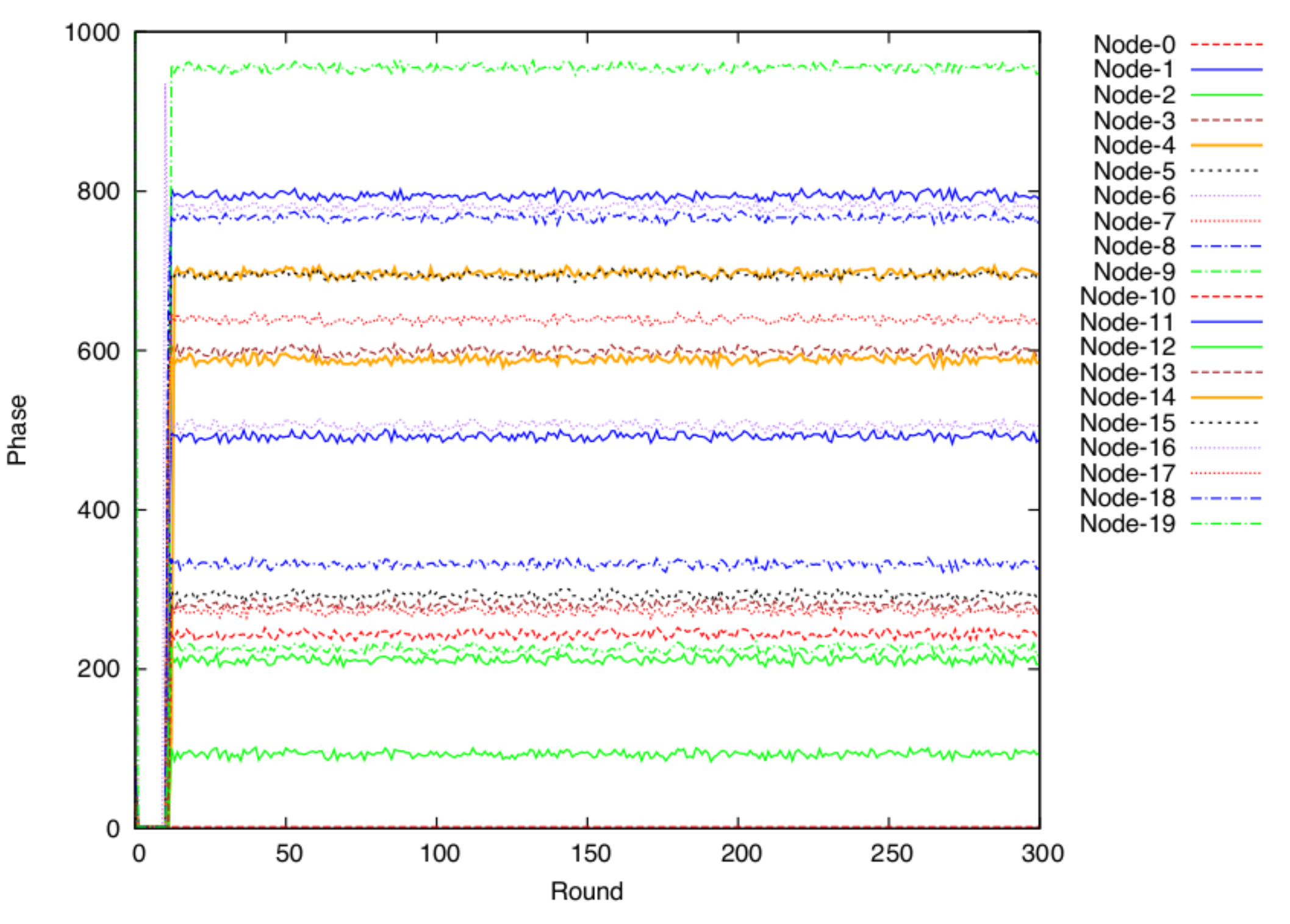}%
			\label{fig:20nodes-star-result-light-bad}}
	}
	\caption{20-node star topology evaluation (problematic case).}
	\label{fig:20nodes-star-result-bad}
	
\end{figure*}
Figure \ref{fig:6nodes-star-result-bad} and \ref{fig:20nodes-star-result-bad} shows the problematic cases. For M-DWARF, there is no problem in the sparse networks, but, in the dense networks, nodes take more time to become stable. The reason is that, when the network is dense, the usable time interval gap for each node is shortened. Therefore, if several nodes start closely to each other at the initial configuration, their colliding packets make them initially become unaware of each other and their highly absorbed forces demand more time to separate away from each other and desynchronize. In the case of EXT-DESYNC, because each node relies on only information of two phase neighbors, the error from one node can propagate throughout the networks and cause all the nodes' phases to fluctuate all the time.
\paragraph{Chain Topology}
The chain topology is one of the simplest multi-hop topologies. Every node is lined up without a loop. The 3-node and 10-node chain topologies are depicted in Figure \ref{fig:3nodes-chain-eval} and \ref{fig:10nodes-chain-eval}. 

\begin{figure*}
	\centering{
		\subfloat[M-DWARF]{\includegraphics[scale=0.27]{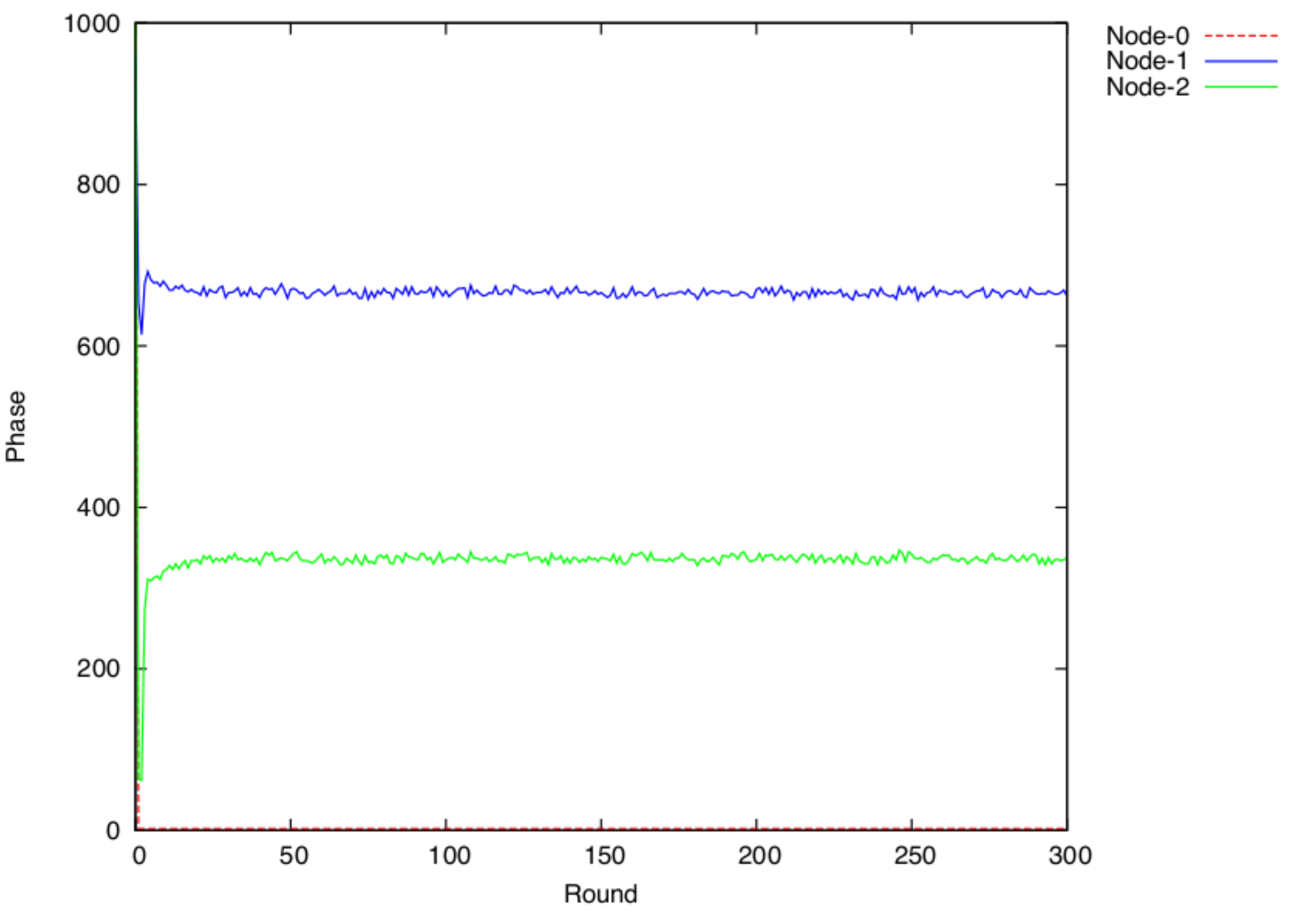}%
			\label{fig:3nodes-chain-result-mdwarf-good}}
		\subfloat[EXT-DESYNC]{\includegraphics[scale=0.27]{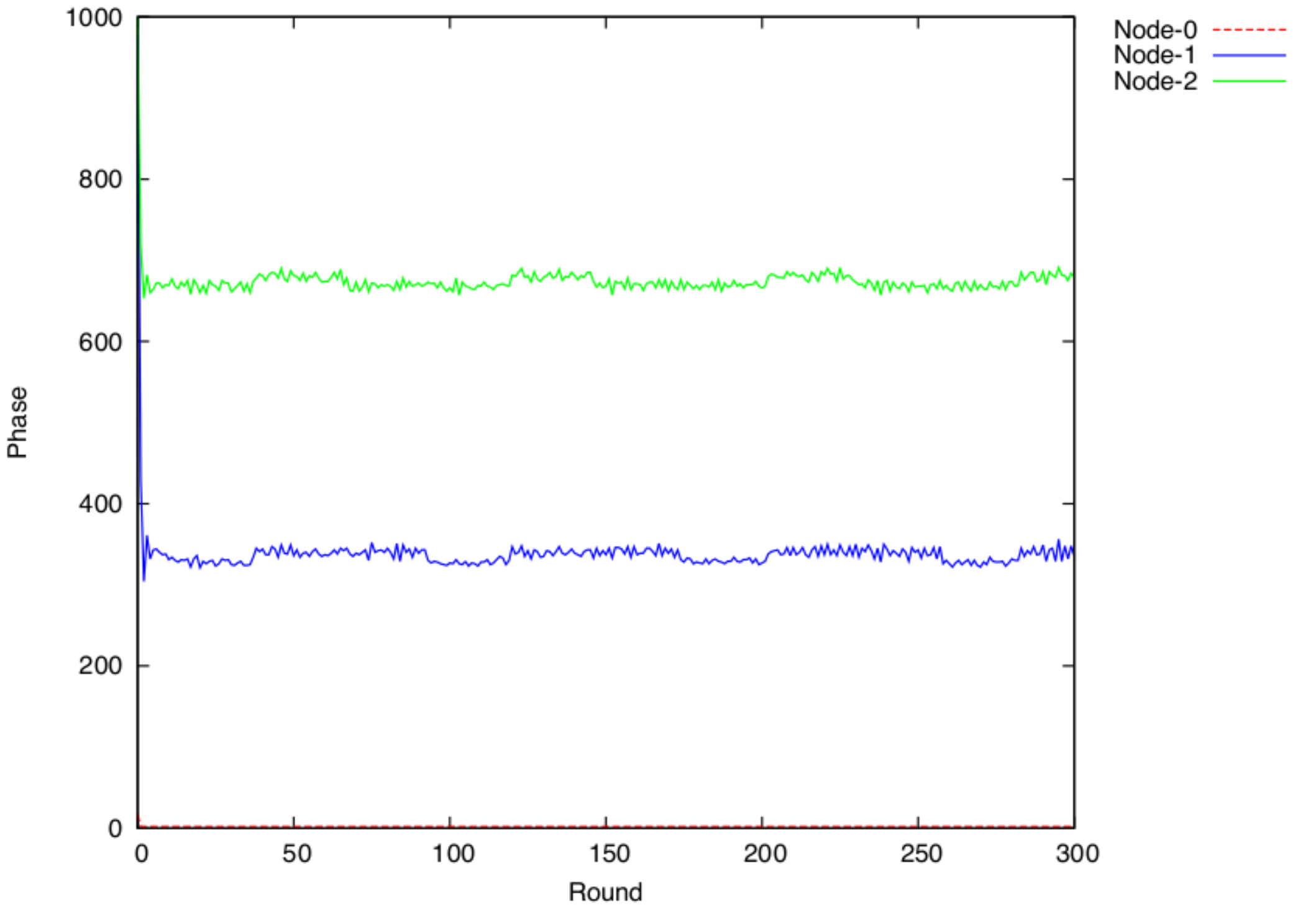}%
			\label{fig:3nodes-chain-result-extdesync-good}}
		\subfloat[LIGHTWEIGHT]{\includegraphics[scale=0.27]{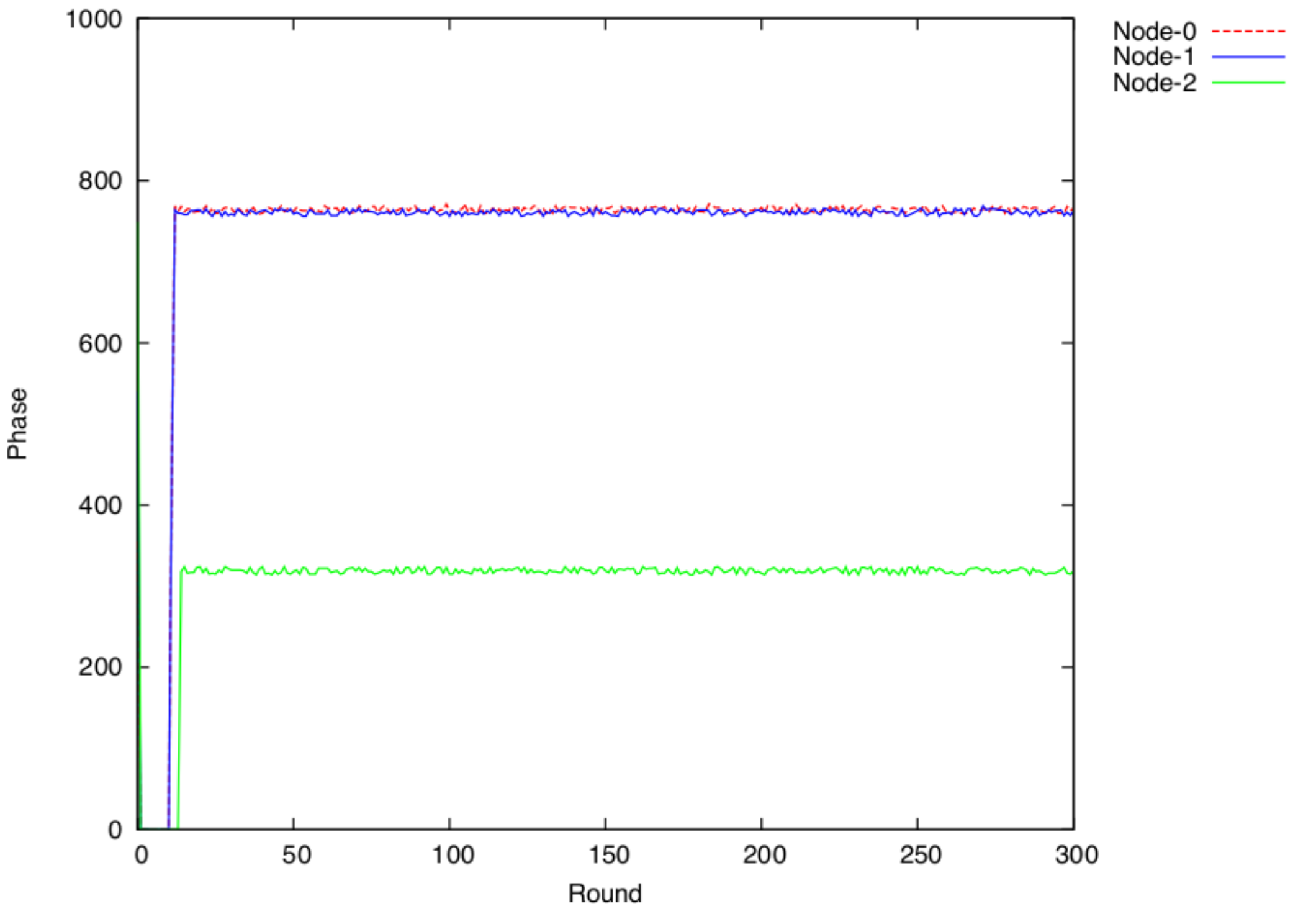}%
			\label{fig:3nodes-chain-result-light-good}}
	}
	\caption{3-node chain topology evaluation (average case).}
	\label{fig:3nodes-chain-result-good}
	
\end{figure*}
\begin{figure*}
	\centering{
		\subfloat[M-DWARF]{\includegraphics[scale=0.27]{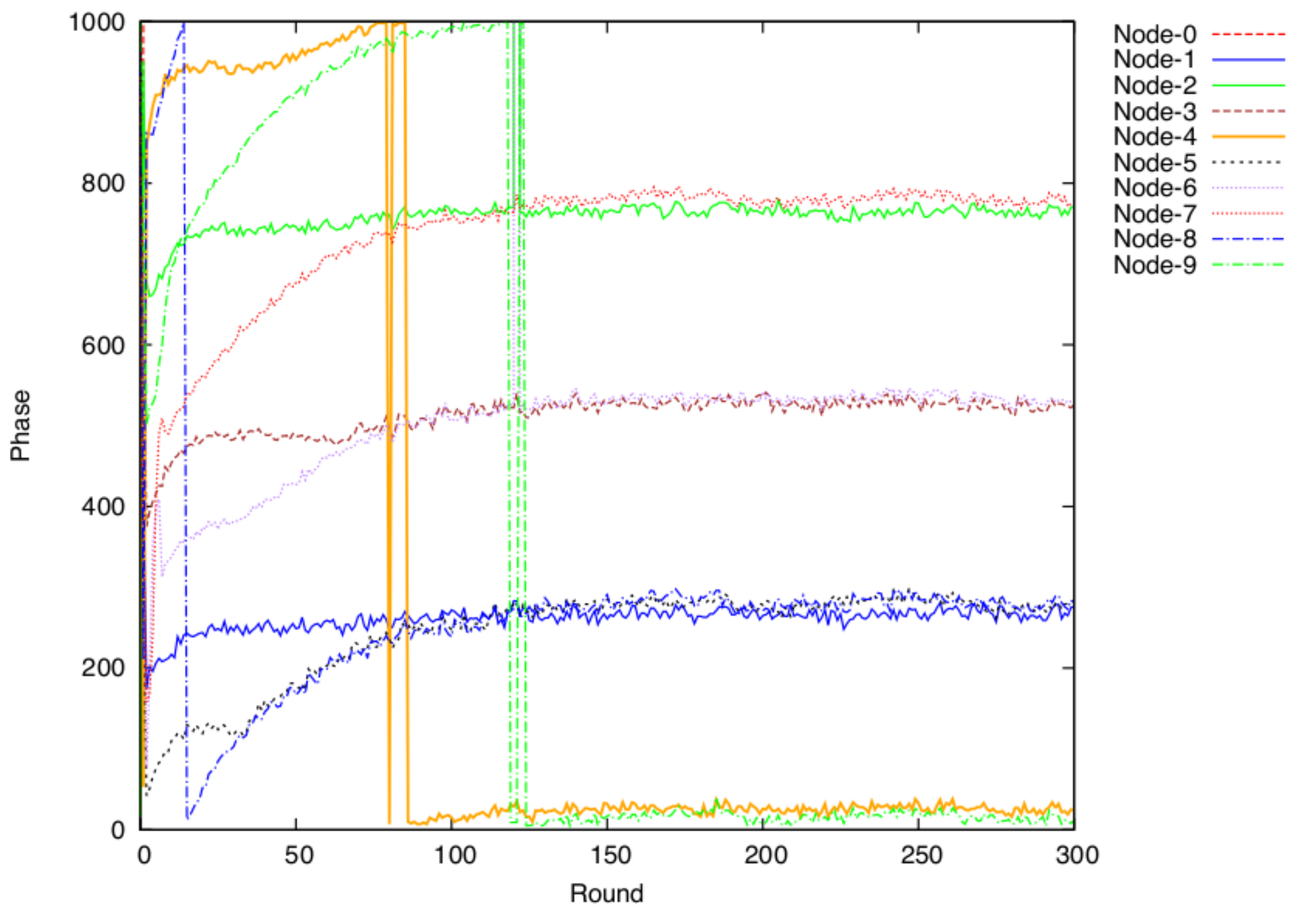}%
			\label{fig:10nodes-chain-result-mdwarf-good}}
		\subfloat[EXT-DESYNC]{\includegraphics[scale=0.27]{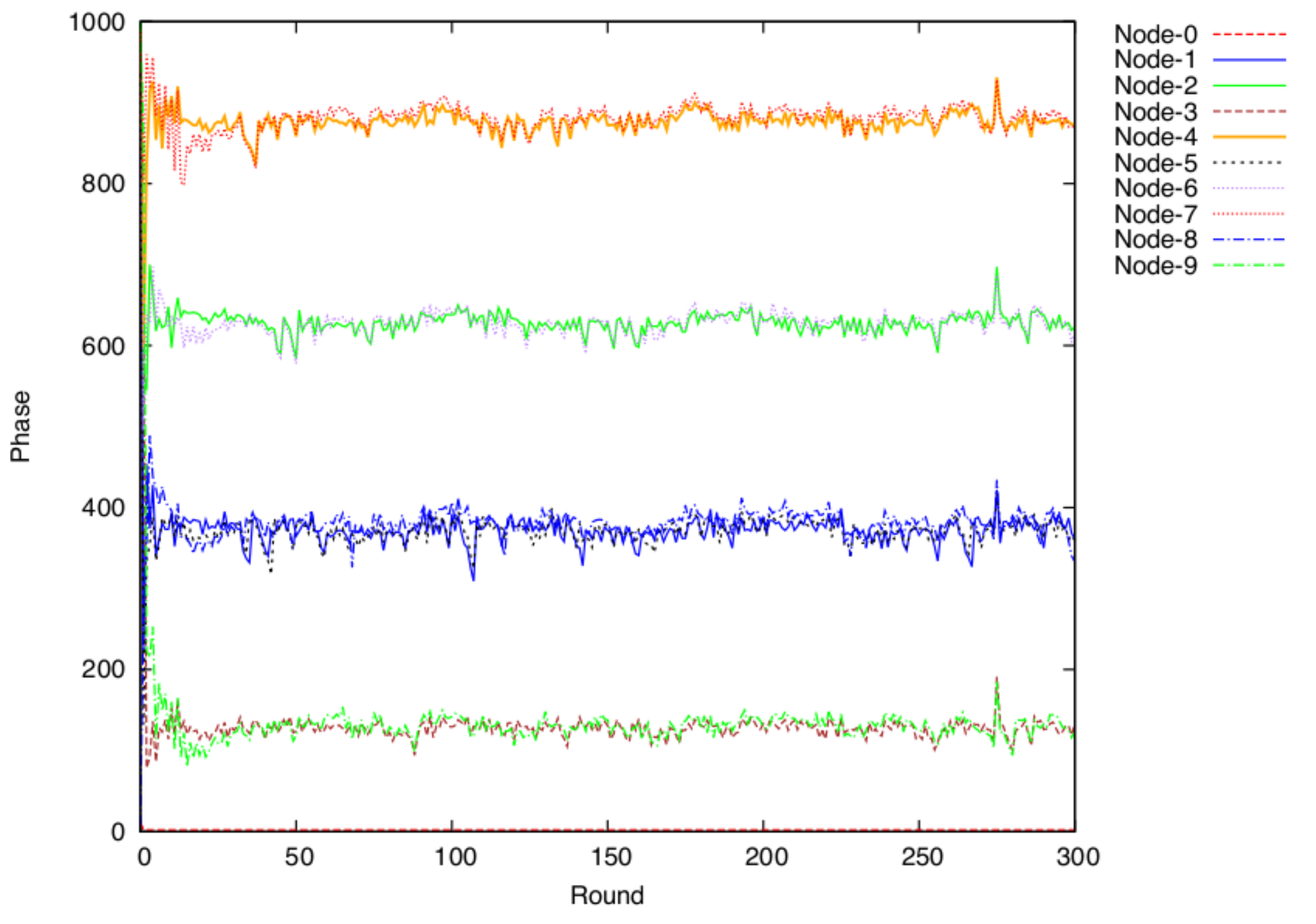}%
			\label{fig:10nodes-chain-result-extdesync-good}}
		\subfloat[LIGHTWEIGHT]{\includegraphics[scale=0.27]{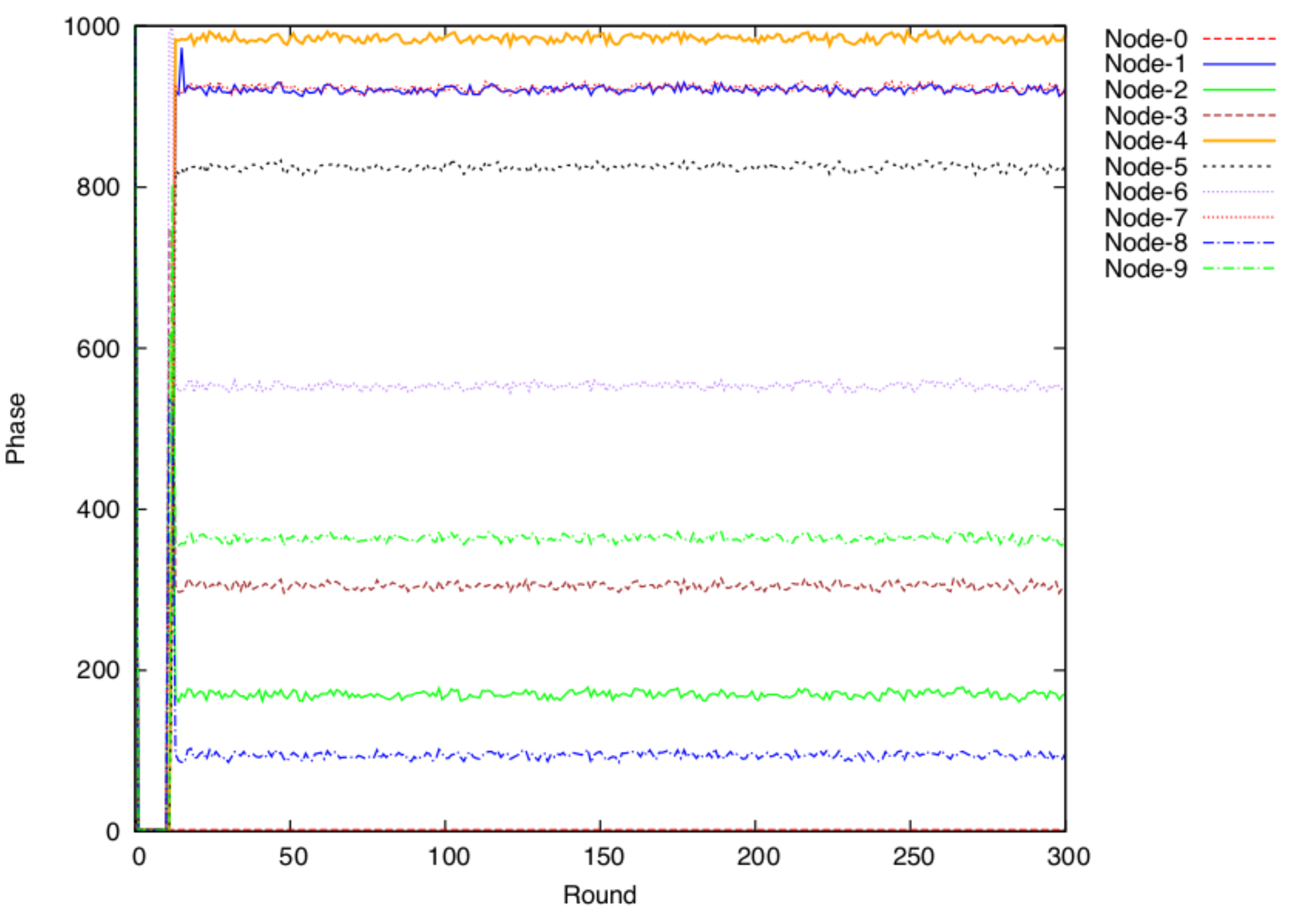}%
			\label{fig:10nodes-chain-result-light-good}}
	}
	\caption{10-node chain topology evaluation (average case).}
	\label{fig:10nodes-chain-result-good}
	
\end{figure*}
The case of 3-node chain topologies is simple because all three nodes simply cannot share the same phase.
From the simulation results shown in Figure \ref{fig:3nodes-chain-result-good}, both M-DWARF and EXT-DESYNC perform very well in terms of fairness because all nodes' phases are equivalently separated. However, the nodes' phases in M-DWARF is slightly more stable than those in EXT-DESYNC. 

For 10-node topologies, some nodes that are farther than two hops are able to occupy the same phase position. There are several optimal solutions. For example, the first optimal solution is that there are four groups of nodes that use different phases. Each node within the same group can use the same phase. These four groups are $\lbrace 0, 3, 6, 9 \rbrace , \lbrace 1, 4, 7 \rbrace, \lbrace 2, 5, 8 \rbrace$, and $\lbrace 6 \rbrace$. Another optimal solution is $\lbrace 0, 4, 9 \rbrace, \lbrace 1, 5, 8 \rbrace, \lbrace 3, 6 \rbrace,$ and $\lbrace 2, 7 \rbrace$.
We note that, in our 30 simulations, M-DWARF could achieve the perfect desynchrony state with 4 slots as in Figure \ref{fig:10nodes-chain-result-mdwarf-good}, but EXTENDED-DEESYNC never achieve the perfect desynchrony state (there are 5 slots in Figure \ref{fig:10nodes-chain-result-extdesync-good} because node 0 is at phase 0).

\begin{figure*}
	\centering{
		\subfloat[M-DWARF]{\includegraphics[scale=0.27]{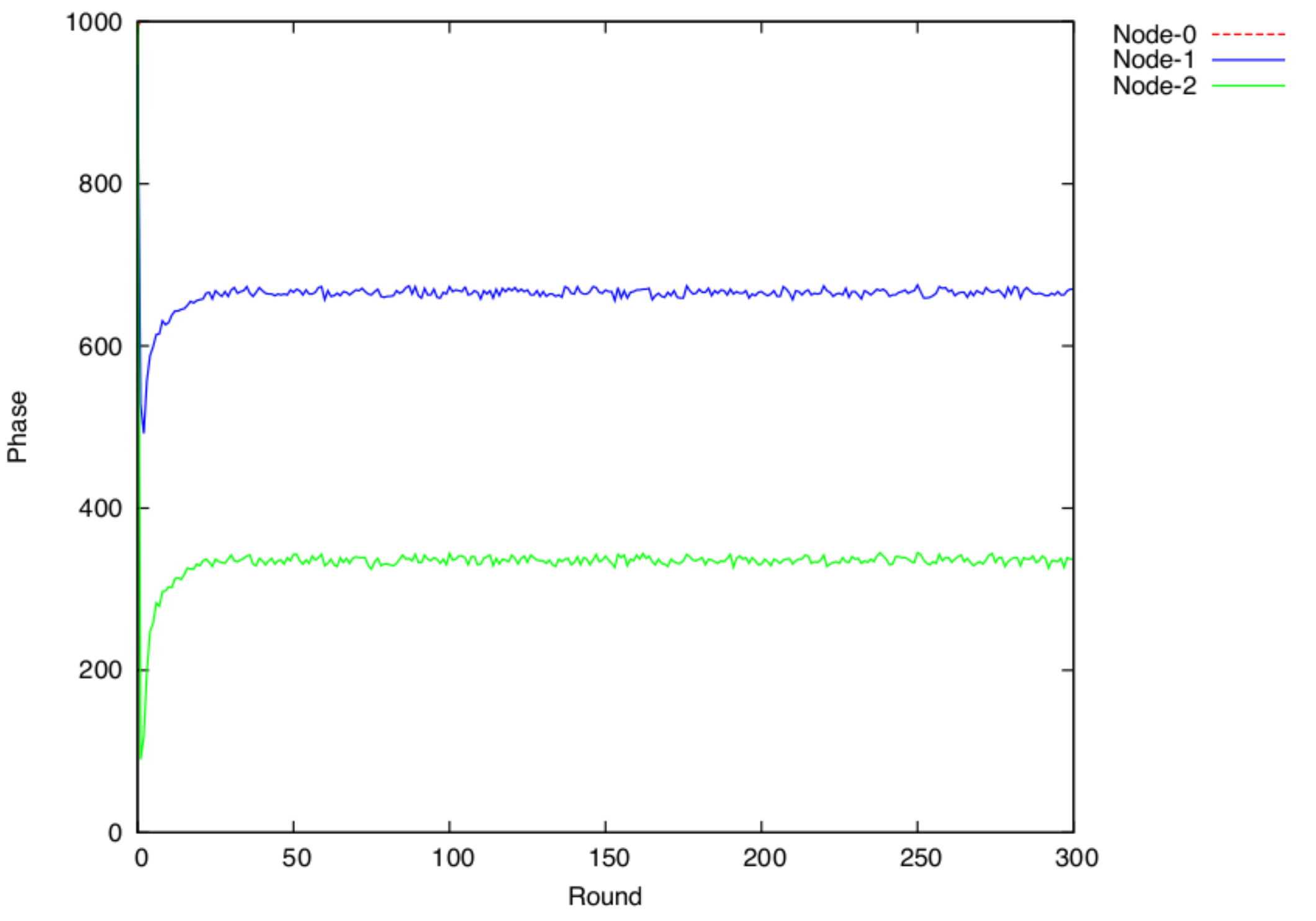}%
			\label{fig:3nodes-chain-result-mdwarf-bad}}
		\subfloat[EXT-DESYNC]{\includegraphics[scale=0.27]{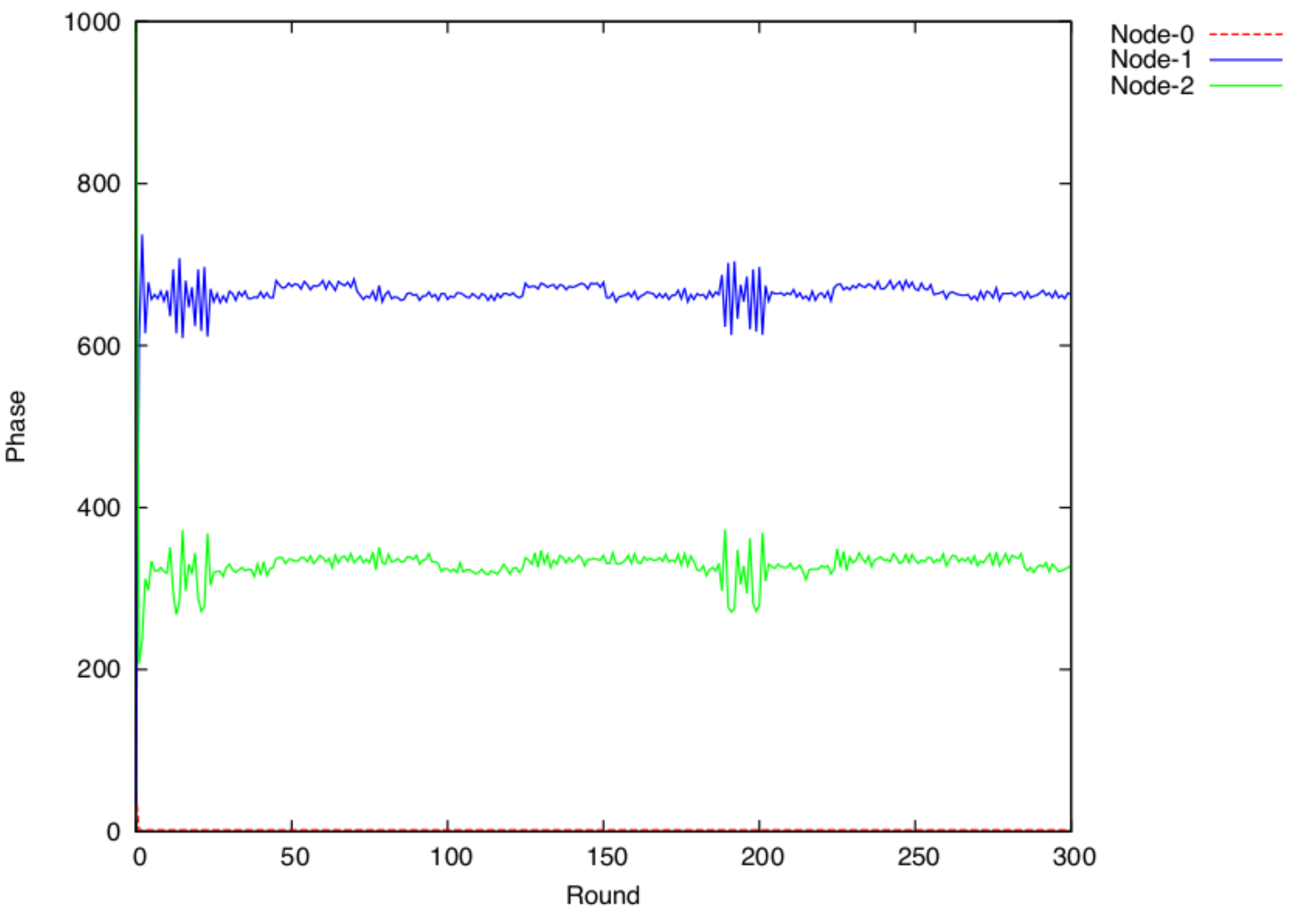}%
			\label{fig:3nodes-chain-result-extdesync-bad}}
		\subfloat[LIGHTWEIGHT]{\includegraphics[scale=0.27]{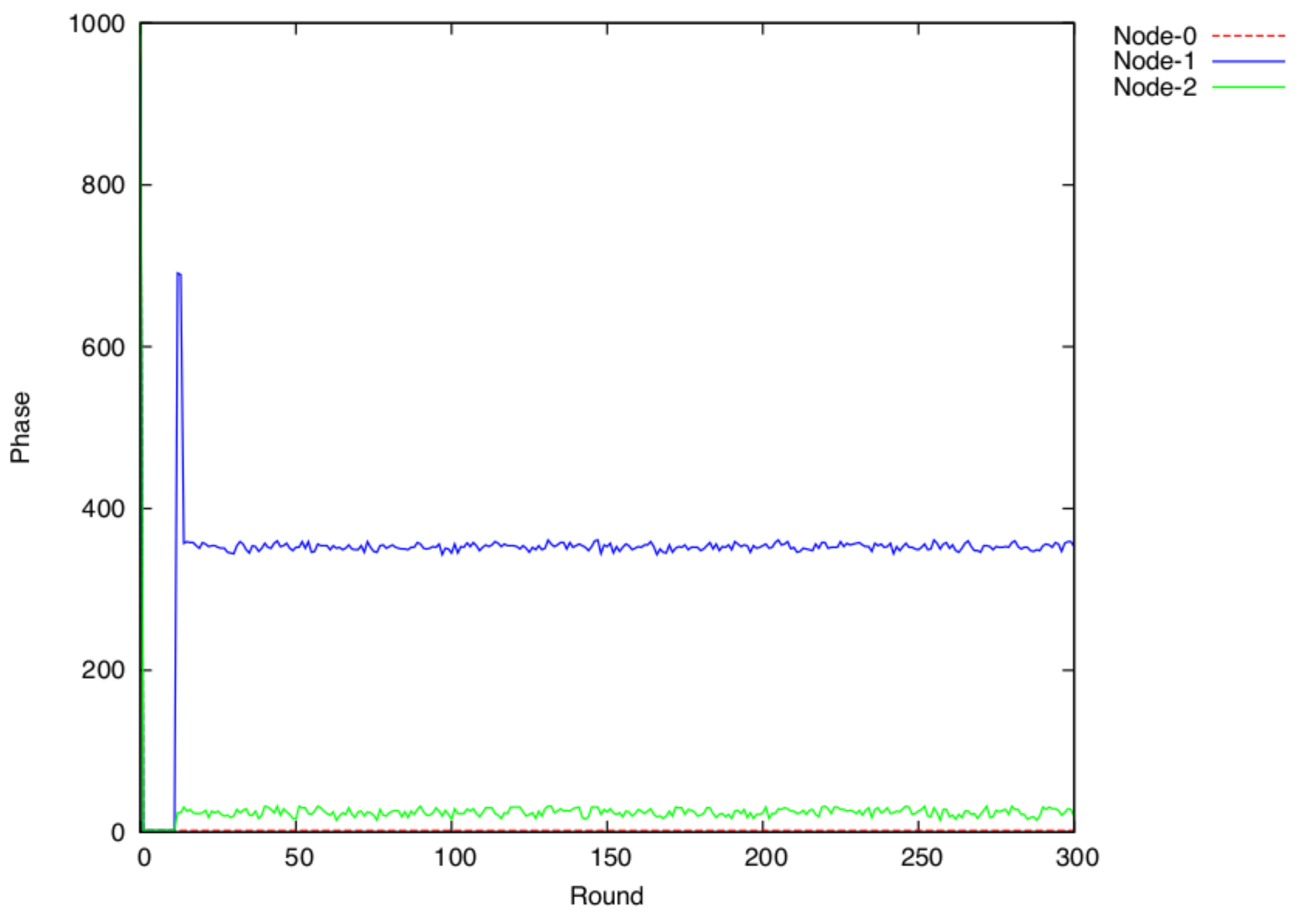}%
			\label{fig:3nodes-chain-result-light-bad}}
	}
	\caption{3-node chain topology evaluation (problematic case).}
	\label{fig:3nodes-chain-result-bad}
	
\end{figure*}
\begin{figure*}
	\centering{
		\subfloat[M-DWARF]{\includegraphics[scale=0.27]{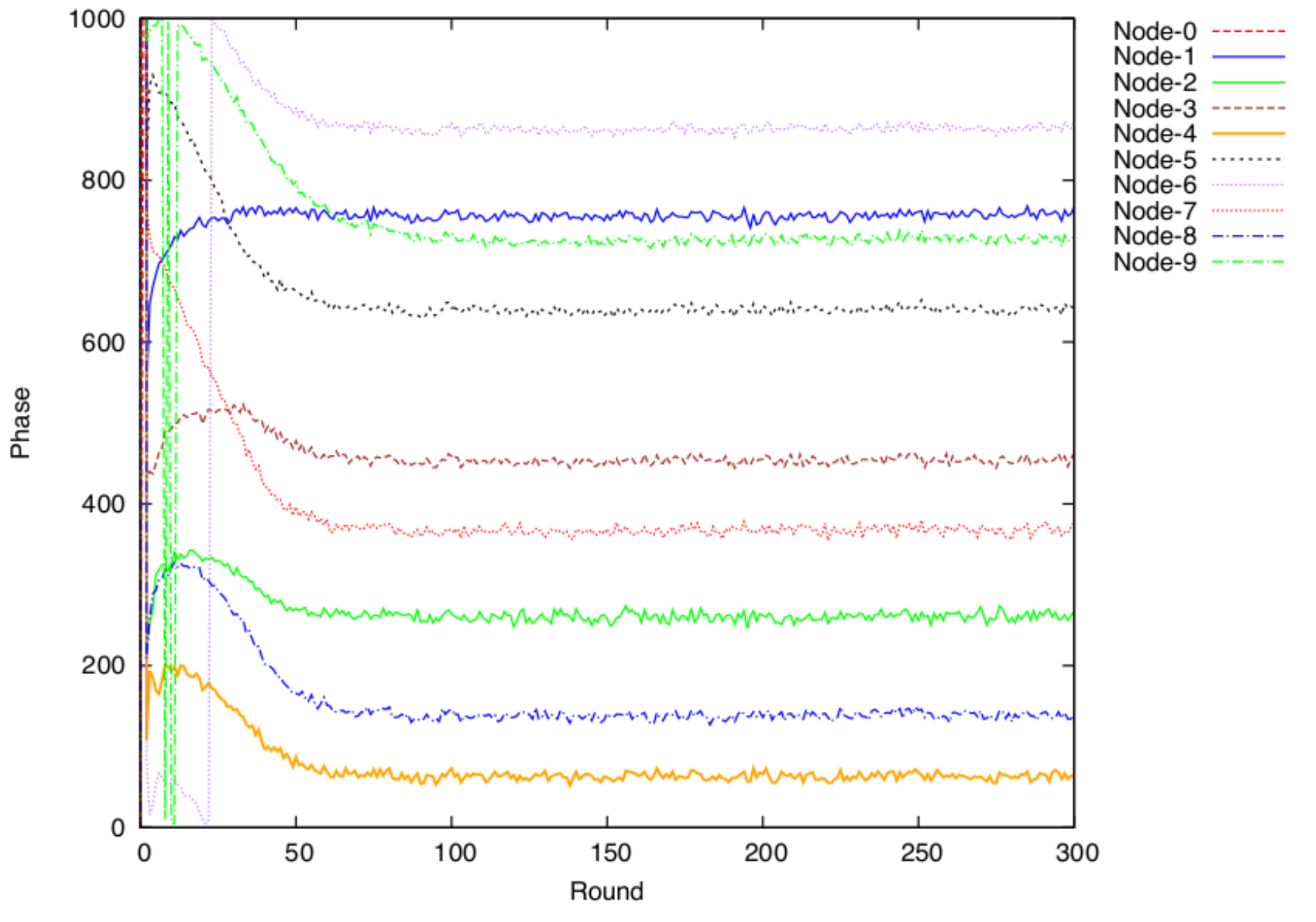}%
			\label{fig:10nodes-chain-result-mdwarf-bad}}
		\subfloat[EXT-DESYNC]{\includegraphics[scale=0.27]{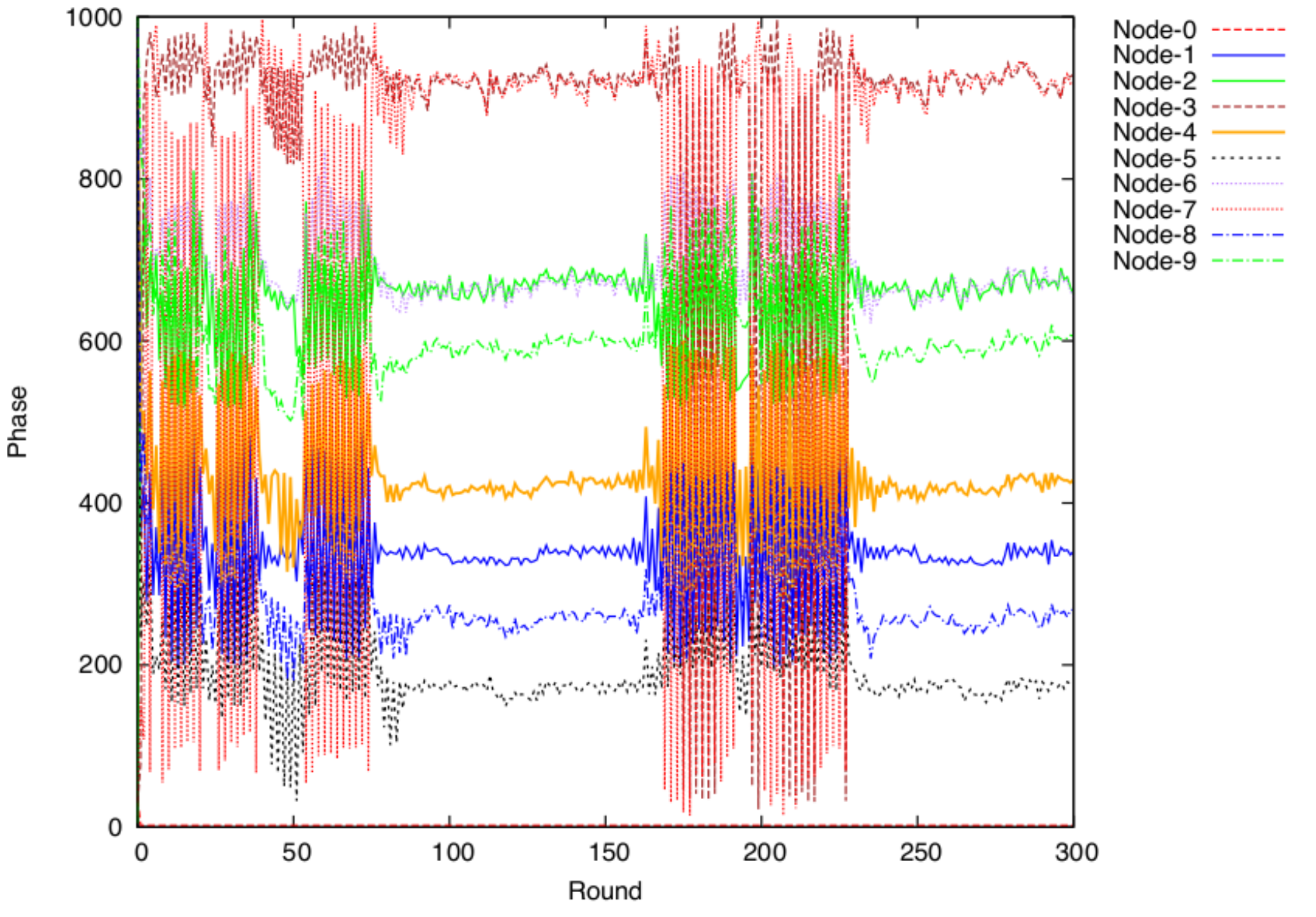}%
			\label{fig:10nodes-chain-result-extdesync-bad}}
		\subfloat[LIGHTWEIGHT]{\includegraphics[scale=0.27]{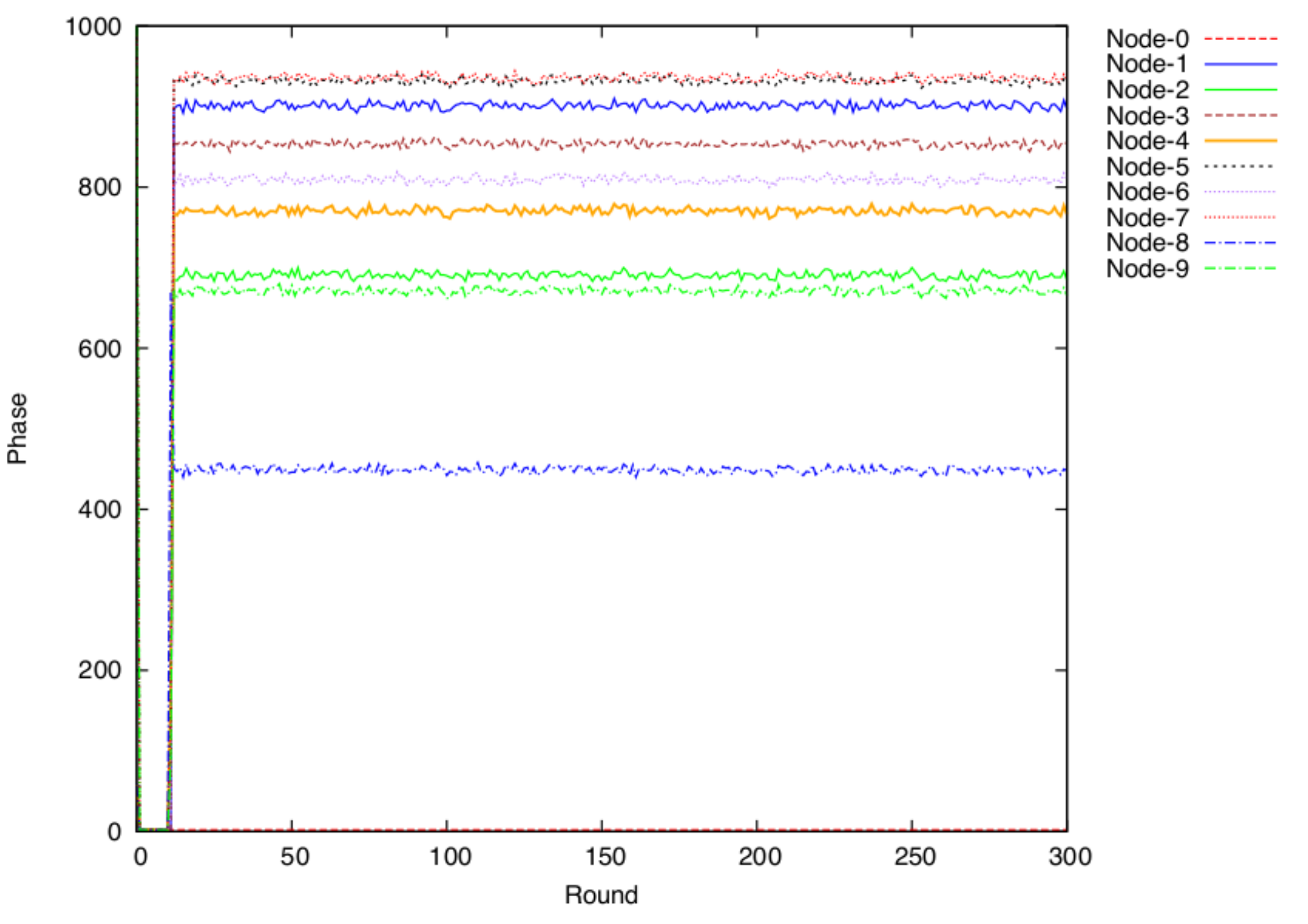}%
			\label{fig:10nodes-chain-result-light-bad}}
	}
	\caption{10-node chain topology evaluation (problematic case).}
	\label{fig:10nodes-chain-result-bad}
	
\end{figure*}
In some problematic cases, the phases of the EXT-DESYNC slightly fluctuate in the sparse networks (Figure \ref{fig:3nodes-chain-result-extdesync-bad}) and highly fluctuate in the dense networks (Figure \ref{fig:10nodes-chain-result-extdesync-bad}). Additionally, for both EXT-DESYNC and M-DWARF, each node attempts to gradually adjust its phase in the range between previous and next phase neighbors. Therefore, if the nodes' initial phrases are not proper, they may be not able to adjust their phases over the range of such neighbors and the networks cannot get into the perfect desynchrony state.

\paragraph{Cycle Topology}
The cycle topology is simlar to the chain topology. All nodes are lined up but the first and last nodes are connected together to form a closed loop. Figure \ref{fig:4nodes-cycle-eval} and \ref{fig:10nodes-cycle-eval} depict 4-node and 10-node cycle topologies.

\begin{figure*}
	\centering{
		\subfloat[M-DWARF]{\includegraphics[scale=0.27]{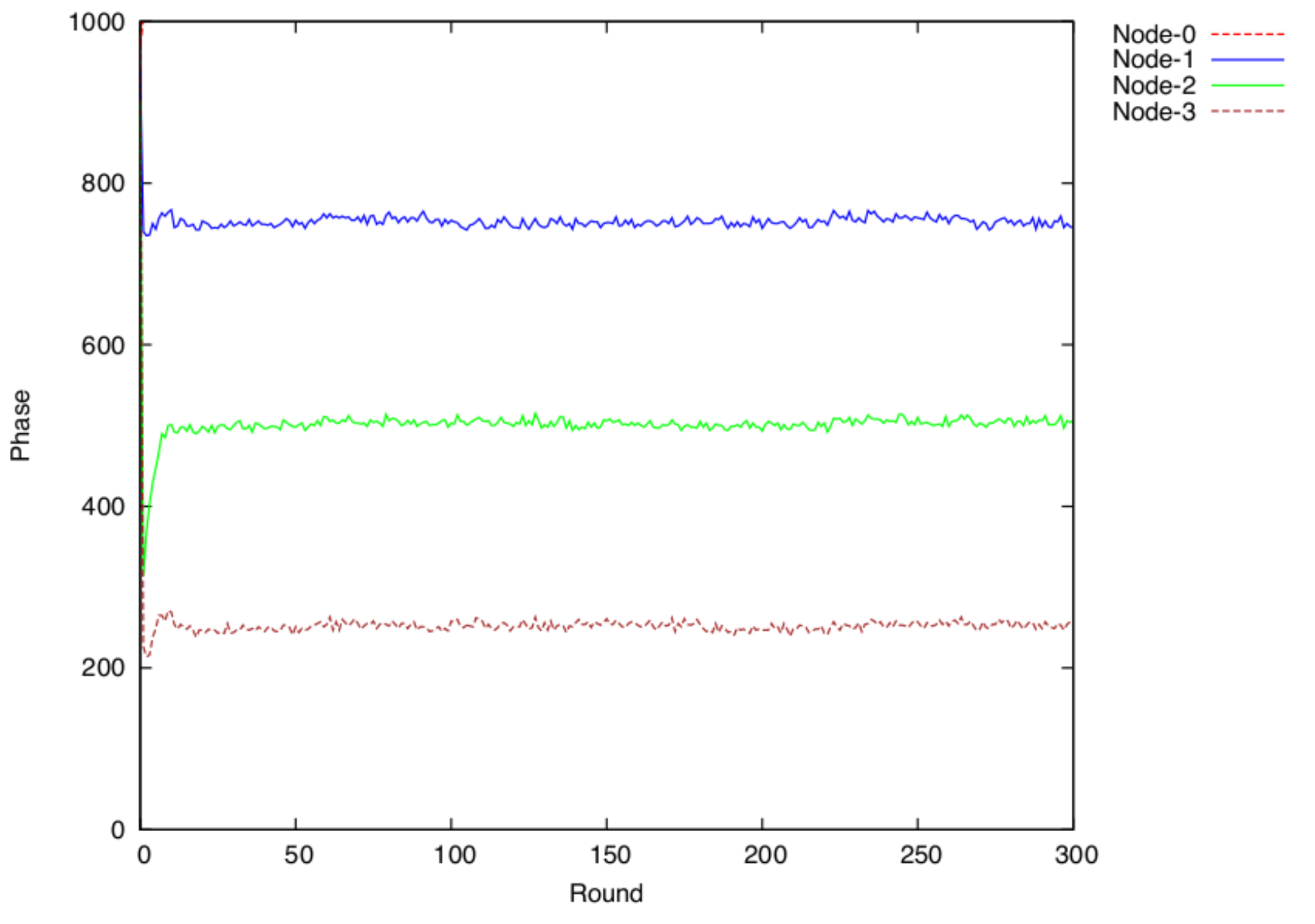}
			\label{fig:4nodes-cycle-result-mdwarf-good}}
		\subfloat[EXT-DESYNC]{\includegraphics[scale=0.27]{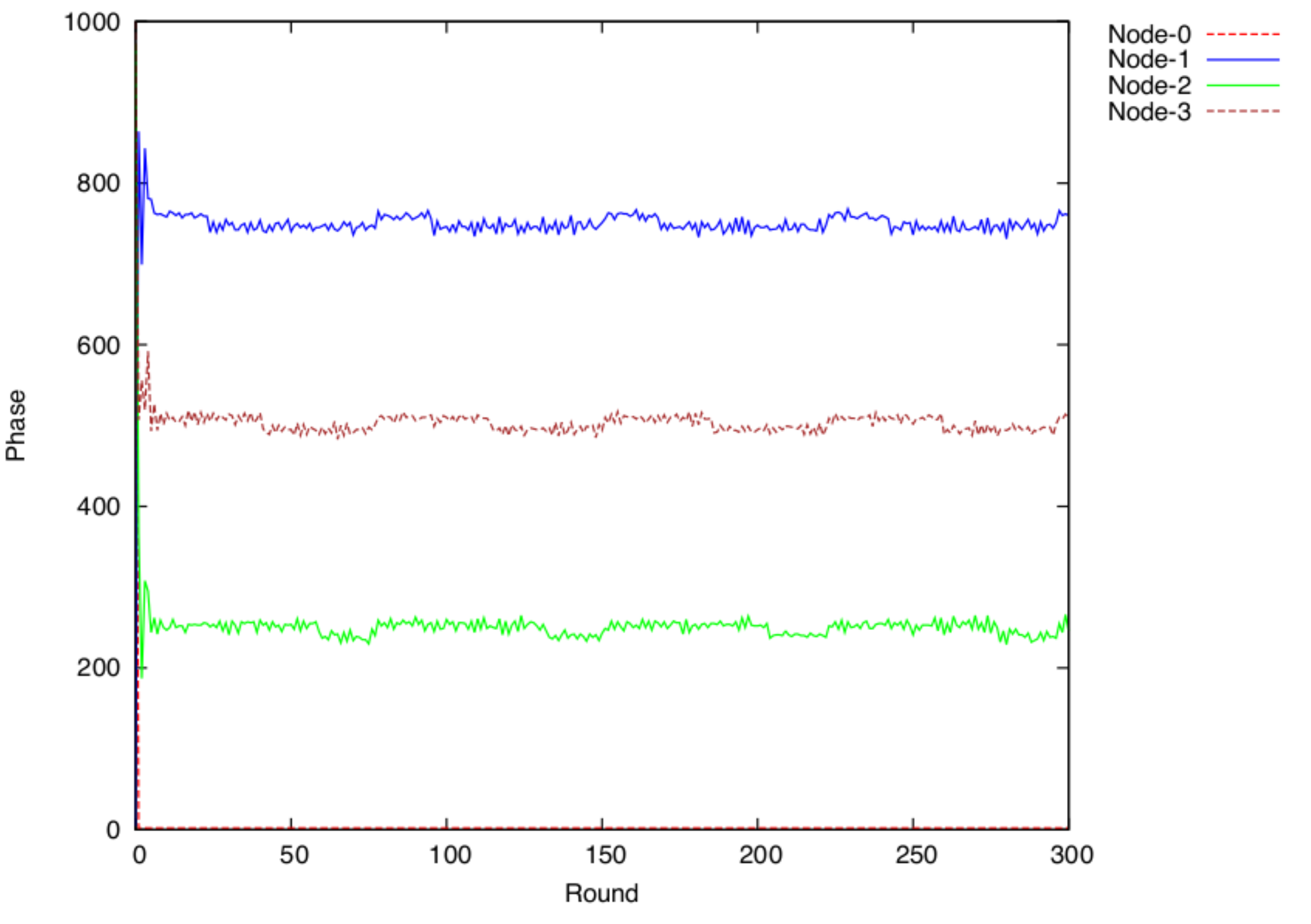}
			\label{fig:4nodes-cycle-result-extdesync-good}}
		\subfloat[LIGHTWEIGHT]{\includegraphics[scale=0.27]{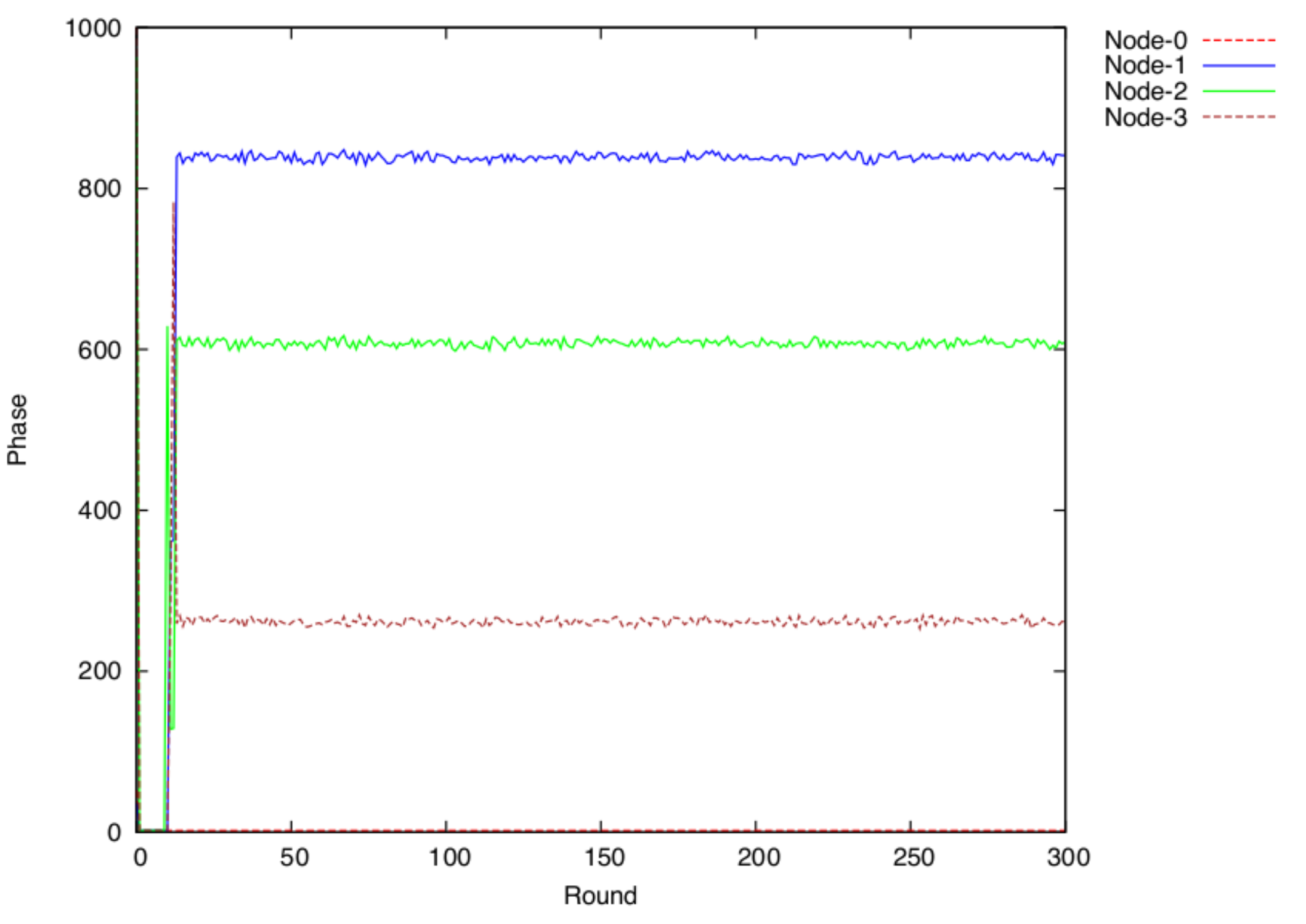}
			\label{fig:4nodes-cycle-result-light-good}}
	}
	\caption{4-node cycle topology evaluation (average case)}
	\label{fig:4nodes-cycle-result-good}
\end{figure*}
\begin{figure*}
\centering{
		\subfloat[M-DWARF]{\includegraphics[scale=0.27]{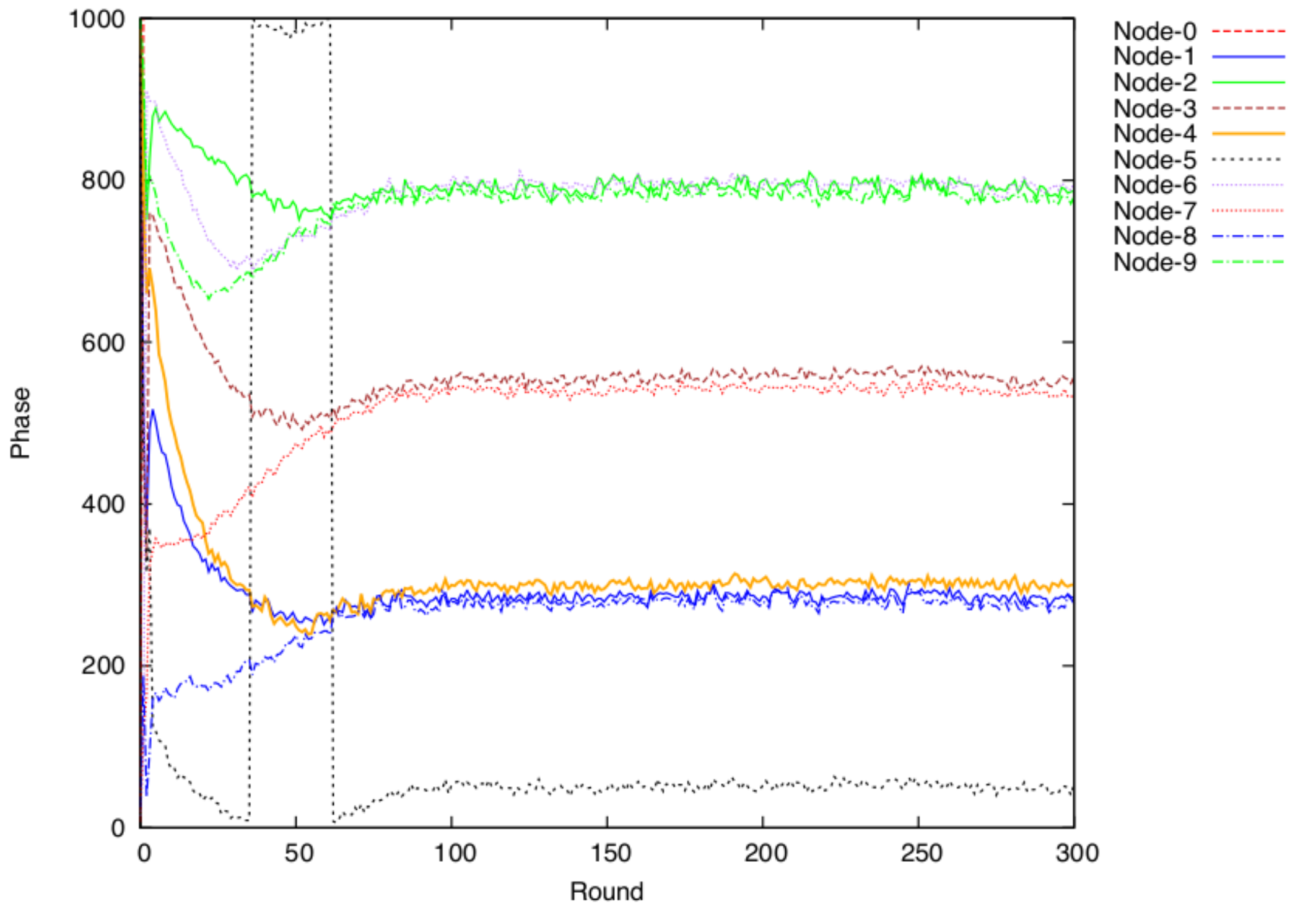}
			\label{fig:10nodes-cycle-result-mdwarf-good}}
		\subfloat[EXT-DESYNC]{\includegraphics[scale=0.27]{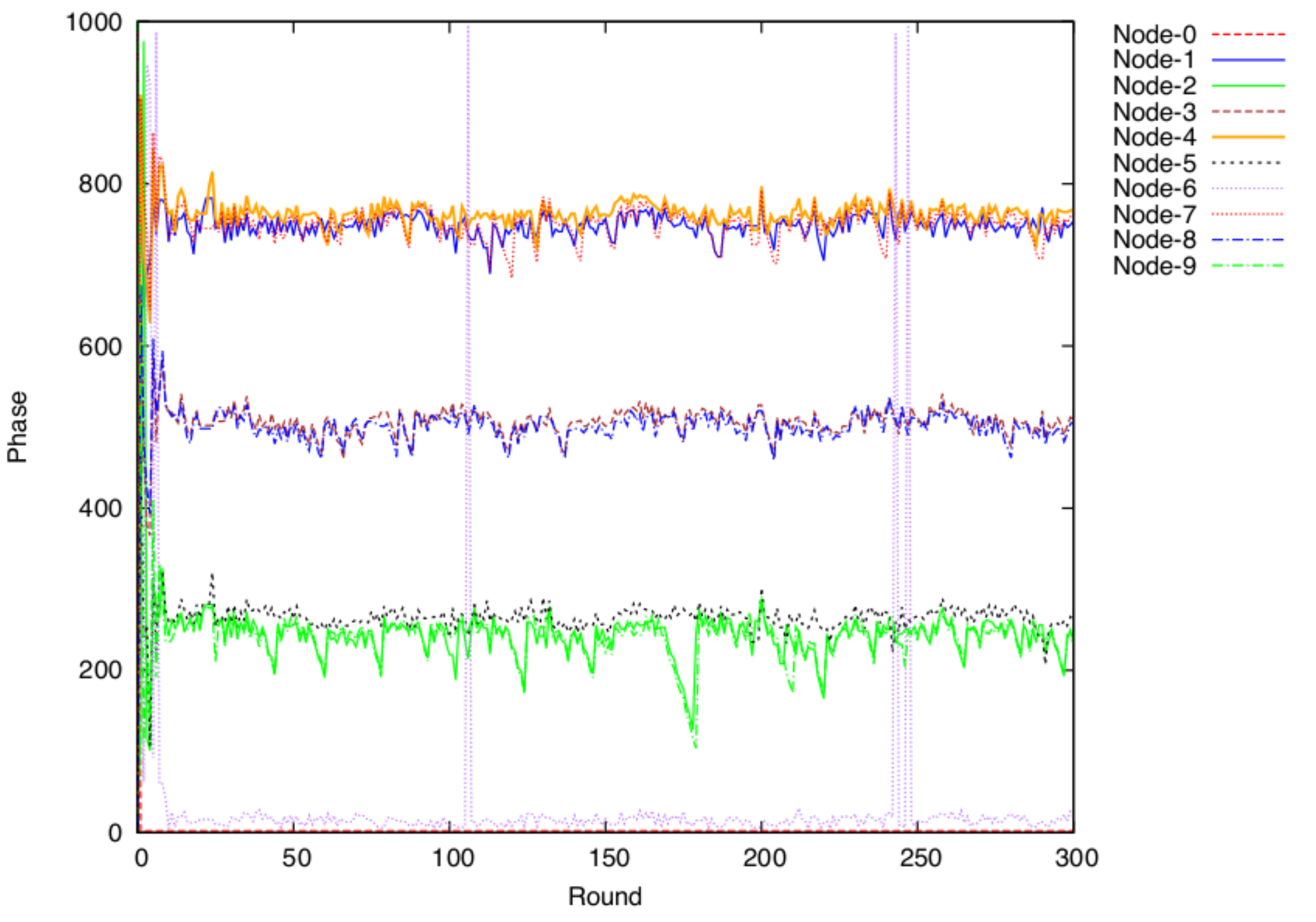}
			\label{fig:10nodes-cycle-result-extdesync-good}}
		\subfloat[LIGHTWEIGHT]{\includegraphics[scale=0.27]{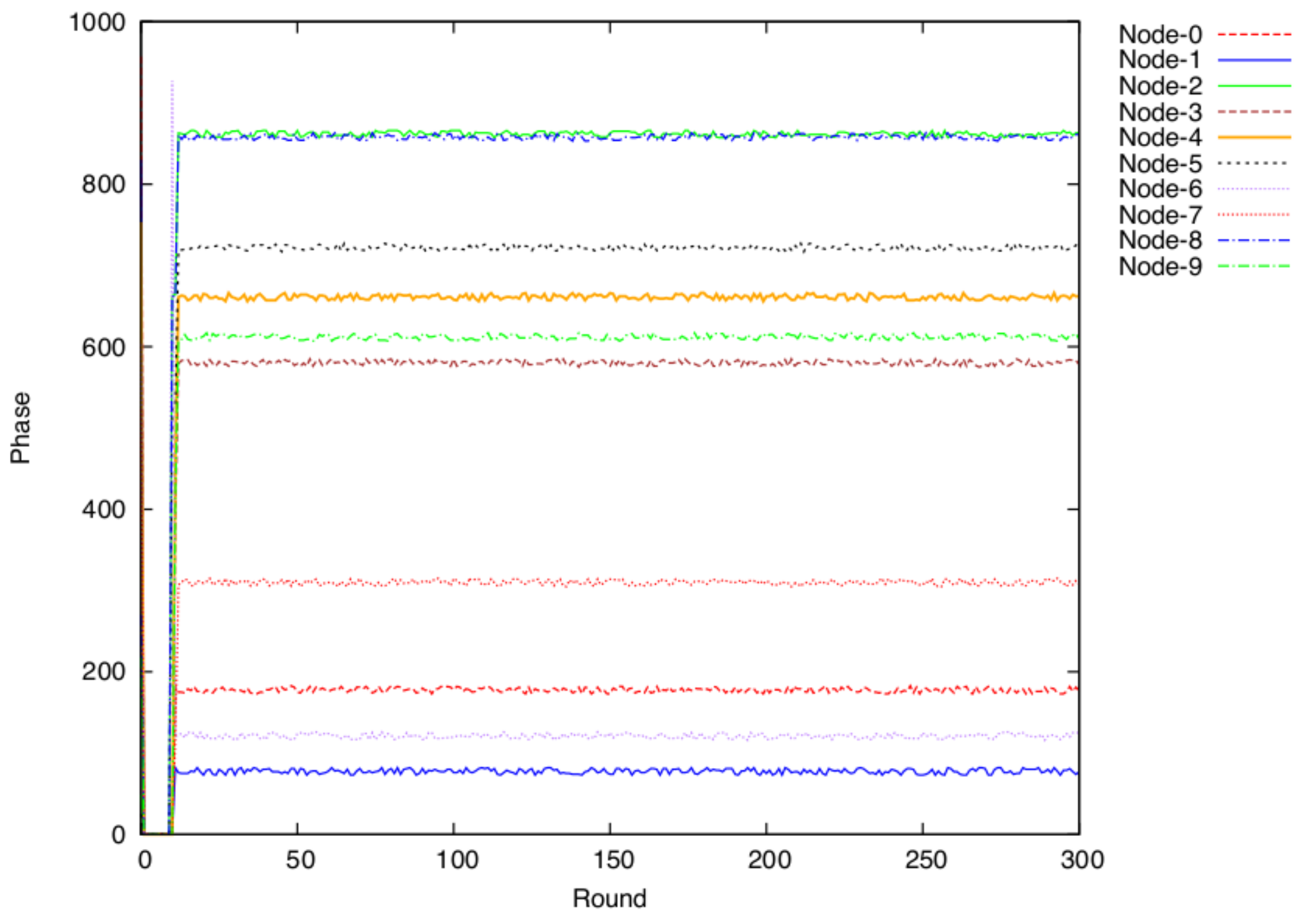}
			\label{fig:10nodes-cycle-result-light-good}}
}
\caption{10-node cycle topology evaluation (average case)}
\label{fig:10nodes-cycle-result-good}	
\end{figure*}

For 4-node cycle networks, all nodes are within two hops from each other. Therefore, every node has known phases of other nodes and no node can share the same phase. On average, both the algorithms, M-DWARF and EXT-DESYNC, achieve the perfect desynchrony state as shown in Figure \ref{fig:4nodes-cycle-result-mdwarf-good} and \ref{fig:4nodes-cycle-result-extdesync-good}. We note that, for 10-node networks, the optimal solution in the cycle topology is different from the optimal solution in the chain topology because node 0 and 9 cannot share the same phase in the cycle topology. Both M-DWARF and EXT-DESYNCE can achieve the perfect desynchrony state in this case (see Figure \ref{fig:10nodes-cycle-result-mdwarf-good} and \ref{fig:10nodes-cycle-result-extdesync-good}). However, repeatedly, the phases of EXT-DESYNC are not stable.

\begin{figure*}
	\centering{
		\subfloat[M-DWARF]{\includegraphics[scale=0.27]{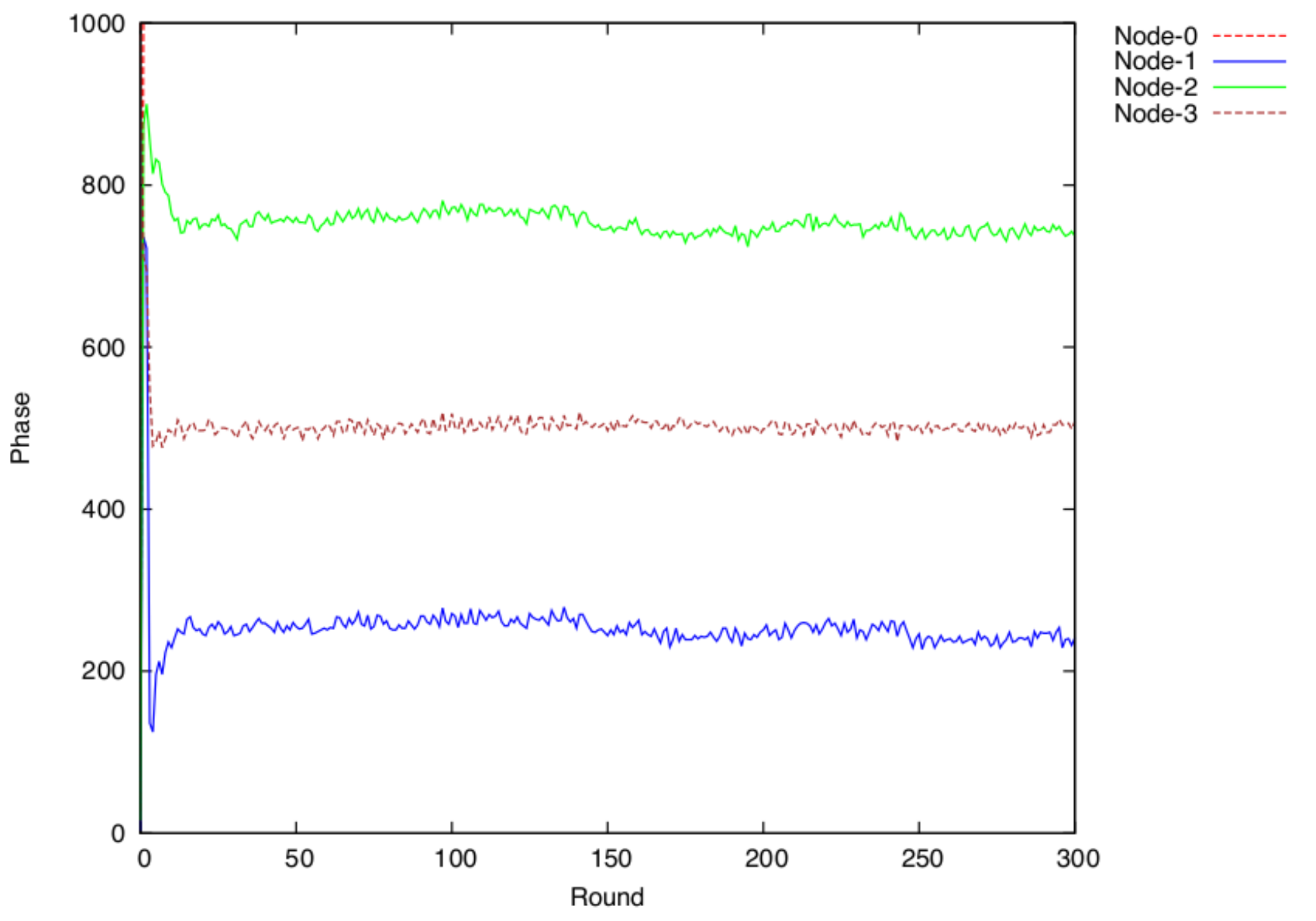}%
			\label{fig:4nodes-cycle-result-mdwarf-bad}}
		\subfloat[EXT-DESYNC]{\includegraphics[scale=0.27]{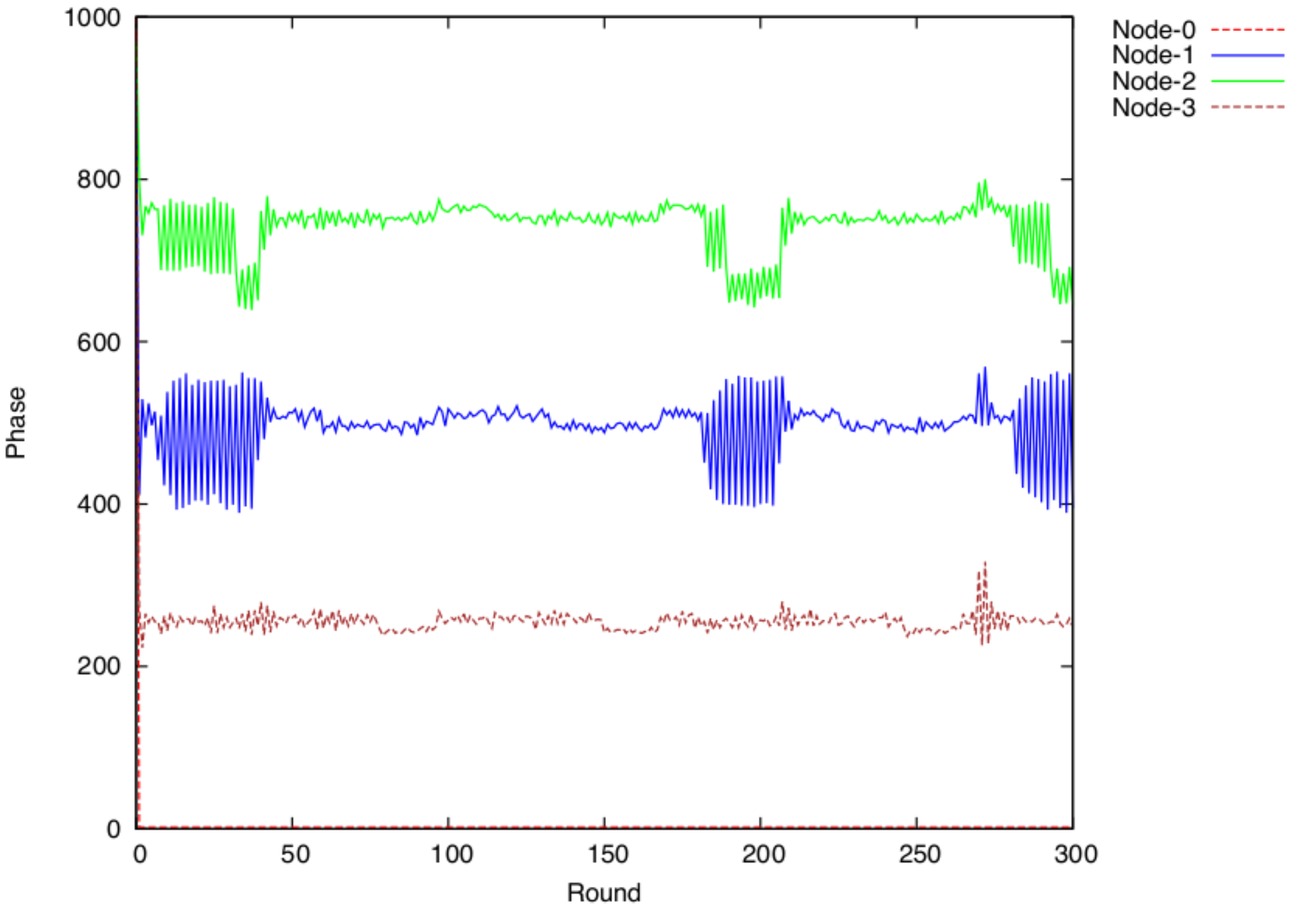}%
			\label{fig:4nodes-cycle-result-extdesync-bad}}
		\subfloat[LIGHTWEIGHT]{\includegraphics[scale=0.27]{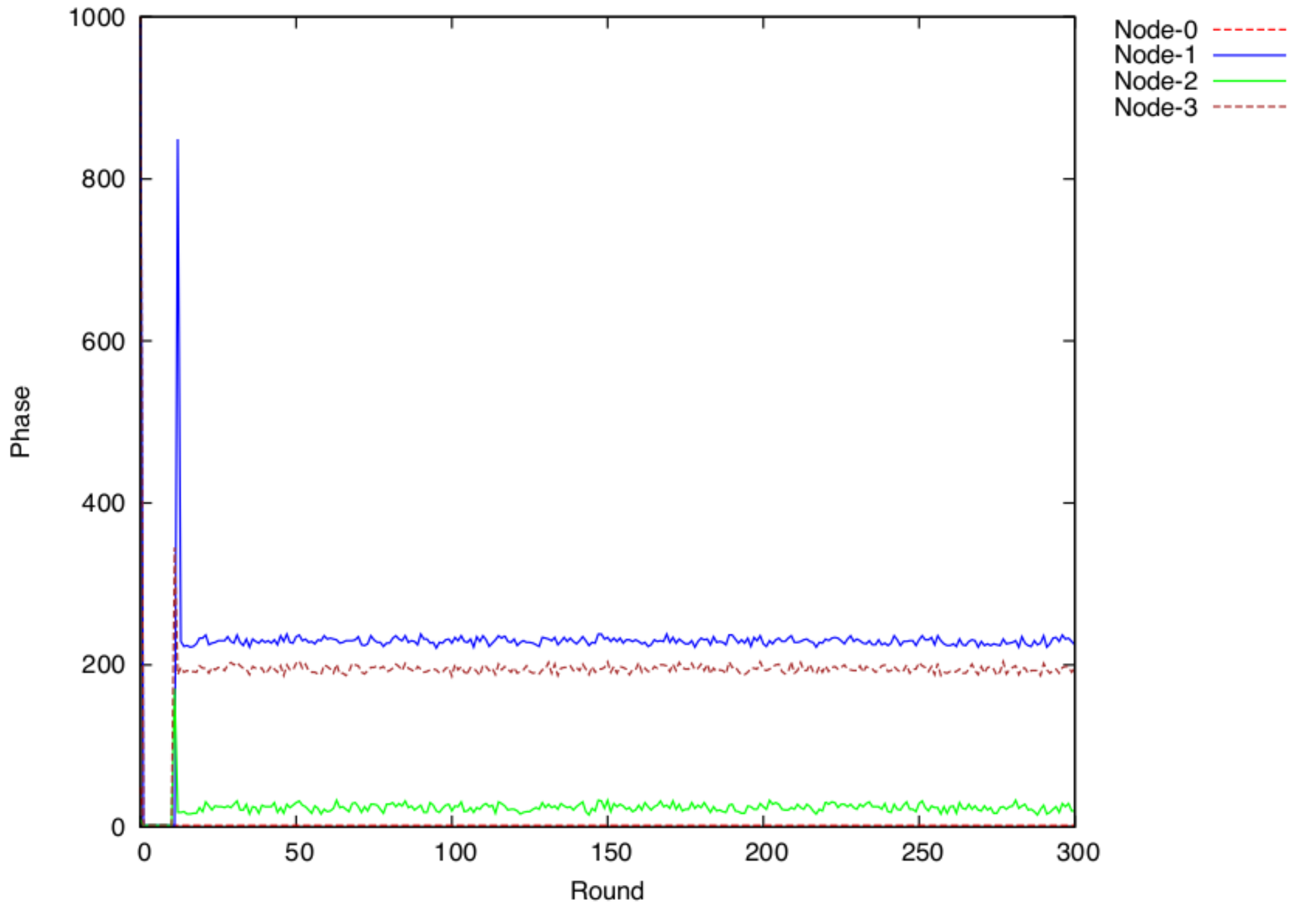}%
			\label{fig:4nodes-cycle-result-light-bad}}
	}
	\caption{4-node cycle topology evaluation (problematic case).}
	\label{fig:4nodes-cycle-result-bad}
	
\end{figure*}

\begin{figure*}
	\centering{
		\subfloat[M-DWARF]{\includegraphics[scale=0.27]{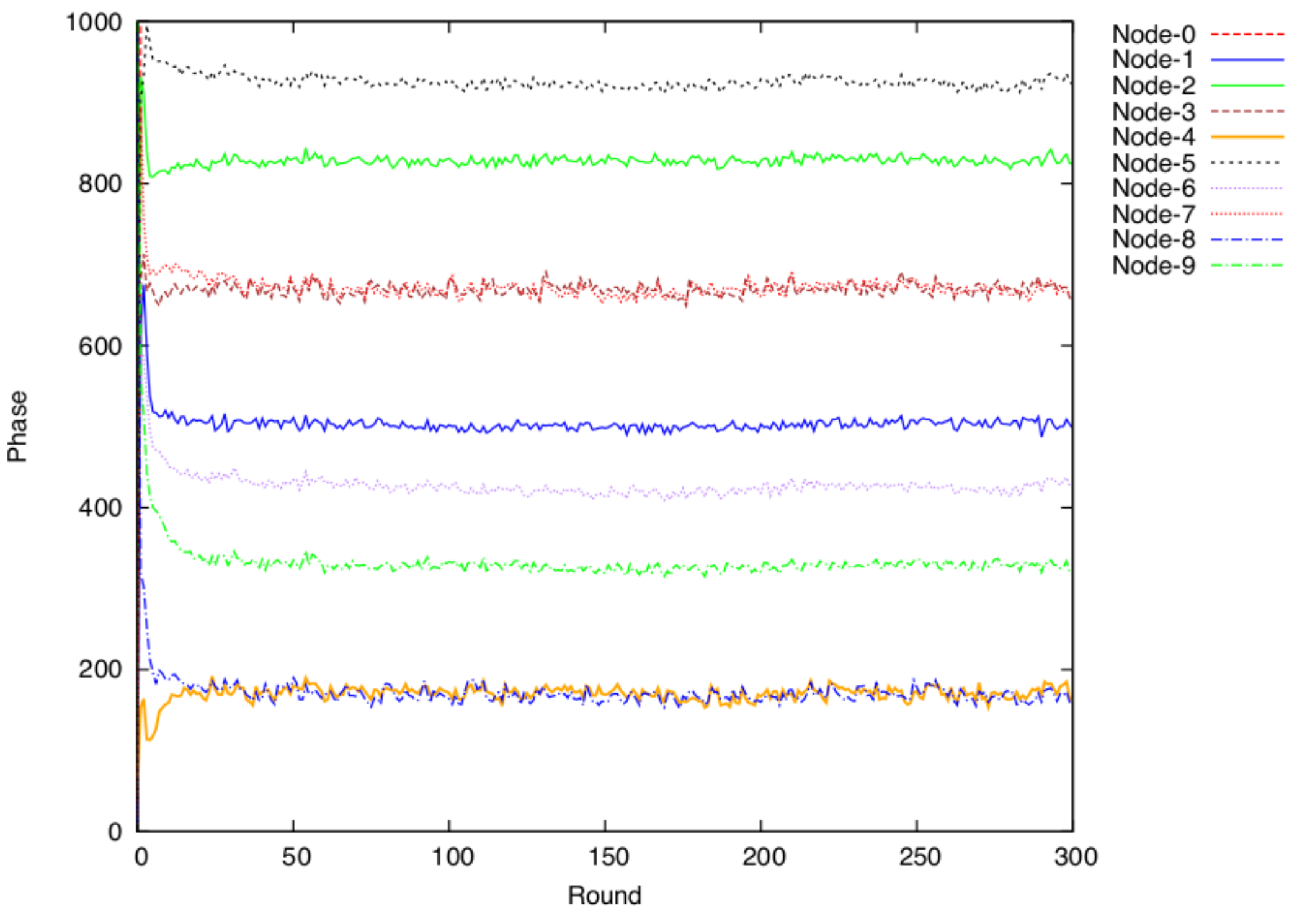}%
			\label{fig:10nodes-cycle-result-mdwarf-bad}}
		\subfloat[EXT-DESYNC]{\includegraphics[scale=0.27]{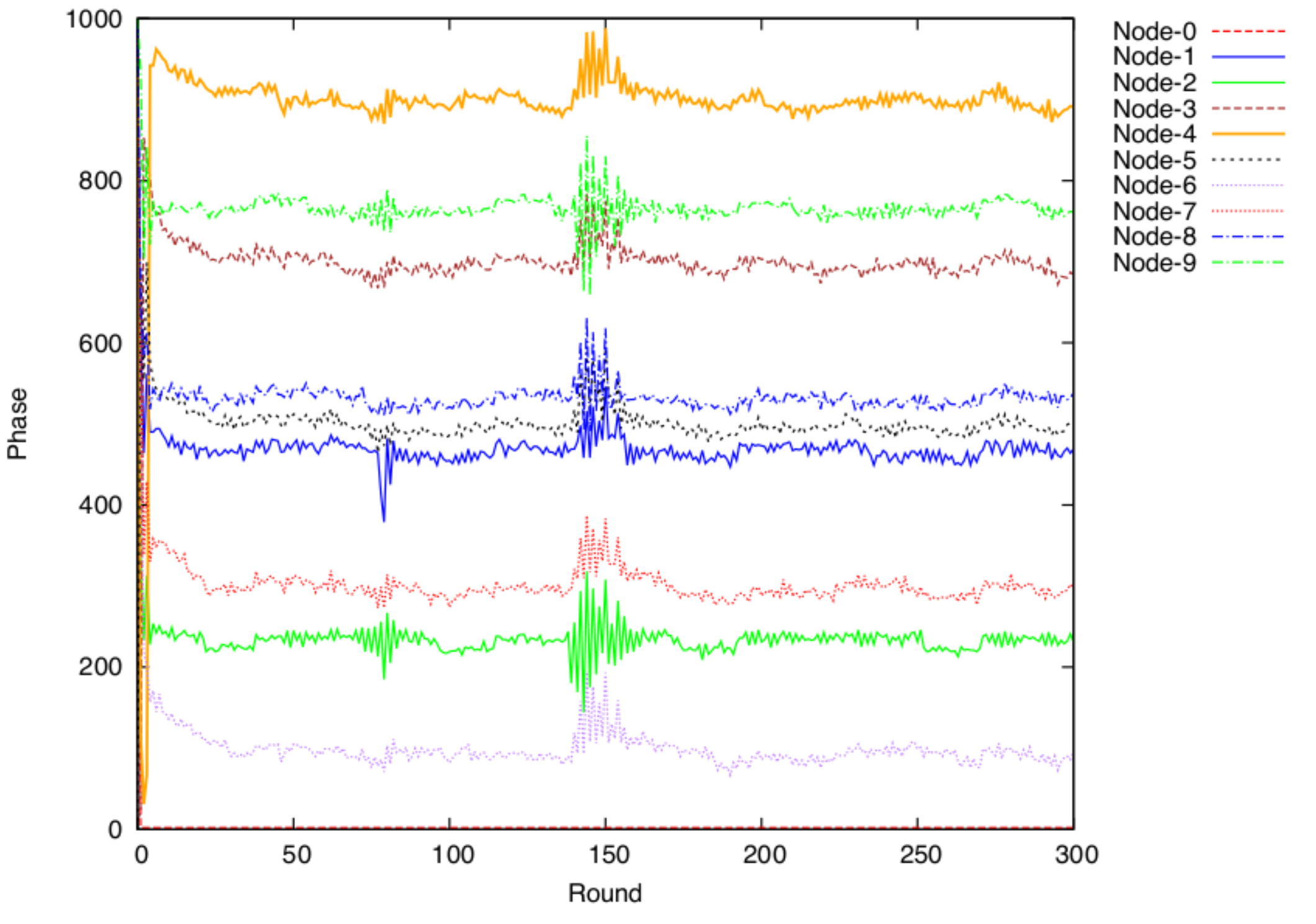}%
			\label{fig:10nodes-cycle-result-extdesync-bad}}
		\subfloat[LIGHTWEIGHT]{\includegraphics[scale=0.27]{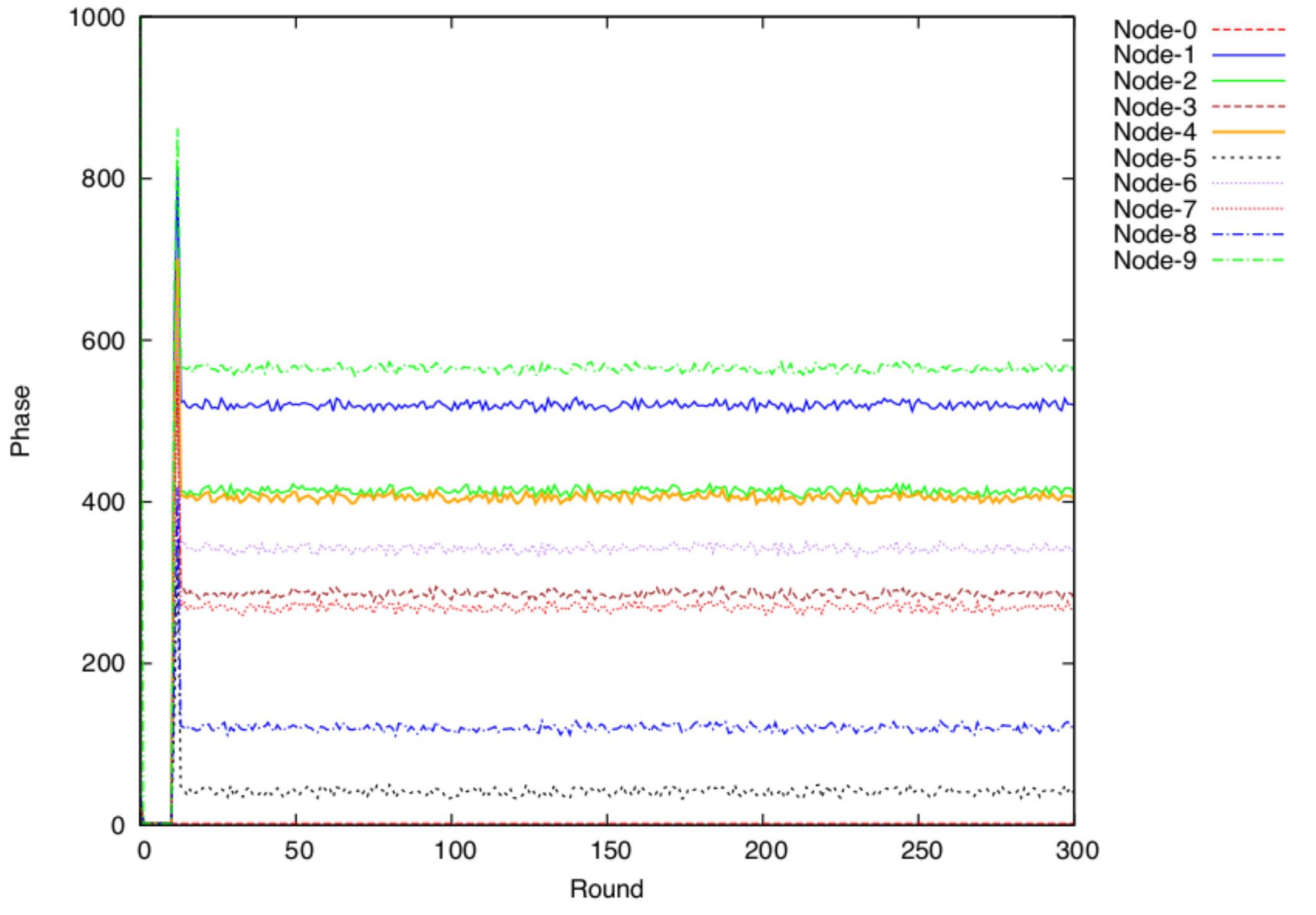}%
			\label{fig:10nodes-cycle-result-light-bad}}
	}
	\caption{10-node cycle topology evaluation (problematic case).}
	\label{fig:10nodes-cycle-result-bad}
	
\end{figure*}

In some problematic cases, the result is similar to the problematic result of the chain topology with the same reason that the initial configuration affects how nodes adjust theirs phases to the proper phases. Figure \ref{fig:4nodes-cycle-result-mdwarf-bad}, \ref{fig:4nodes-cycle-result-extdesync-bad}, \ref{fig:10nodes-cycle-result-mdwarf-bad}, and \ref{fig:10nodes-cycle-result-extdesync-bad} illustrate the problem.

\paragraph{Dumbbell Topology}
Dumbbell is a network  topology commonly found in practice. It is one form of transit-stub networks. In a dumbbell network, there are a number of nodes connected to a single relay node which is connected to another relay node at another side to which another group of nodes are connected, to create a dumbbell topology. Figure \ref{fig:6nodes-butterfly-eval} and \ref{fig:20nodes-butterfly-eval} depict 6-node and 20-node dumbbell networks.

\begin{figure*}
	\centerline{
		\subfloat[M-DWARF]{\includegraphics[scale=0.27]{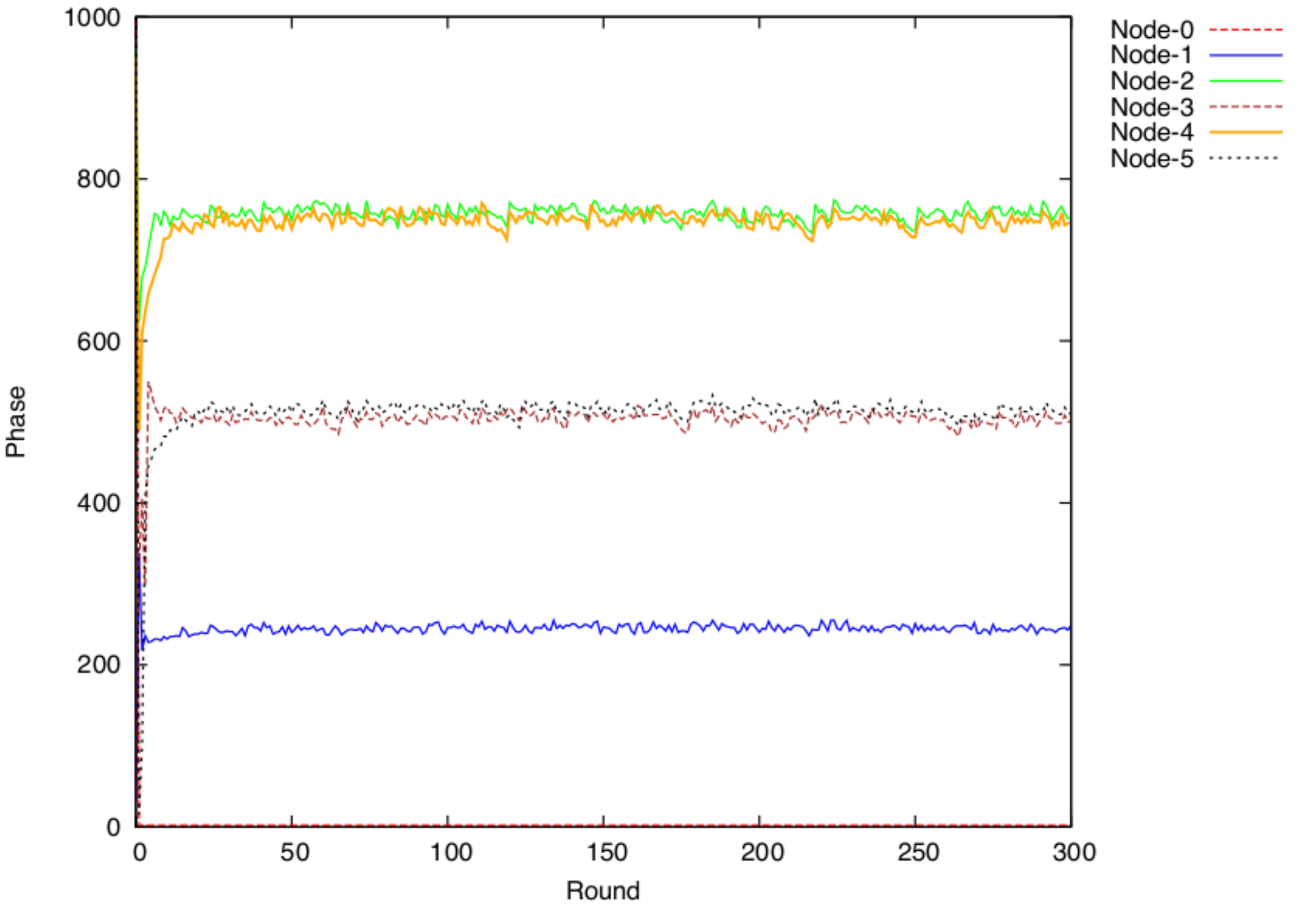}%
			\label{fig:6nodes-butterfly-result-mdwarf-good}}
		\subfloat[EXT-DESYNC]{\includegraphics[scale=0.27]{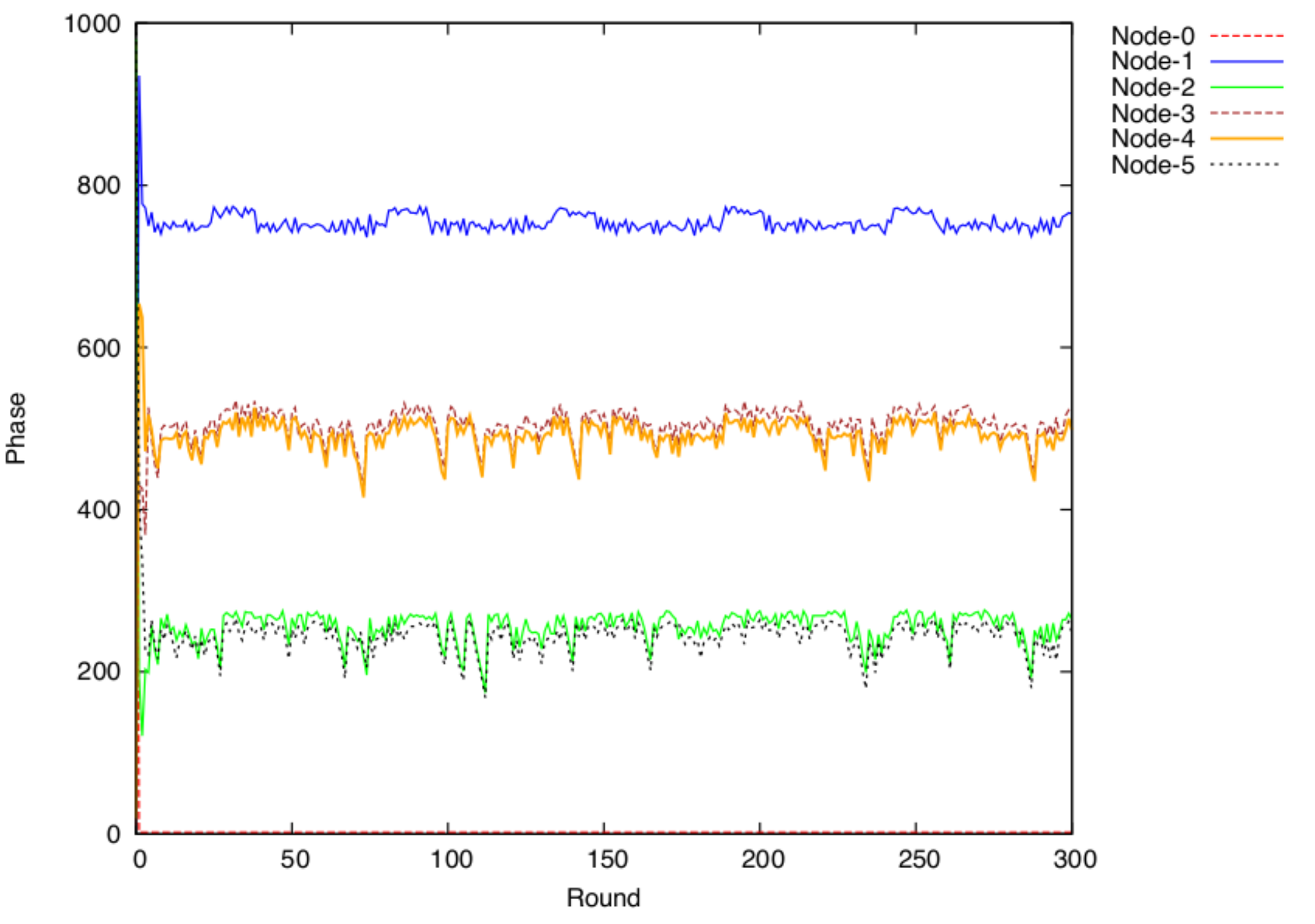}%
			\label{fig:6nodes-butterfly-result-extdesync-good}}
		\subfloat[LIGHTWEIGHT]{\includegraphics[scale=0.27]{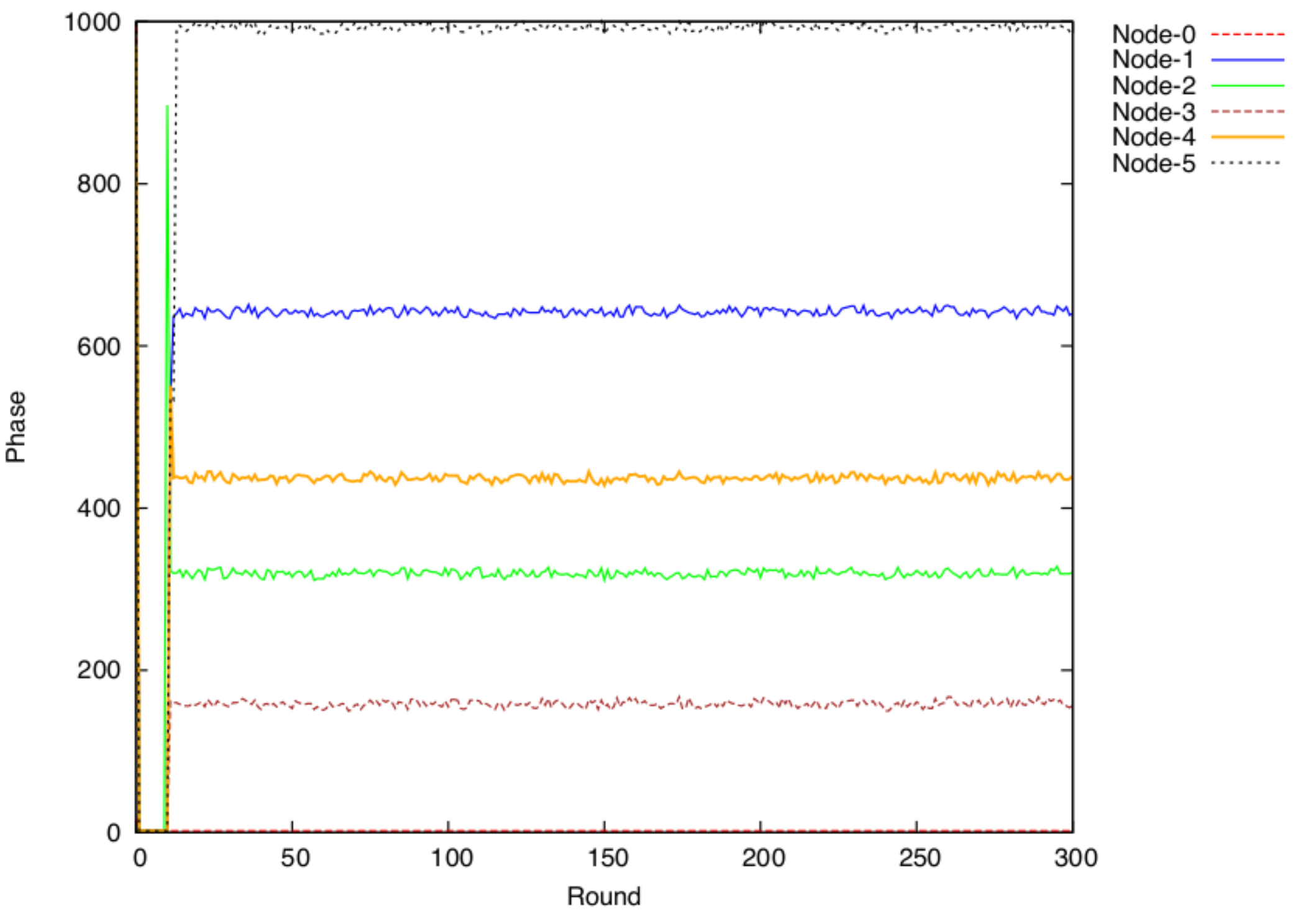}%
			\label{fig:6nodes-butterfly-result-light-good}}
	}
	\caption{6-node dumbbell topology evaluation (average case).}
	\label{fig:6nodes-butterfly-result-good}
	
\end{figure*}

\begin{figure*}
	\centerline{
		\subfloat[M-DWARF]{\includegraphics[scale=0.27]{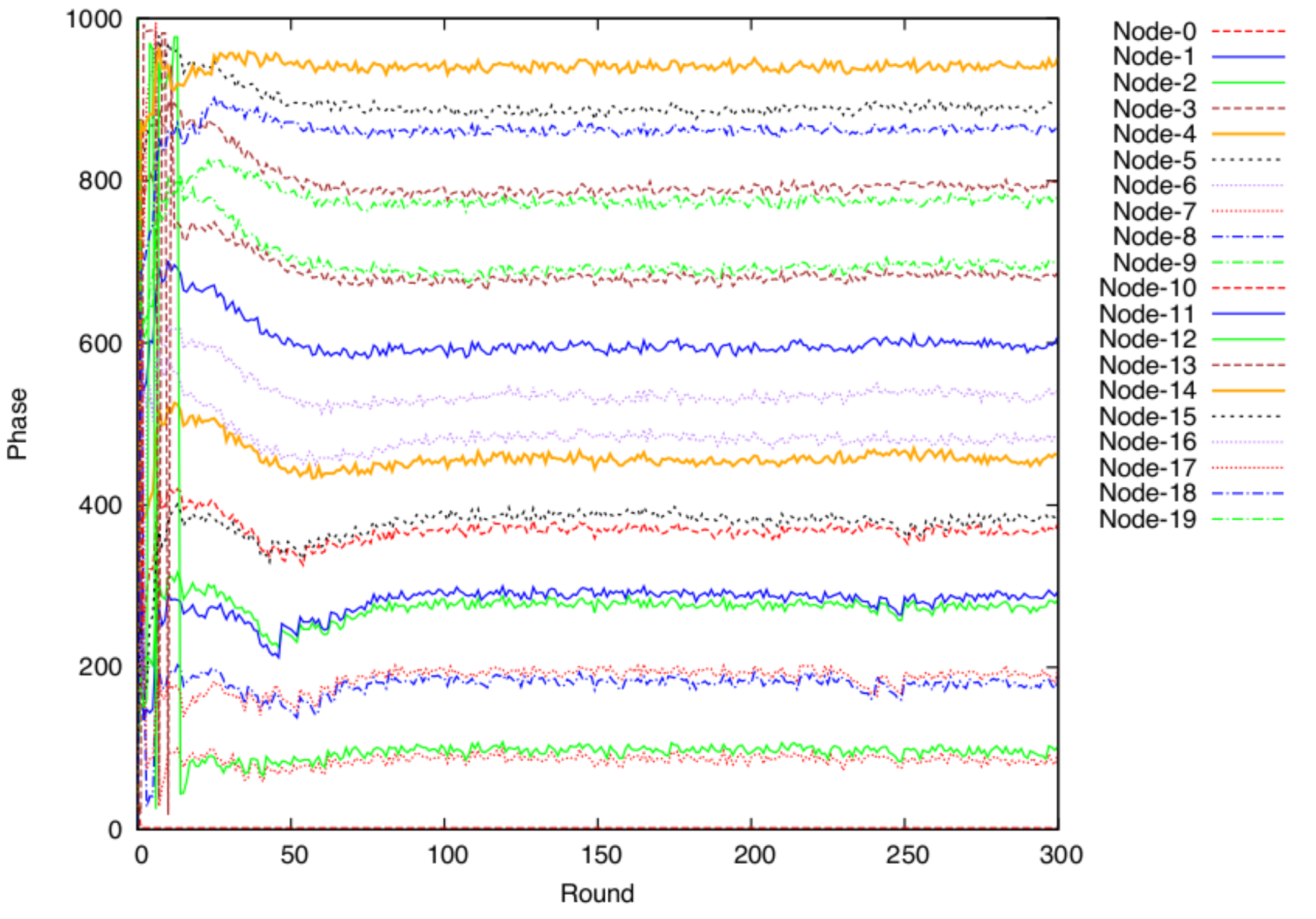}%
			\label{fig:20nodes-butterfly-result-mdwarf-good}}
		\subfloat[EXT-DESYNC]{\includegraphics[scale=0.27]{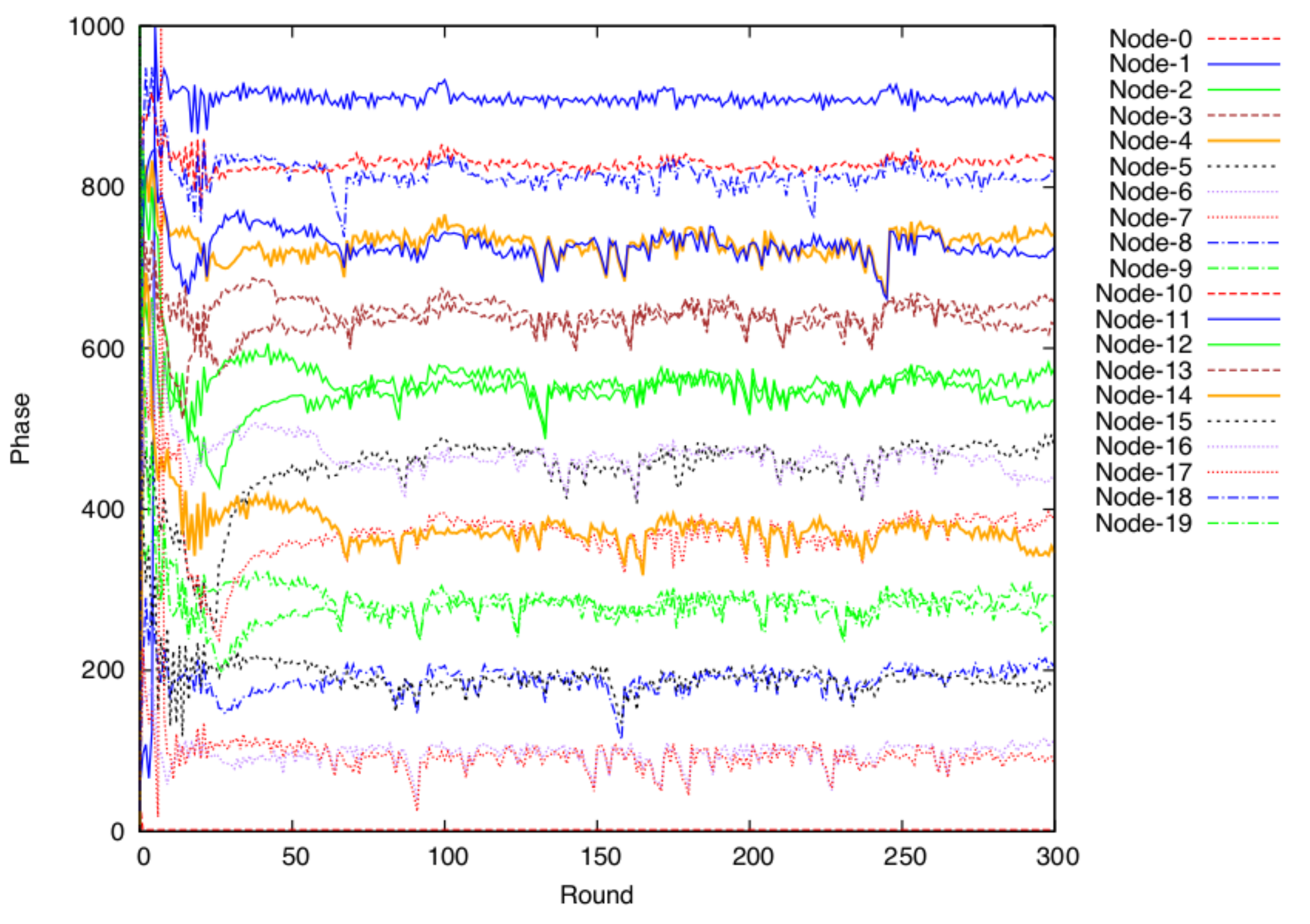}%
			\label{fig:20nodes-butterfly-result-extdesync-good}}
		\subfloat[LIGHTWEIGHT]{\includegraphics[scale=0.27]{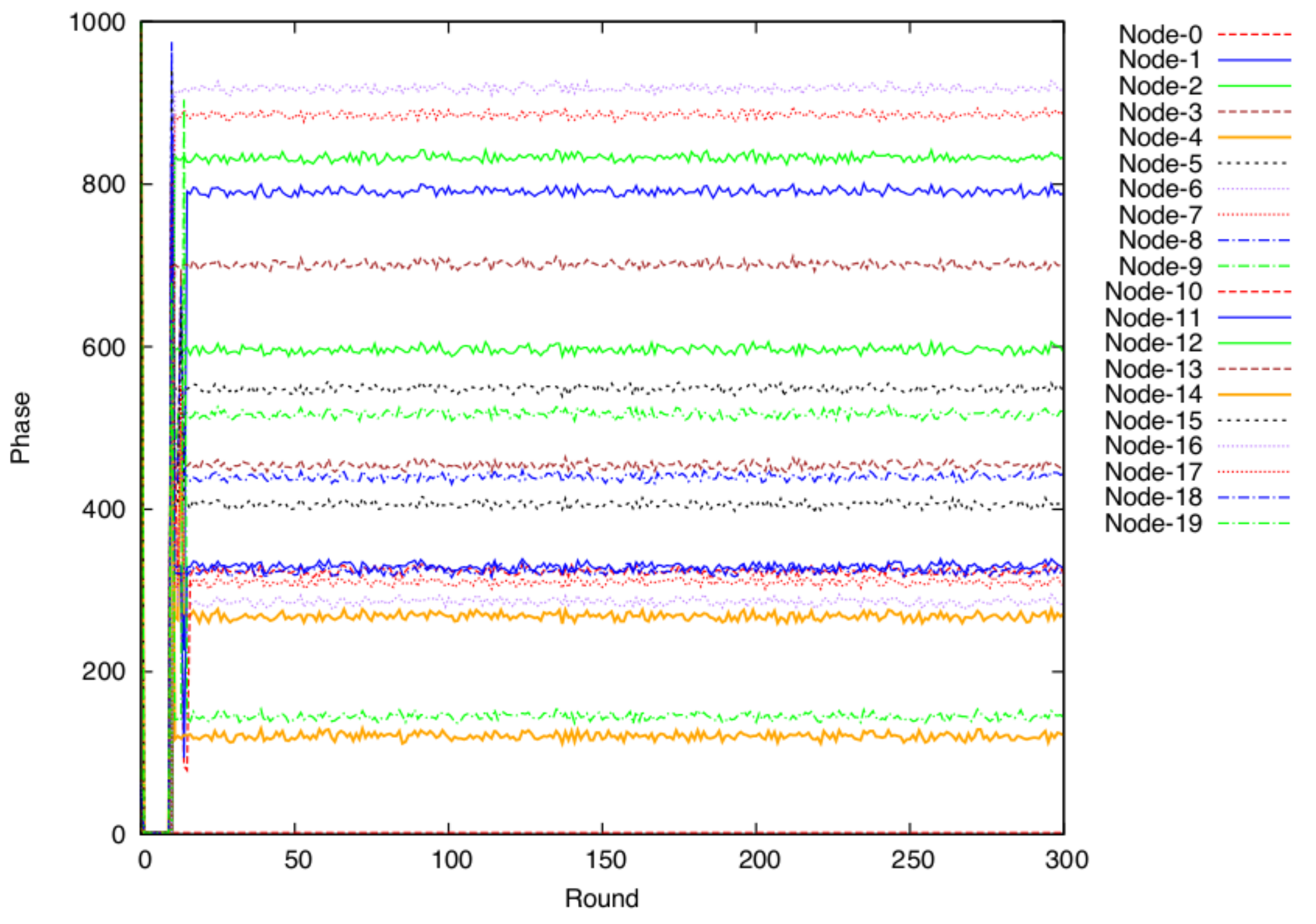}%
			\label{fig:20nodes-butterfly-result-light-good}}
	}
	\caption{20-node dumbbell topology evaluation (average case).}
	\label{fig:20nodes-butterfly-result-good}
	
\end{figure*}

Obviously, a node at one side can share the same phase with another node at the other because they are far from each other beyond two hops. Therefore, for 6-node networks, the optimal solution requires 4 slots. Each slot interval is $T/4$. For 20-node networks, the optimal solution requires 11 slots. Each slot interval is $T/11$.
In Figure \ref{fig:6nodes-butterfly-result-mdwarf-good} and \ref{fig:6nodes-butterfly-result-extdesync-good}, both M-DWARF and EXT-DESYNC achieve the optimal solution for 6-node dumbbell networks. However, M-DWARF is again more stable than EXT-DESYNC.

\begin{figure*}
	\centerline{
		\subfloat[M-DWARF]{\includegraphics[scale=0.27]{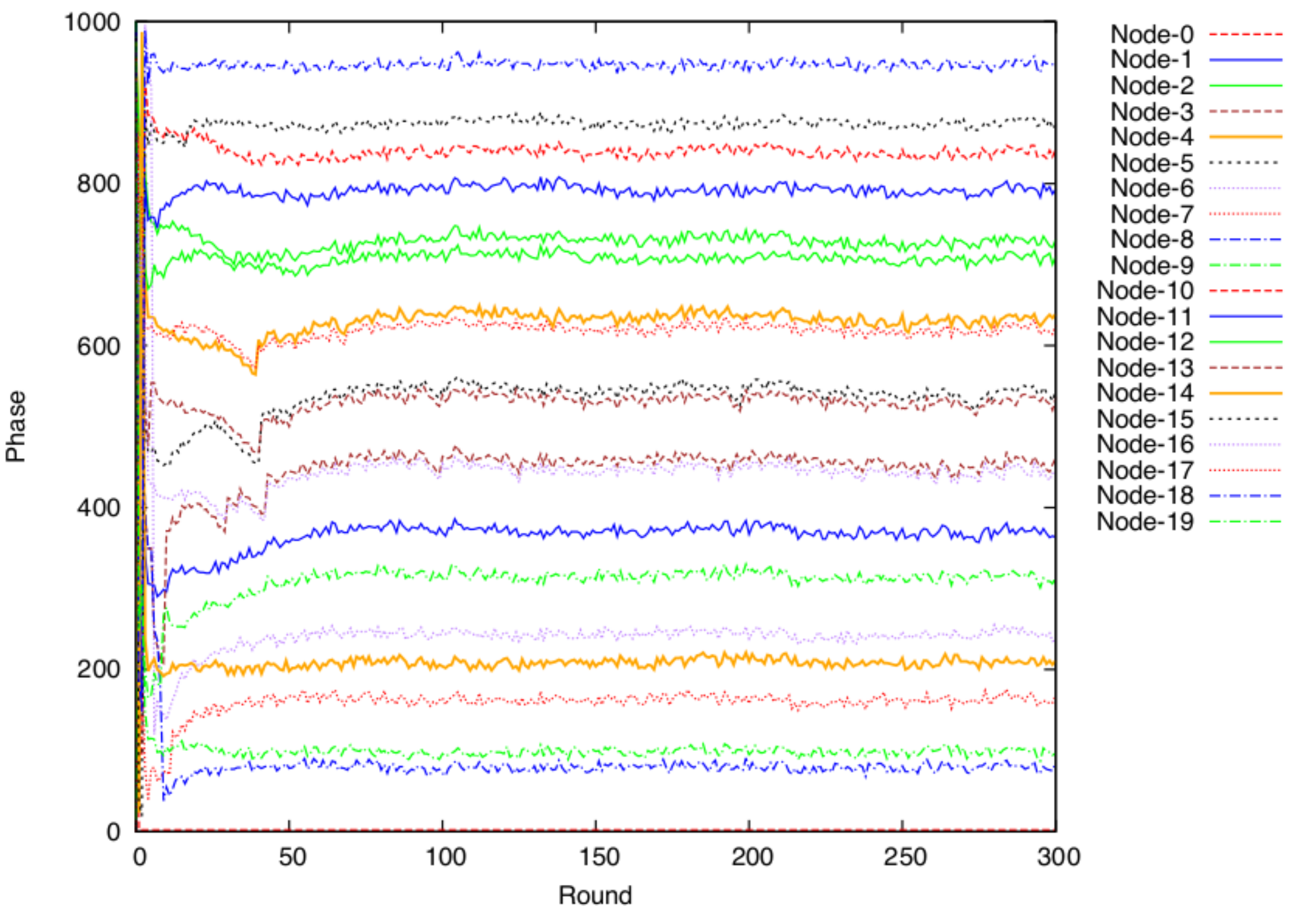}%
			\label{fig:20nodes-butterfly-result-mdwarf-bad}}
		\subfloat[EXT-DESYNC]{\includegraphics[scale=0.27]{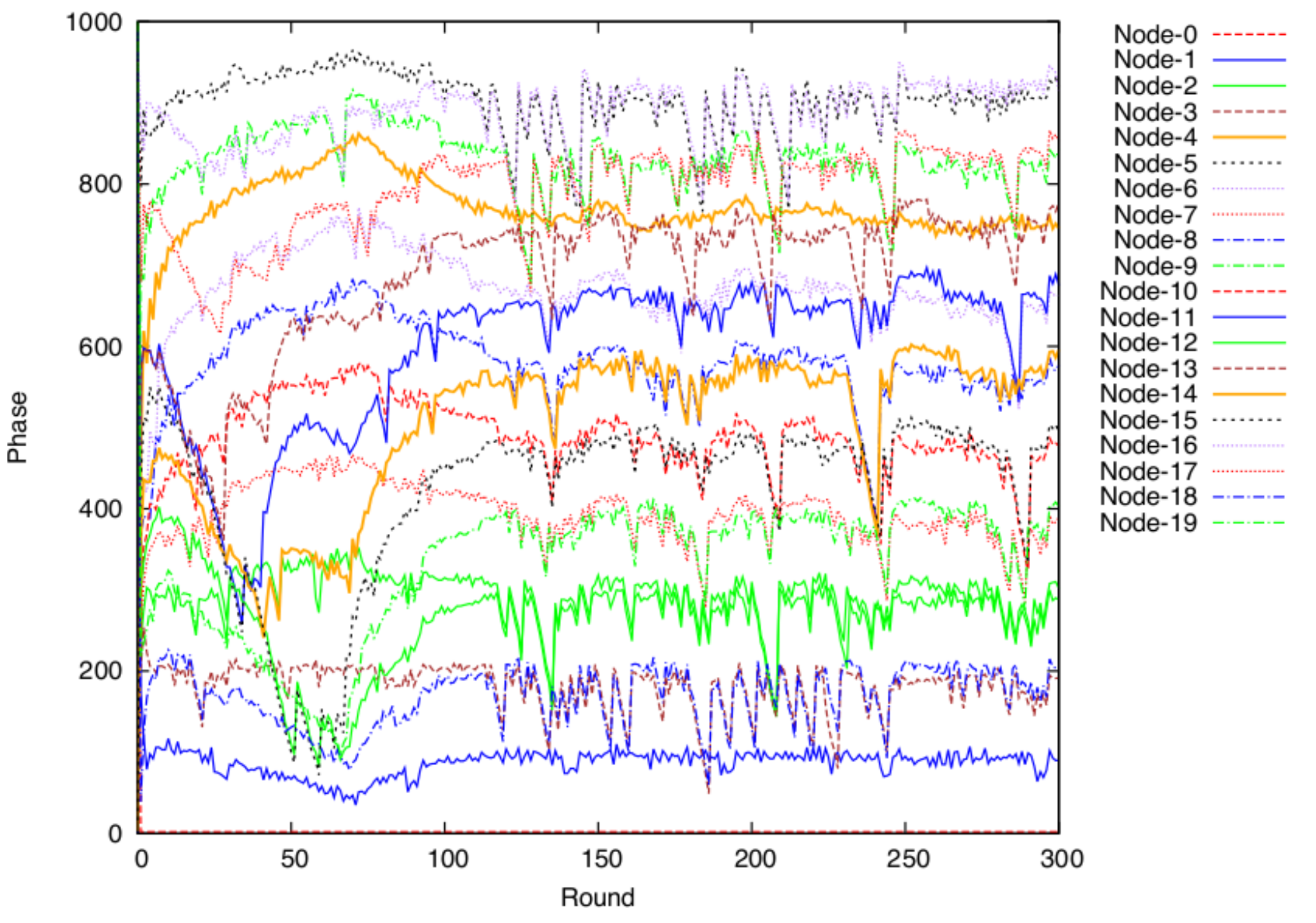}%
			\label{fig:20nodes-butterfly-result-extdesync-bad}}
		\subfloat[LIGHTWEIGHT]{\includegraphics[scale=0.27]{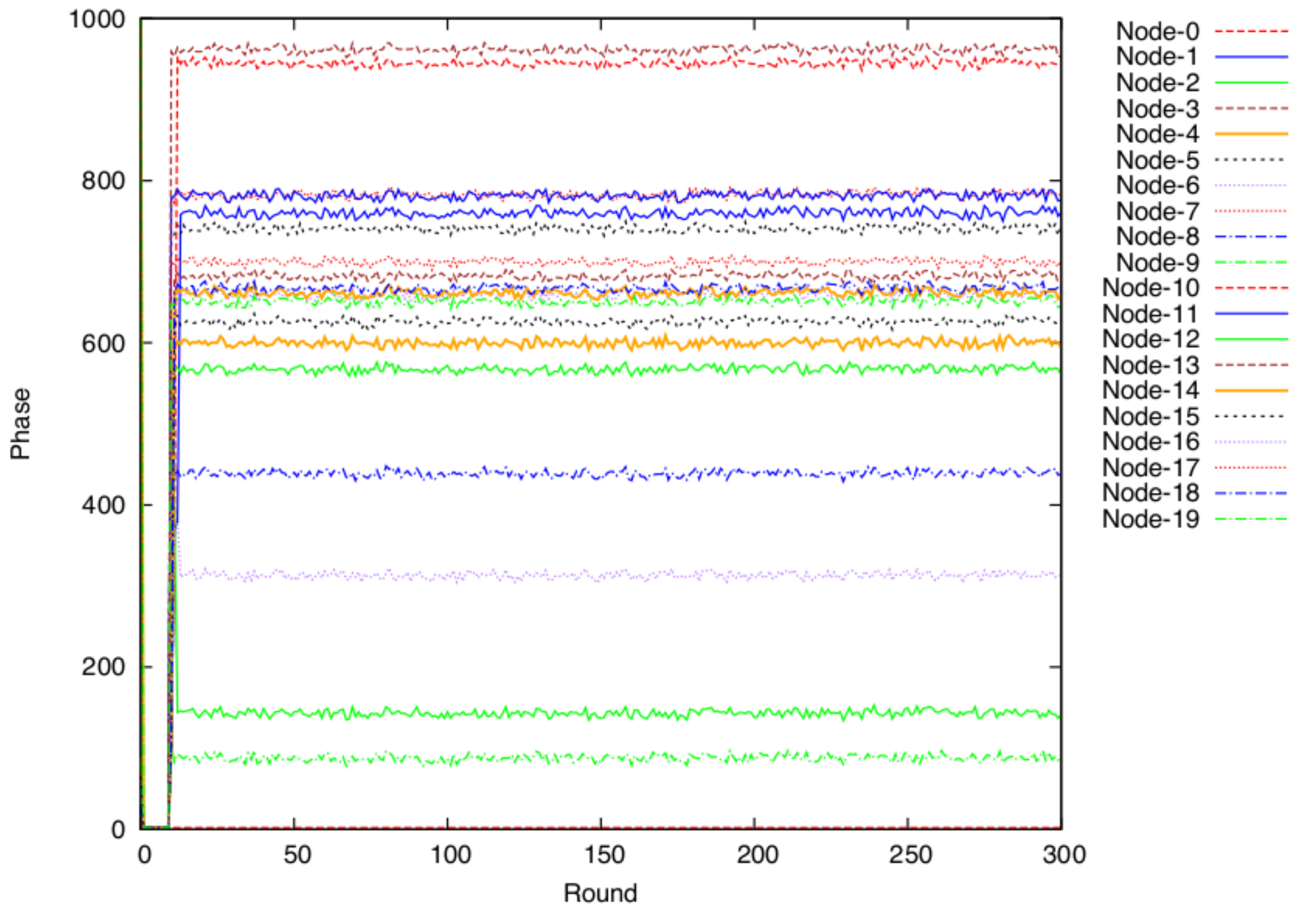}%
			\label{fig:20nodes-butterfly-result-light-bad}}
	}
	\caption{20-node dumbbell topology evaluation (problematic case).}
	\label{fig:20nodes-butterfly-result-bad}
	
\end{figure*}

When the network is dense, EXT-DESYNC highly fluctuates (Figure \ref{fig:20nodes-butterfly-result-extdesync-good}). In problematic cases, if one node adapts their phase too fast, several nodes are affected because there are many two-hop neighbors in the dumbbell networks as illustrated in Figure \ref{fig:20nodes-butterfly-result-extdesync-bad}.  In contrast, in M-DWARF, the fast adaptation of one node does not highly impact to the other nodes because they do not rely on only their two phase neighbors but use all information from neighbors within two hops as the force function.

\begin{figure*}
	\centerline{
		\subfloat[M-DWARF]{\includegraphics[scale=0.27]{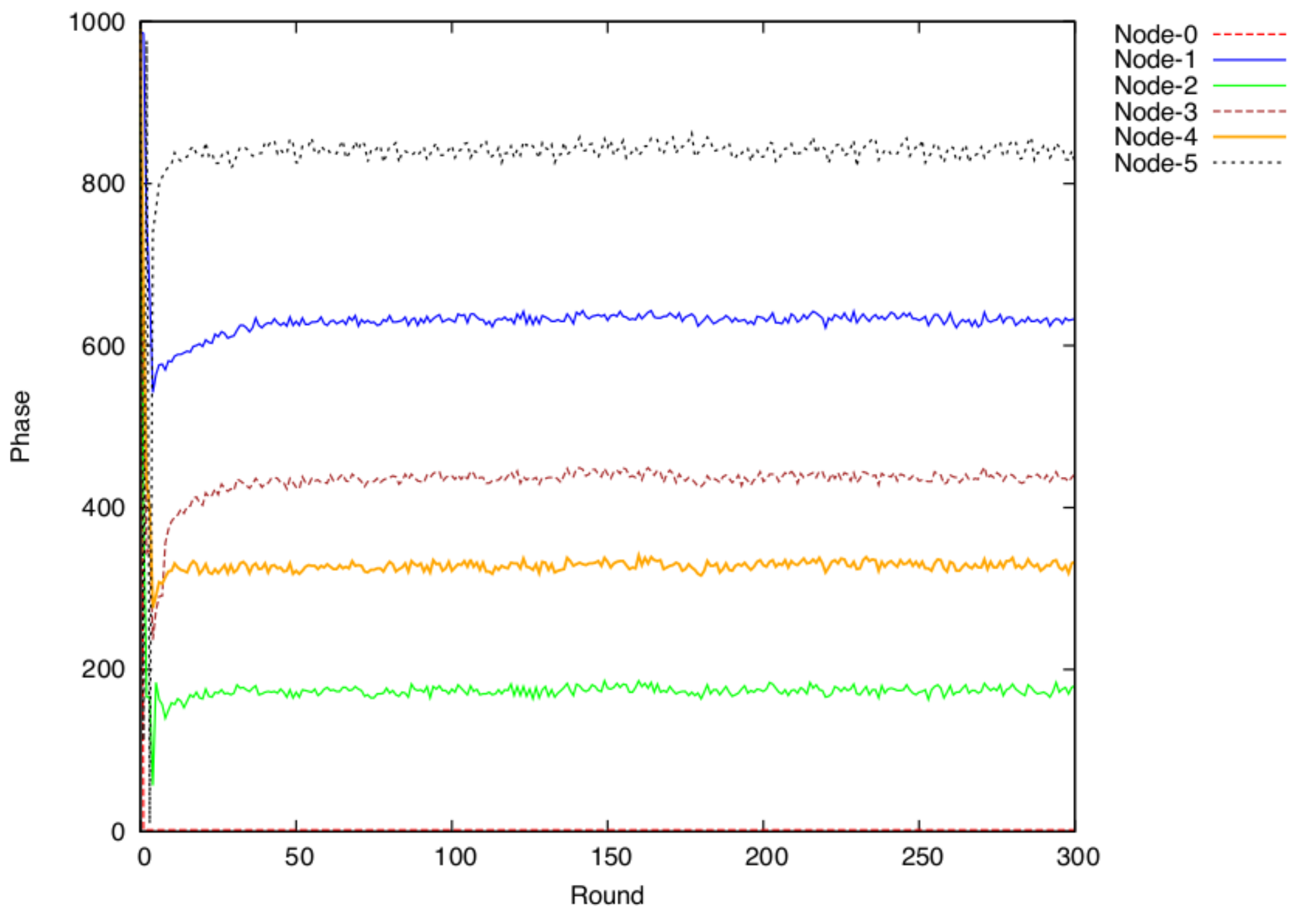}%
			\label{fig:6nodes-butterfly-result-mdwarf-bad}}
		\subfloat[EXT-DESYNC]{\includegraphics[scale=0.27]{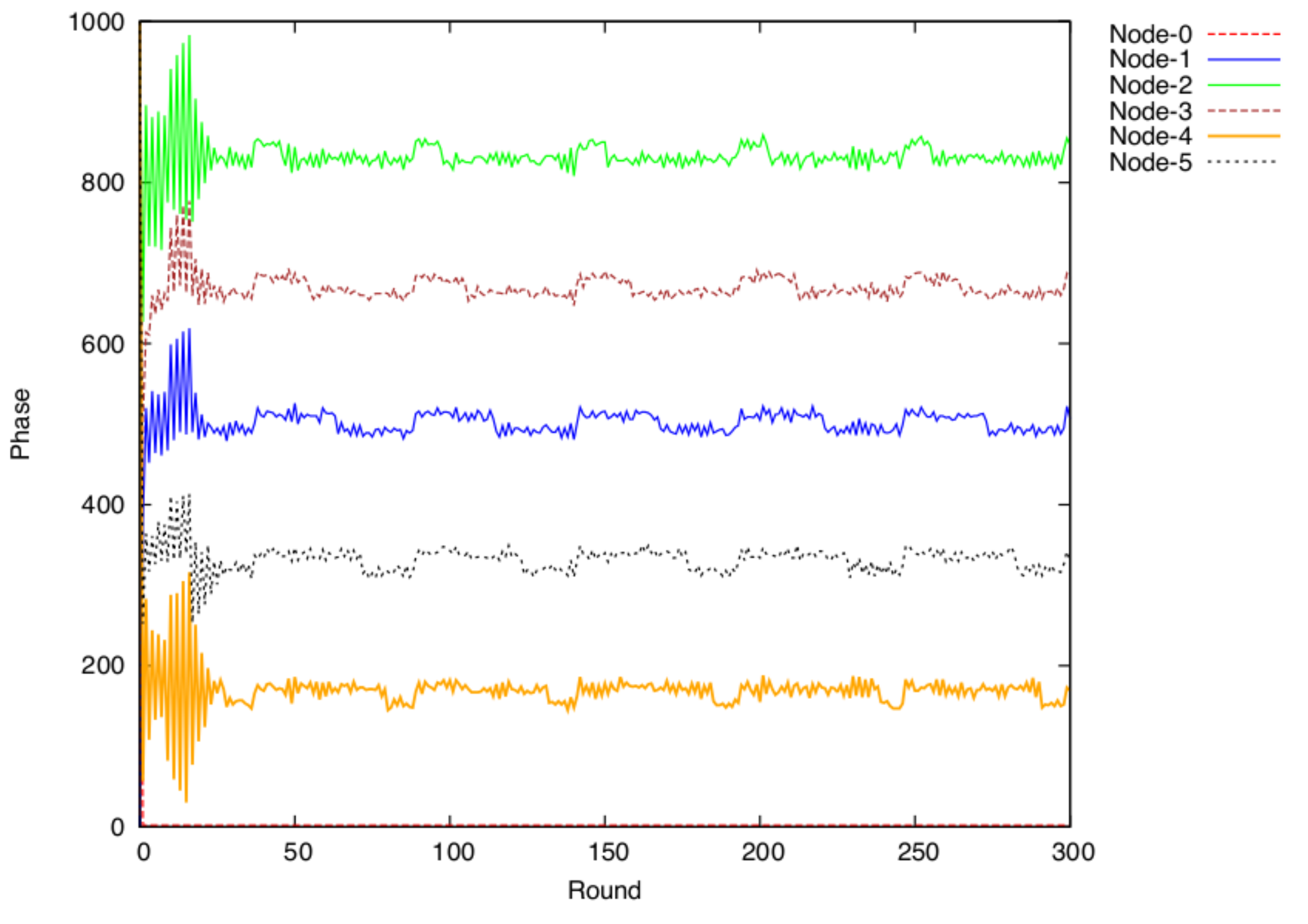}%
			\label{fig:6nodes-butterfly-result-extdesync-bad}}
		\subfloat[LIGHTWEIGHT]{\includegraphics[scale=0.27]{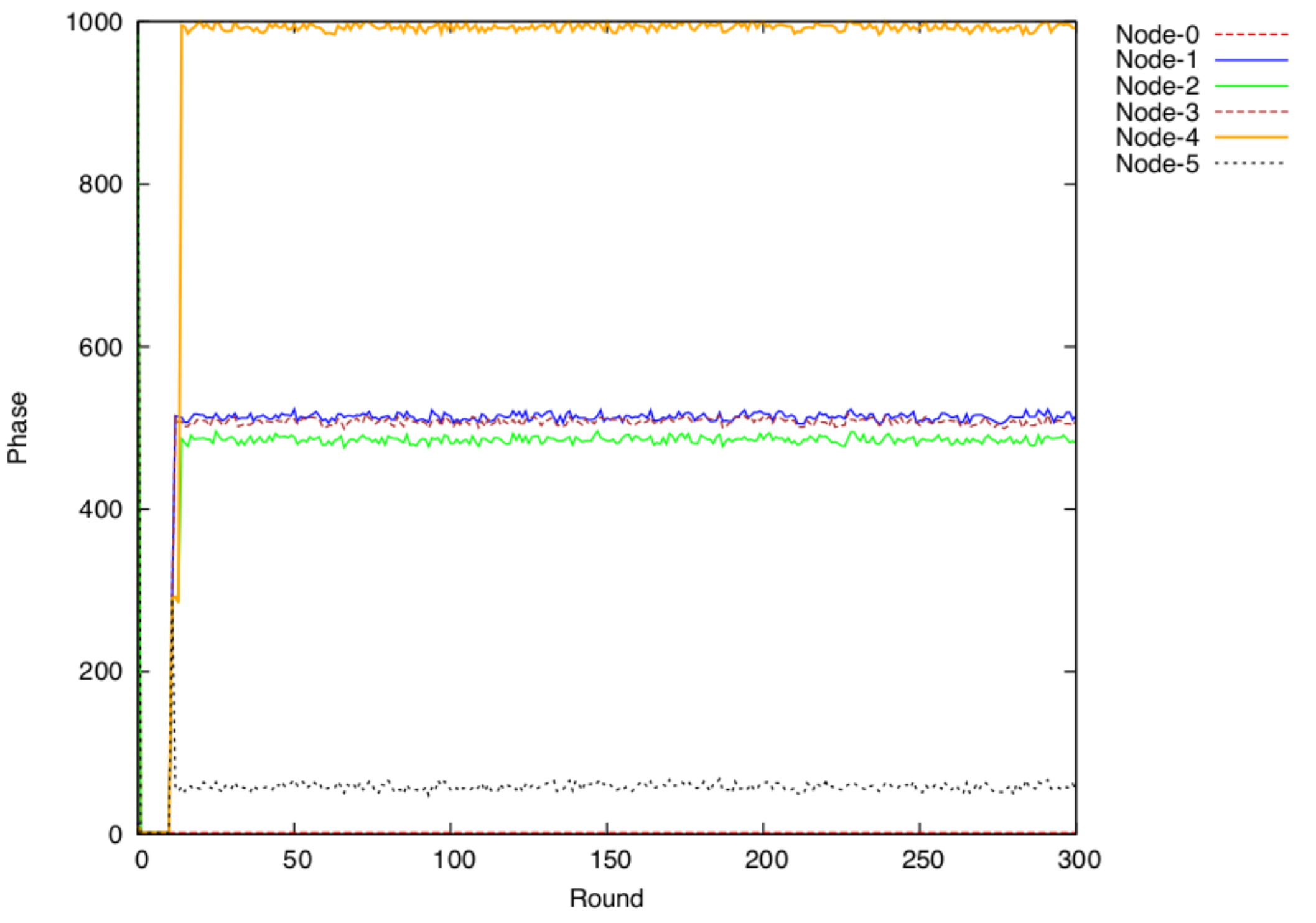}%
			\label{fig:6nodes-butterfly-result-light-bad}}
	}
	\caption{6-node dumbbell topology evaluation (problematic case).}
	\label{fig:6nodes-butterfly-result-bad}
	
\end{figure*}

\begin{figure*}
	\centerline{
		\subfloat[M-DWARF]{\includegraphics[scale=0.27]{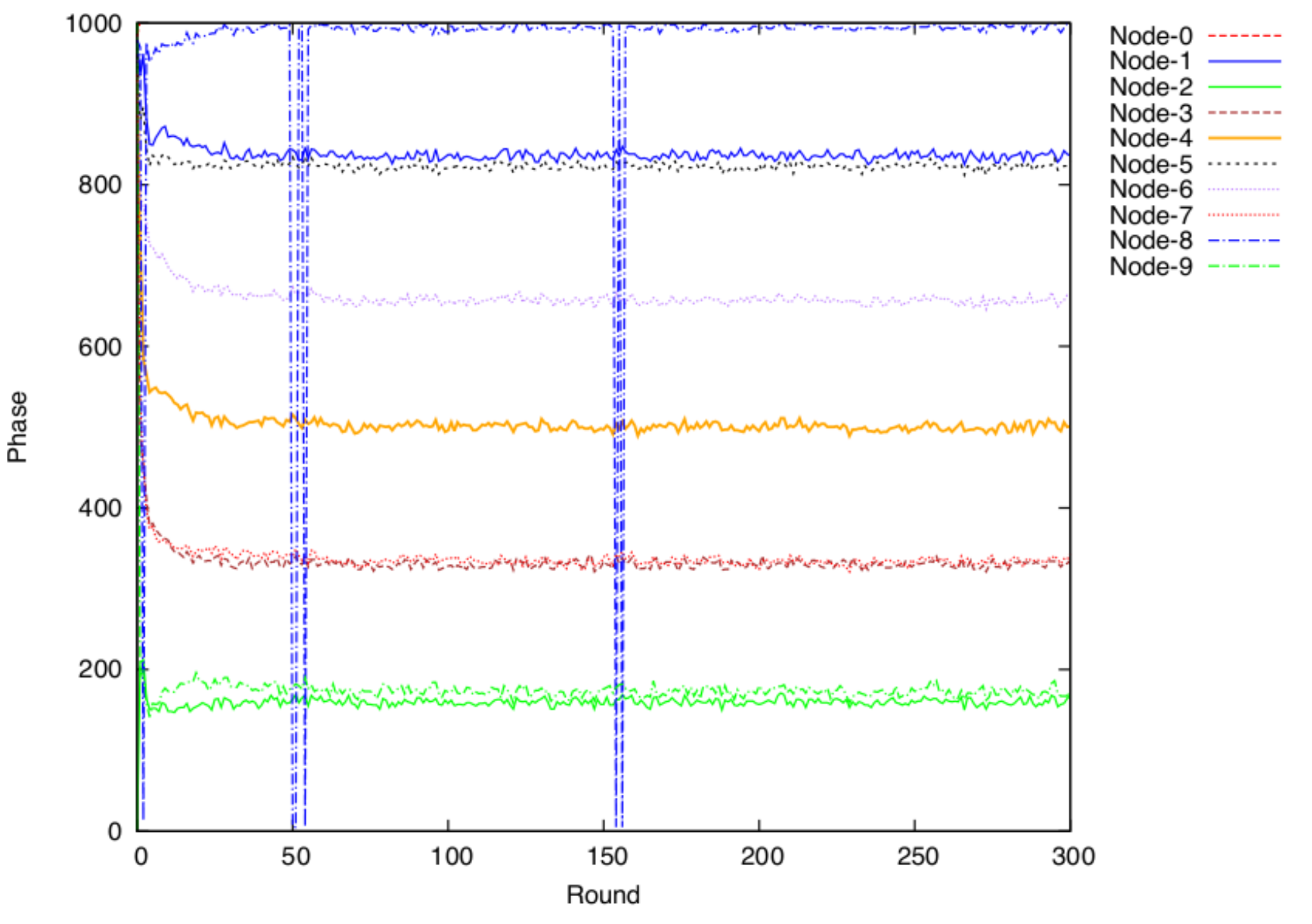}%
			\label{fig:10nodes-mesh-result-mdwarf-good}}
		\subfloat[EXT-DESYNC]{\includegraphics[scale=0.27]{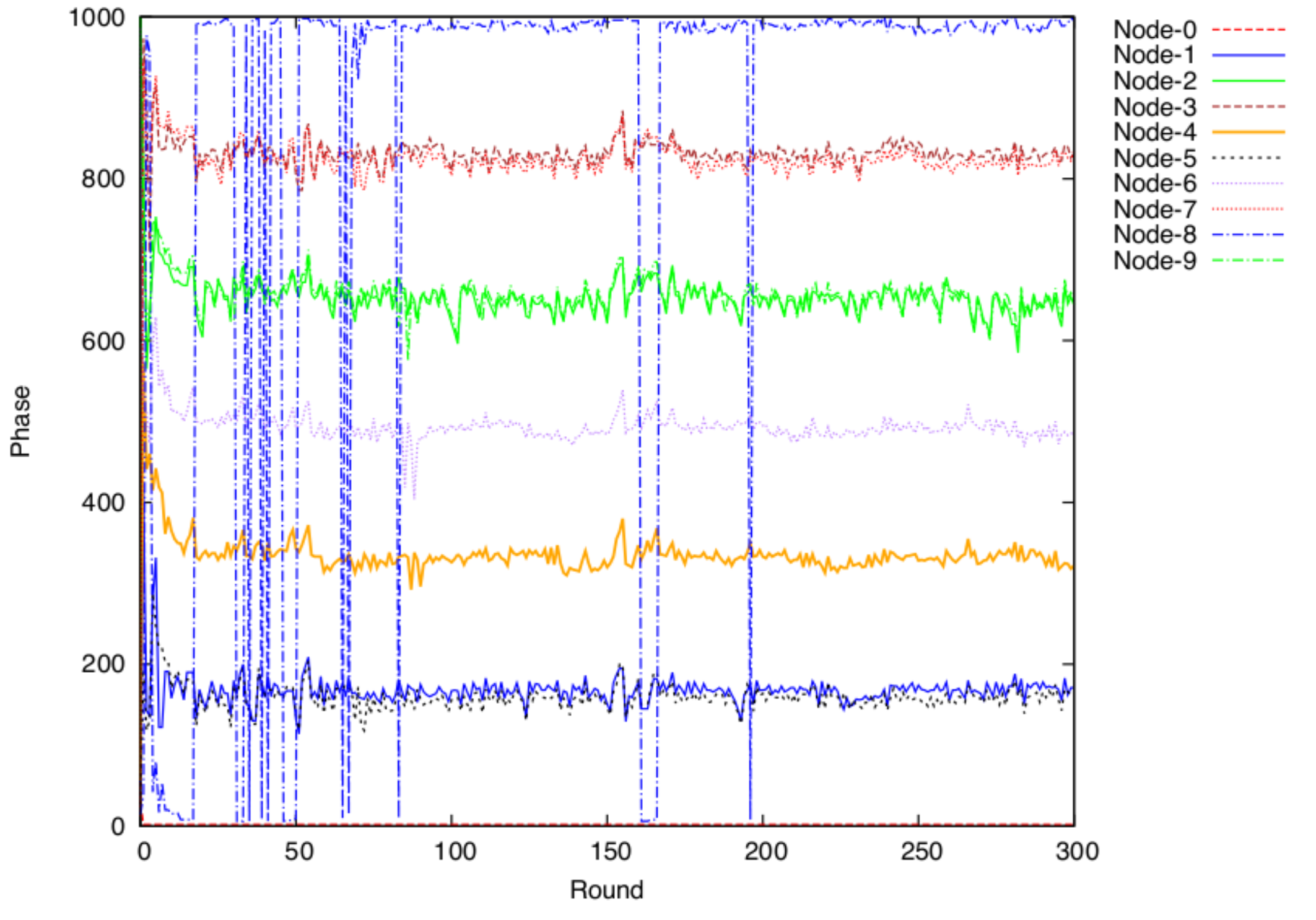}%
			\label{fig:10nodes-mesh-result-extdesync-good}}
		\subfloat[LIGHTWEIGHT]{\includegraphics[scale=0.27]{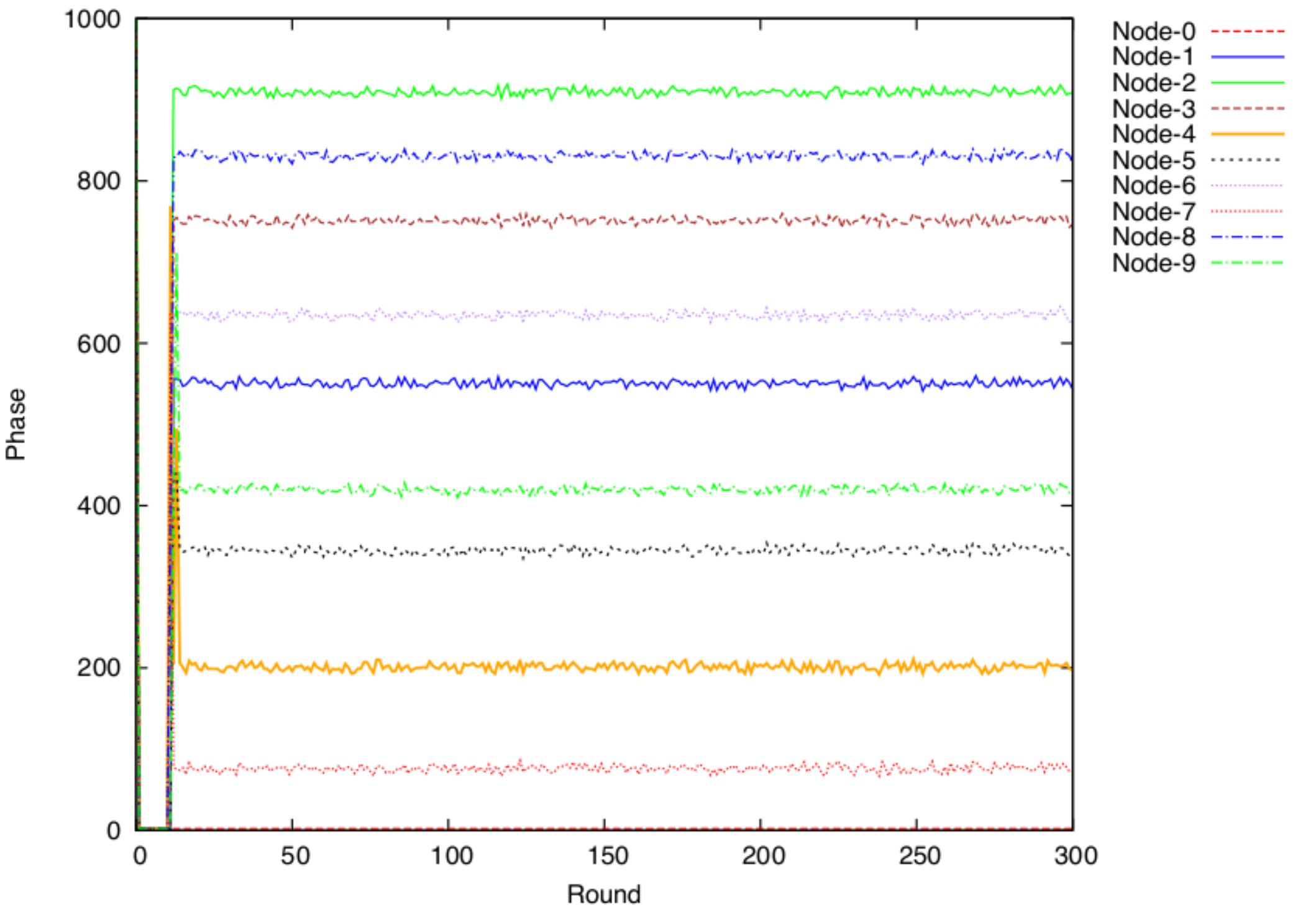}%
			\label{fig:10nodes-mesh-result-light-good}}
	}
	\caption{10-node mesh topology evaluation (average case).}
	\label{fig:10nodes-mesh-result-good}
	
\end{figure*}

\begin{figure*}
	\centerline{
		\subfloat[M-DWARF]{\includegraphics[scale=0.27]{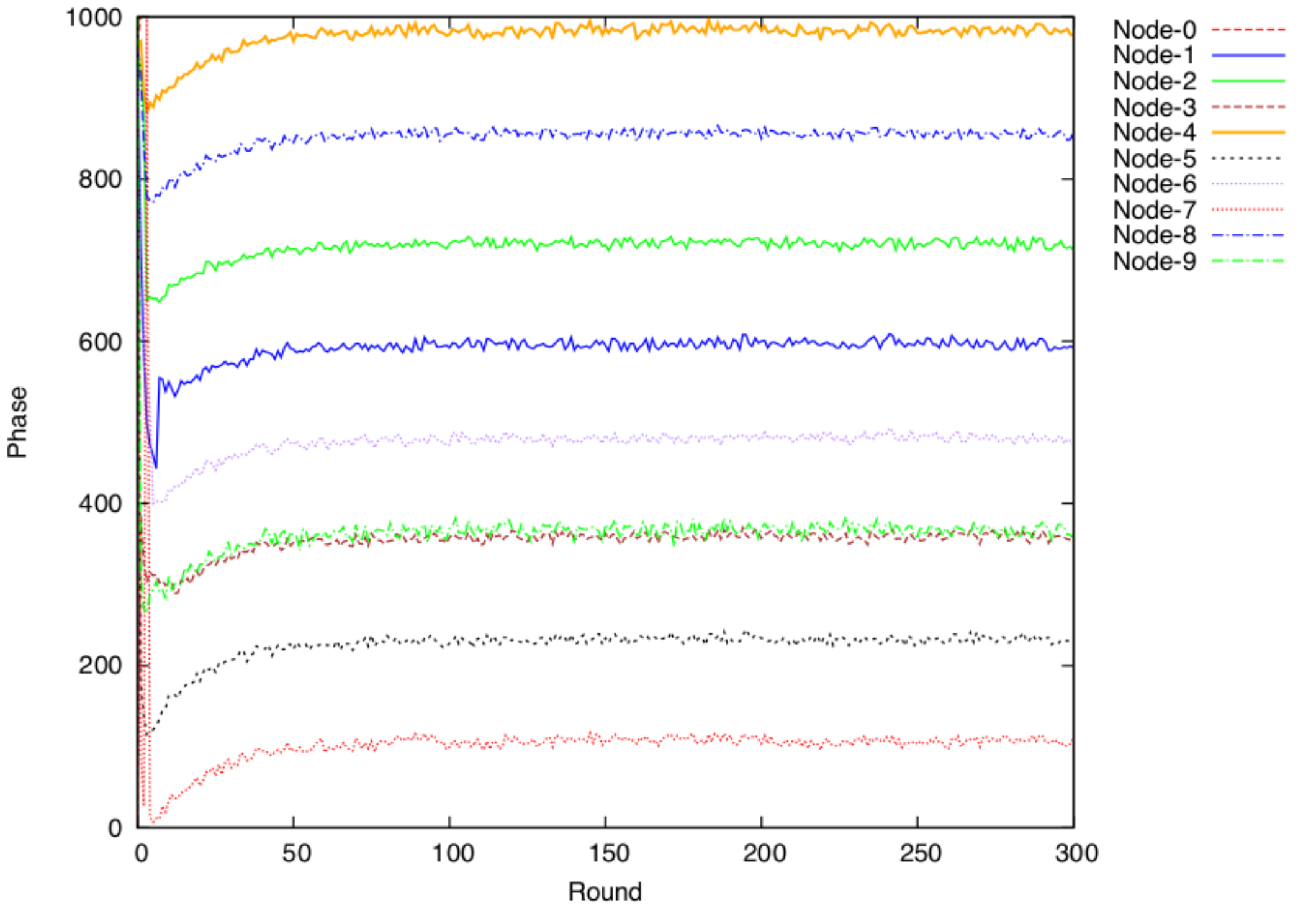}%
			\label{fig:10nodes-mesh-result-mdwarf-bad}}
		\subfloat[EXT-DESYNC]{\includegraphics[scale=0.27]{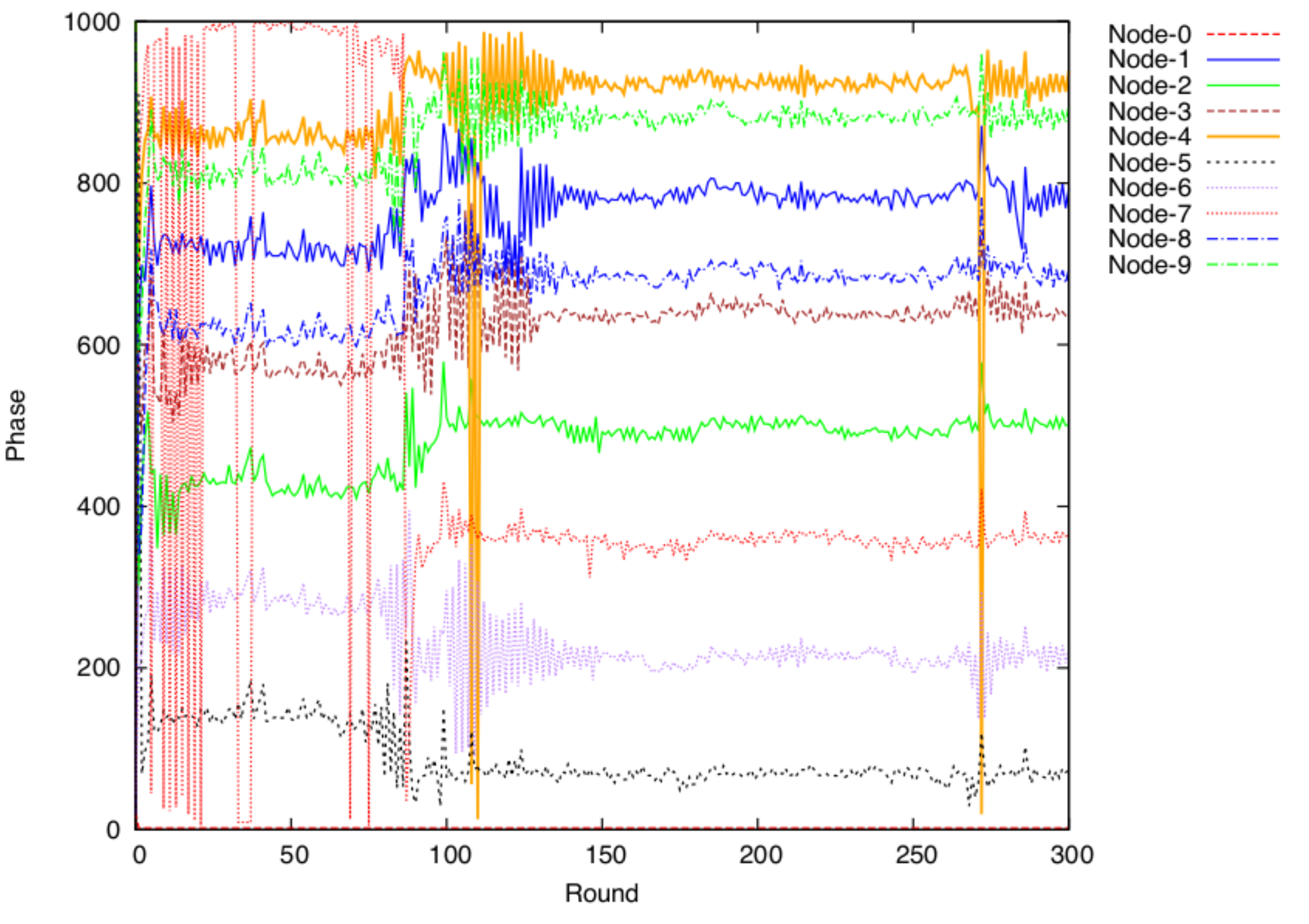}%
			\label{fig:10nodes-mesh-result-extdesync-bad}}
		\subfloat[LIGHTWEIGHT]{\includegraphics[scale=0.27]{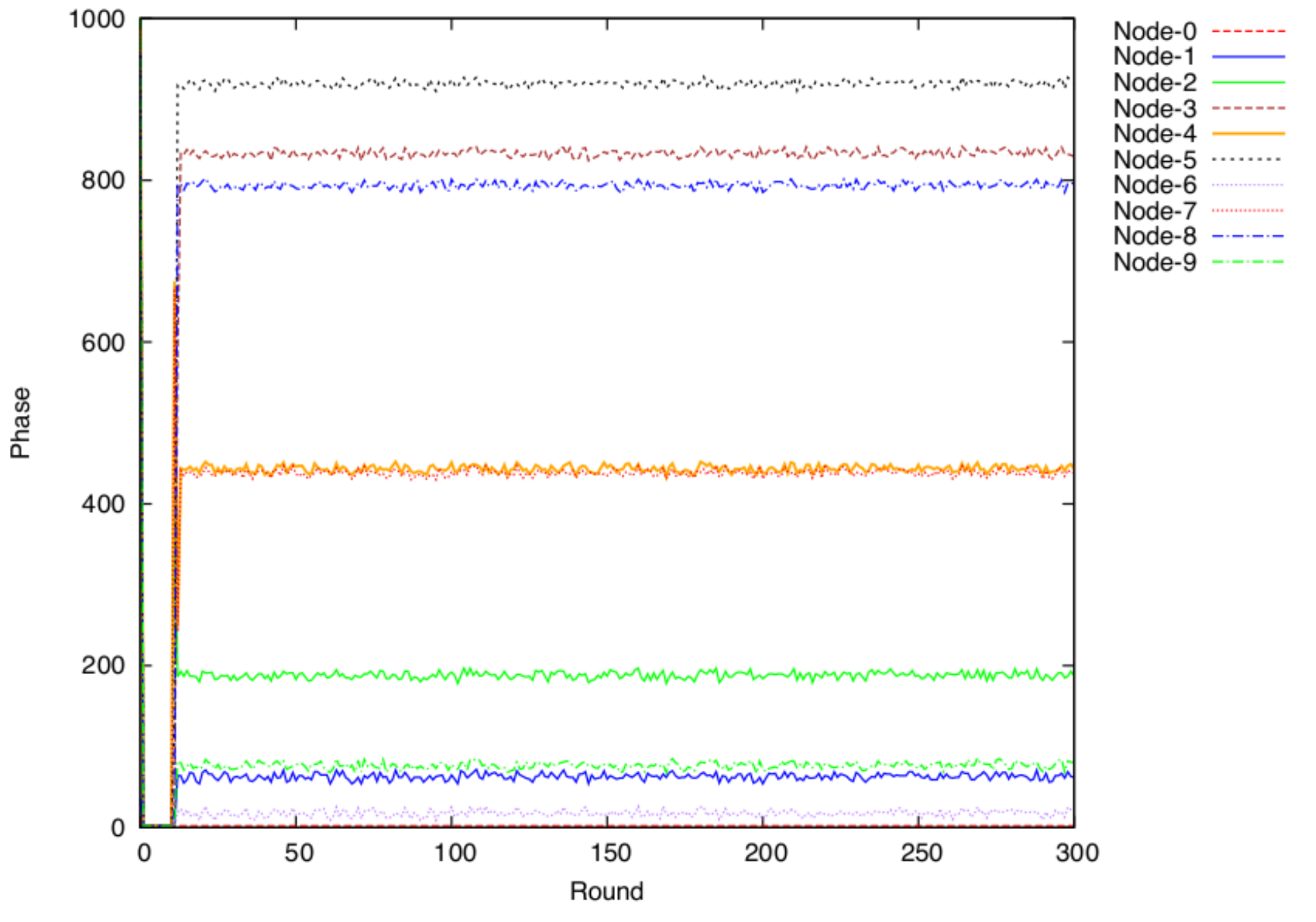}%
			\label{fig:10nodes-mesh-result-light-bad}}
	}
	\caption{10-node mesh topology evaluation (problematic case).}
	\label{fig:10nodes-mesh-result-bad}
	
\end{figure*}

\paragraph{Mesh Topology}
We use the same 10-node mesh topology as evaluated in \cite{4663417}; Figure \ref{fig:10nodes-topo} depicts such a topology. A solid line represents one-hop connectivity whereas a dash line represents two-hop connectivity. In \cite{4663417}, they state the optimal solution of this topology requires eight slots. However, in our simulation, we discover a better solution that requires only six slots, as illustrated in Figure \ref{fig:10nodes-mesh-result-mdwarf-good} for M-DWARF and Figure \ref{fig:10nodes-mesh-result-extdesync-good} for EXT-DESYNC. The problematic case for both M-DWARF and EXT-DESYNC is that the initial configuration is not proper, which will be explained in Problematic Cases below. However, the results show that while the phases of nodes in EXT-DESYNC are highly affected from high fluctuation (Figure \ref{fig:10nodes-mesh-result-extdesync-good} and \ref{fig:10nodes-mesh-result-extdesync-bad}), those in M-DWARF are comparatively much more stable (Figure \ref{fig:10nodes-mesh-result-mdwarf-good} and \ref{fig:10nodes-mesh-result-mdwarf-bad}).

\paragraph{A Summary Result of Average Cases}
From Table \ref{tab:simresults}, for small network sizes ($n<=6$), we can observe that all the algorithms easily achieve fast convergence rates. However, on medium-sized networks ($n>6$), M-DWARF and EXT-DESYNC alternate to have medium and fast convergence because nodes are more likely to have collision at the beginning. In average cases of all topologies, LIGHTWEIGHT produces quite similar results. For LIGHTWEIGHT, each node randomly chooses the beginning of a time slot by avoiding the collision with one-hop neighbors. Therefore, on average, LIGHTWEIGHT brings about several inequivalent interval gaps between two consecutive phase neighbor nodes, while M-DWARF and EXT-DESYNC attempt to equalize phase intervals. Therefore, LIGHTWEIGHT fails at channel utilization fairness. On the other hand, EXT-DESYNC suffers from poor phase stability because of its initial design to use phase information only from previous and next phase neighbors. In contrast, because M-DWARF uses phase information from all the phase neighbors, it has much better stability, which leads nodes to acquiring firmer time slots with smaller time guards and causing less message interference and collision. In conclusion, of the three algorithms, M-DWARF is the only one that produces fast convergence, attains high stability and maintains channel utilization fairness.

\begin{table*}[t]
\centering
\renewcommand{\arraystretch}{1.2}
\caption{Average-Case Comparison of Desynchronization Algorithms on Various Multihop Topologies}
{
\begin{tabular}{|l|l|l|l|l|l|}
\hline
\multicolumn{1}{|c|}{\multirow{2}{*}{Topology}} & \multicolumn{1}{c|}{\multirow{2}{*}{NetworkSize}} & \multicolumn{1}{c|}{\multirow{2}{*}{Matrix}} & \multicolumn{3}{c|}{Algorithms}    \\ \cline{4-6} 
\multicolumn{1}{|c|}{}                          & \multicolumn{1}{c|}{}                             & \multicolumn{1}{c|}{}                        & M-DWARF & \small EXT-DESYNC & LIGHTWEIGHT \\ \hline \hline
\multirow{6}{*}{Star}                           & \multirow{3}{*}{6 nodes (Figure \ref{fig:6nodes-star-eval})}                          & Convergence                                  &  	Fast    & Fast       & Fast        \\ \cline{3-6} 
                                                &                                                   & Stability                                    & High    & Low        & High        \\ \cline{3-6} 
                                                &                                                   & Fairness                                     & Yes    & Yes       & No        \\ \cline{2-6} 
                                                & \multirow{3}{*}{20 nodes (Figure \ref{fig:20nodes-star-eval})}                         & Convergence                                  & Fast        &  Medium          & Fast            \\ \cline{3-6} 
                                                &                                                   & Stability                                    & High        &    Low        &   High          \\ \cline{3-6} 
                                                &                                                   & Fairness                                     &  Yes       &   Yes         &   No          \\ \hline
\multirow{6}{*}{Chain}                          & \multirow{3}{*}{3 nodes (Figure \ref{fig:3nodes-chain-eval})}                          & Convergence                                  &  Fast       &  Fast          &  Fast           \\ \cline{3-6} 
                                                &                                                   & Stability                                    &  High       &  High          &  High           \\ \cline{3-6} 
                                                &                                                   & Fairness                                     &  Yes       &   Yes         &  No           \\ \cline{2-6} 
                                                & \multirow{3}{*}{10 nodes (Figure \ref{fig:10nodes-chain-eval})}                         & Convergence                                  &  Medium       & Fast           &  Fast           \\ \cline{3-6} 
                                                &                                                   & Stability                                    &  High       &   Medium         &    High         \\ \cline{3-6} 
                                                &                                                   & Fairness                                     & Yes        &  Yes          &  No           \\ \hline
\multirow{6}{*}{Cycle}                          & \multirow{3}{*}{4 nodes (Figure \ref{fig:4nodes-cycle-eval})}                          & Convergence                                  & Fast        & Fast           &   Fast          \\ \cline{3-6} 
                                                &                                                   & Stability                                    &  High       &   Medium         &   High          \\ \cline{3-6} 
                                                &                                                   & Fairness                                     &   Yes      &    Yes        &  No           \\ \cline{2-6} 
                                                & \multirow{3}{*}{10 nodes (Figure \ref{fig:10nodes-cycle-eval})}                         & Convergence                                  &       Medium       & Fast           &  Fast             \\ \cline{3-6} 
                                                &                                                   & Stability                                    &      High       &   Medium         &    High             \\ \cline{3-6} 
                                                &                                                   & Fairness                                     &        Yes        &  Yes          &  No              \\ \hline
\multirow{6}{*}{Dumbbell}                       & \multirow{3}{*}{6 nodes (Figure \ref{fig:6nodes-butterfly-eval})}                          & Convergence                                  &    Fast     &   Fast         &   Fast          \\ \cline{3-6} 
                                                &                                                   & Stability                                    &   High      &   Low         &   High          \\ \cline{3-6} 
                                                &                                                   & Fairness                                     &   Yes      &   Yes         & No            \\ \cline{2-6} 
                                                & \multirow{3}{*}{20 nodes (Figure \ref{fig:20nodes-butterfly-eval})}                         & Convergence                                  & Medium        &   Medium         &   Fast          \\ \cline{3-6} 
                                                &                                                   & Stability                                    &  High       &    Low        &   High          \\ \cline{3-6} 
                                                &                                                   & Fairness                                     &  Yes       &   Yes         &   No          \\ \hline
\multirow{3}{*}{Mesh}                           & \multirow{3}{*}{10 nodes (Figure \ref{fig:10nodes-topo})}                         & Convergence                                  &   Fast      &  Fast          &   Fast          \\ \cline{3-6} 
                                                &                                                   & Stability                                    &  High       &   Low         &   High          \\ \cline{3-6} 
                                                &                                                   & Fairness                                     &  Yes       &   Yes         &   No          \\ \hline
\end{tabular}
}

	\label{tab:simresults}
\end{table*}

\paragraph{A Note on Problematic Cases}
In addition to the average cases reported, problematic cases are also shown in order to spot anomalies during our experimentation. We have found that  these two main problems result from initial, random configuration of some simulation runs. Therefore, they are outliers that are separated from our average cases.

First, we observe \emph{phase fluctuation} problems. This kind of problems likely occurs in a dense network. Because the time gaps between nodes' phases are short and two or more randomly initialized nodes start their transmission at the same time, their packet frames will get collided. Moreover, their frames will keep colliding with each other for the next iterations because, when collision occurs, nodes are unaware of each other's phases. For M-DWARF, this kind of problem leads to a slower convergence rate and nodes' phases turn stable after a state of desynchrony. In contrast, for EXT-DESYNC, because of this problem, nodes' phases are fluctuating all over the simulation period and they never reach a state of desynchrony. The main reason for this is again because EXT-DESYNC uses information only from previous and next phase neighbors and the effects of any disturbance on one node will ripple throughout the network. 

We call the second problem \emph{phase suboptimality}. In some problematic cases of multi-hop experiments, the number of desynchronized time slots is sub-optimal. For example, in the problematic cases of the 10-node chain topology, M-DWARF desynchronizes the network to have ten time slots rather than the optimal six time slots. Another example of sub-optimal desynchronized slots is the problematic cases of the 10-node cycle topology which result in a desynchrony state of eight time slots, not the optimal four. Similarly, the problematic cases of dumbbell and mesh topologies produce sub-optimal six and eight time slots, respectively, with their optimal numbers of four and six time slots, respectively. As a result, a node's slot size will be reduced in percentage as in the equation. 
\begin{equation}
\frac{(\frac{T}{s_{o}} - \frac{T}{s_{d}})}{\frac{T}{s_{o}}}\times 100 = \frac{s_{d} - s_{o}}{s_{d}}\times 100
\end{equation}
It is given that $T$ is the period length (eg. 1,000 ms), $s_{o}$ is the optimal number of time slots according to the particular topology, and $s_{d}$ is the number of slots from desynchronization. For example, in case of the 10-node chain topology, where $s_{d}=10$ and $s_{o}=6$, the slot size shrinks as much as 40\%.
\begin{figure*}
	\centerline{
		\subfloat[]{\includegraphics[scale=0.2]{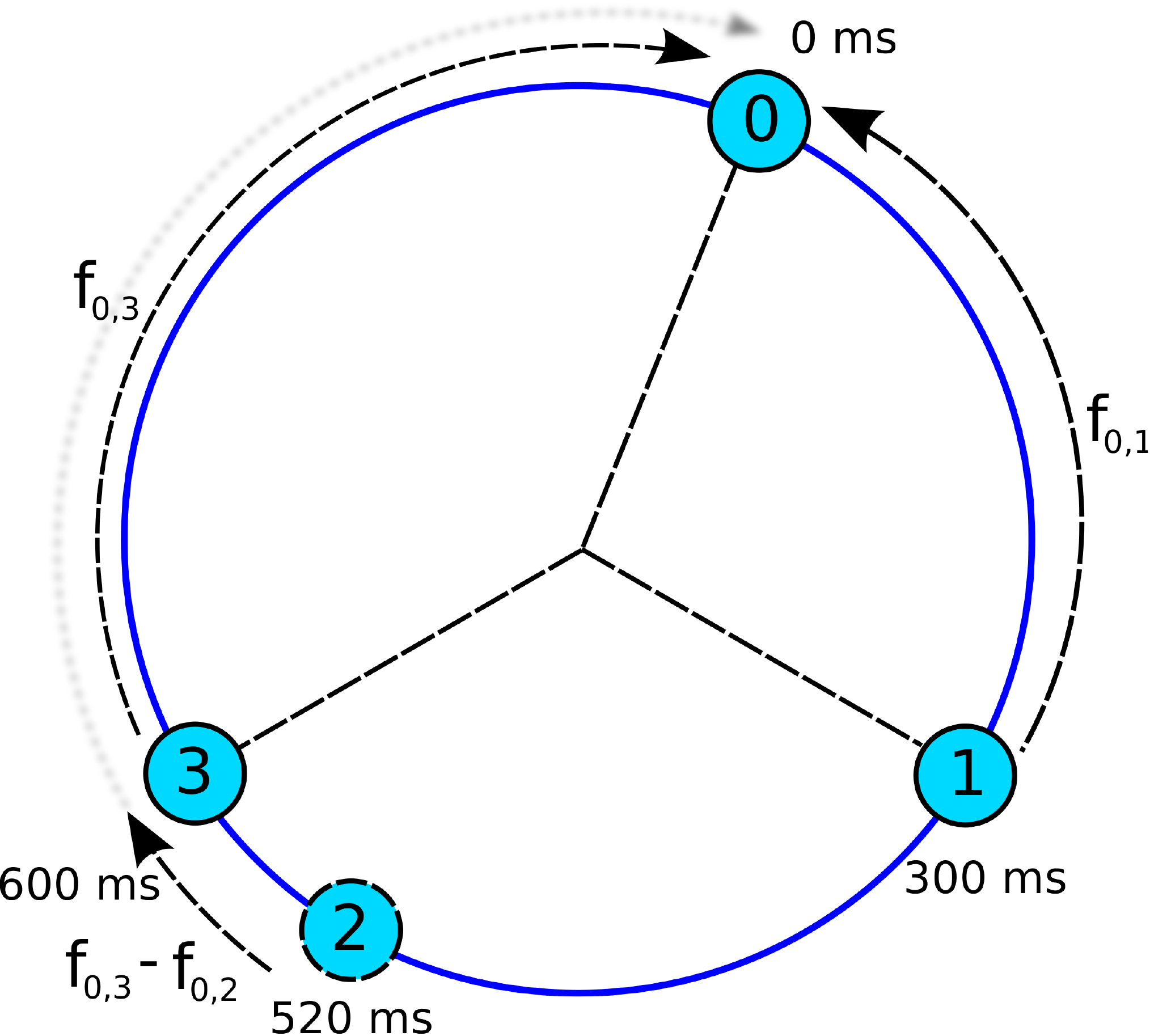}%
			\label{fig:4nodes-chain-dwarf-absorb-split}}
		\hfil
		\subfloat[]{\includegraphics[scale=0.2]{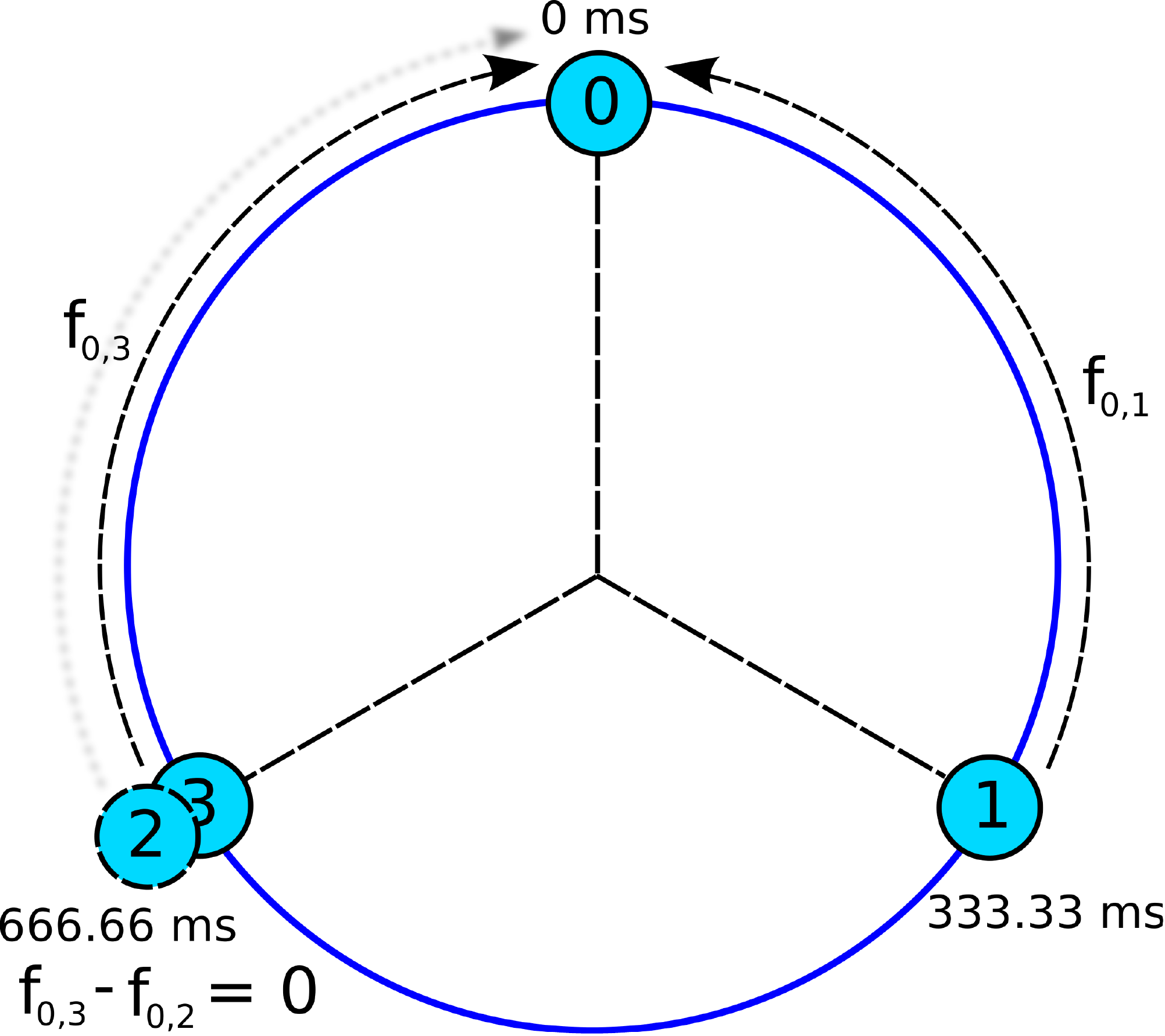}%
			\label{fig:4nodes-chain-dwarf-absorb-perfect}}
	}
	\caption{M-DWARF solves the problem with force absorption; the blur line represented an absorbed force. (a) shows node 2's force is being absorbed by node 3. (b) displays the perfect desynchrony state.}
	\label{fig:4nodes-chain-dwarf-absorb}
	
\end{figure*}

\begin{figure*}
		\centerline{
			\subfloat[]{\includegraphics[scale=0.3]{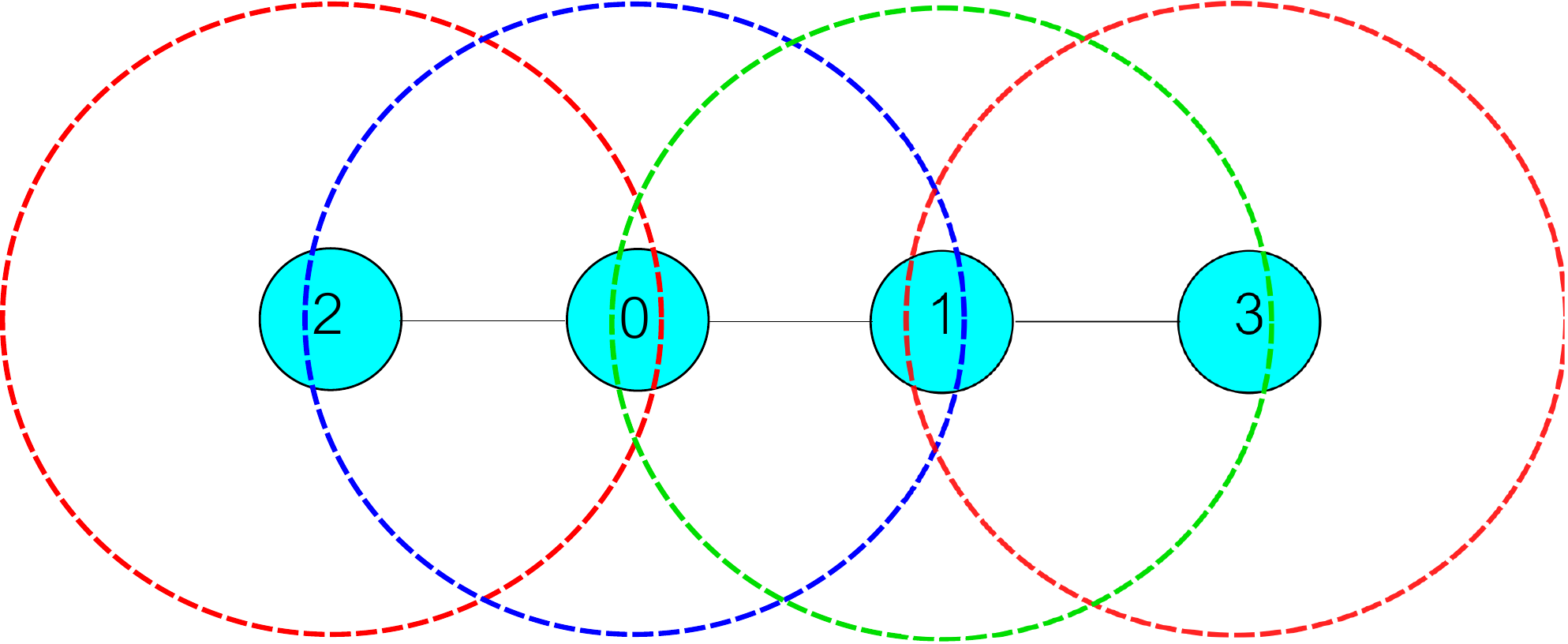}%
				\label{fig:4nodes-limitation-topo}}
			\hfill
			\subfloat[]{\includegraphics[scale=0.9]{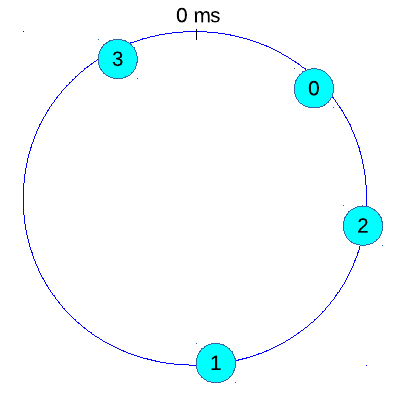}%
				\label{fig:4nodes-limitation-initial}}
			\hfill
			\subfloat[]{\includegraphics[scale=0.9]{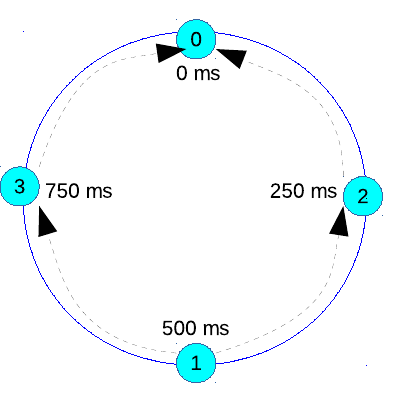}%
				\label{fig:4nodes-limitaion-synchronized}}
		}
		\caption{ (a) is a simple four-node chain topology and (b) is an initial phrase configuration in the local view of node 0 which results in the sub-optimal desynchrony state in (c) because node 2 and 3 are beyond two-hop neighbors but not phase neighbors in the circular framework. As a result, the artificial forces of node 2 and 3 cannot be absorbed.}
		\label{fig:4nodes-limitation} 
\end{figure*}

To figure out what causes these problematic cases, we should take a look at a simple topology of four-node chain networks in Figure \ref{fig:4nodes-limitation}. The initial phases of all the four nodes in the node 0's local view are displayed in Figure \ref{fig:4nodes-limitation-initial} which differs from that of Figure \ref{fig:4nodes-chain-dwarf-absorb} in that the node 2 and 3, which beyond two-hop neighbors, are not the phase neighbors in the circular desynchronization framework. With this initial phase setting, the node 2 and 3's phases cannot be moved to be at the same position because their repulsive forces cannot be absorbed. In sum, none of the forces can be absorbed and none of the nodes' phases can be merged into the same phase position even though node 2 and 3 are beyond two-hop neighbors. This example results in four time slots in four-node chain networks whereas three time slots are optimal. In short, if two nodes are beyond two-hop neighbors but their initial phases are not phase neighbors, the number of time slots in a desynchrony state shall be sub-optimal. However, these problematic cases also affect EXT-DESYNC with even greater phrase fluctuation as mentioned above. 

However, it is shown in \cite{5062256,5062165} that desynchronization is reducible to graph coloring problems  and is well known that k-coloring is an NP-complete problem when $k >= 3$. Moreoever, based on graph coloring, Ergen et al. also proves that the TDMA scheduling problem is NP-complete \cite{Ergen:2010:TDMAAlgo}. As a result, it is even more challenging when one tries to solve this class of problems using distributed algorithms.

\section{Impacts of Period Length (T)}
\label{sec:period_length}
In previous experiments, we set the time period to 1,000 milliseconds. In this experiment, we vary the period from 1,000 to 3,000 milliseconds. To obviously see the impact of period length, we use the star topology which is a multi-hop topology in which every node is a two-hop neighbor to all the other nodes except for the central one which is the only one-hop neighbor of all the nodes (Figure \ref{fig:20nodes-star-eval}). We set the number of nodes to 30.
With this topology and the number of nodes, the network is saturated within the 1,000-millisecond period. 

\begin{figure*}
\centerline{
	\subfloat[$T$ = 1000 ms]{\includegraphics[scale=0.27]{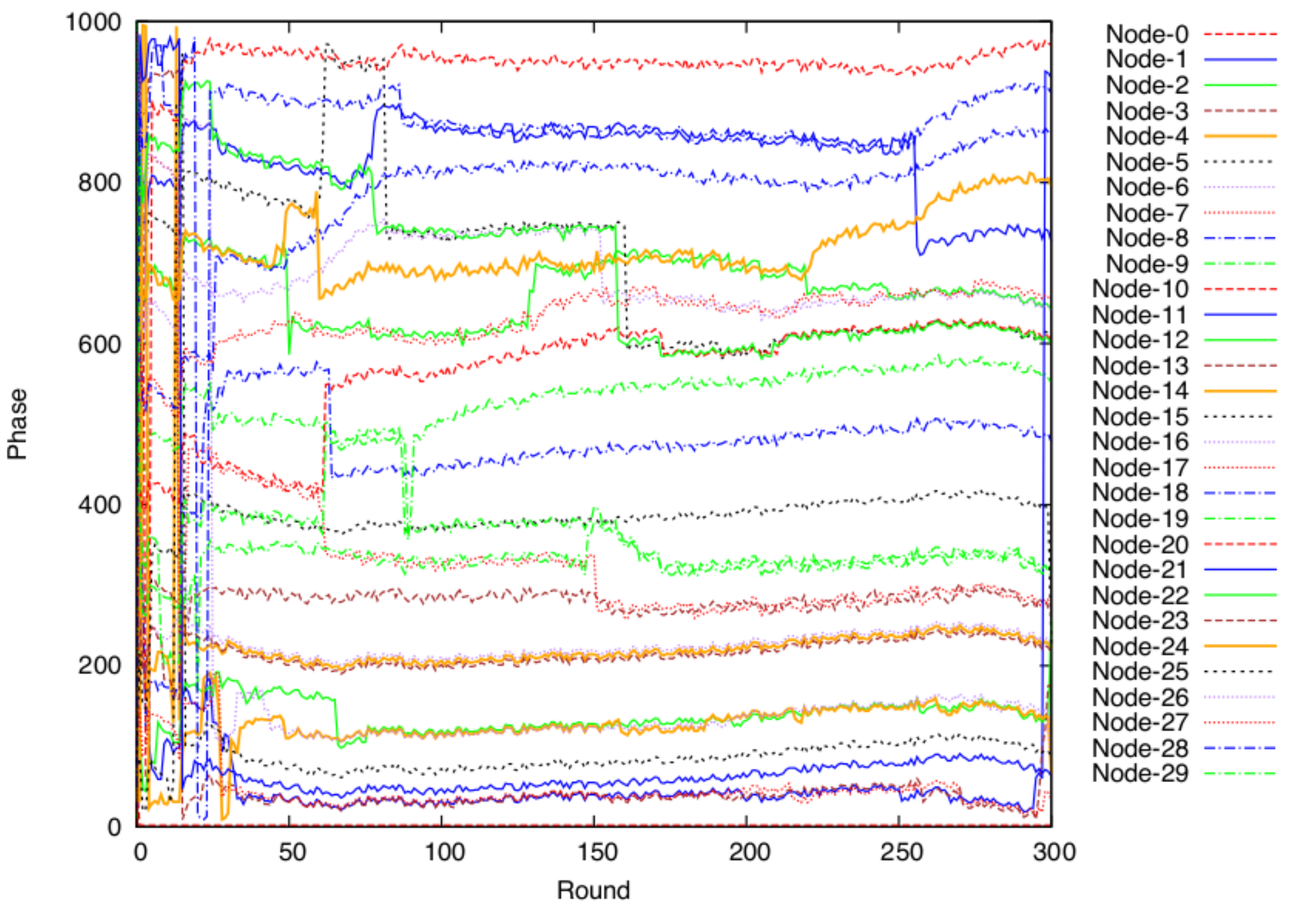}%
	\label{fig:period1000-mdwarf}}
	\subfloat[$T$ = 2000 ms]{\includegraphics[scale=0.27]{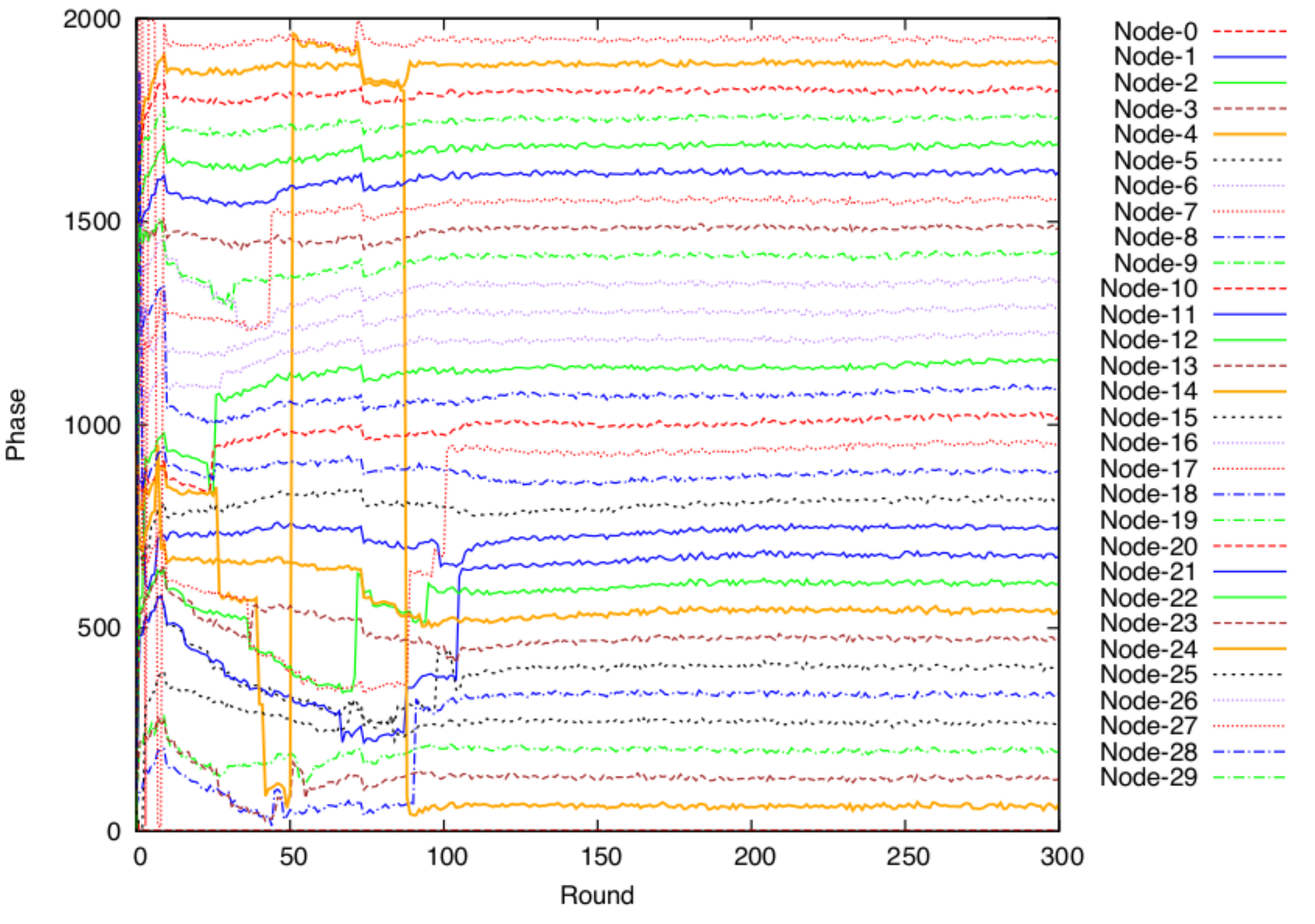}%
	\label{fig:period2000-mdwarf}}
	\subfloat[$T$ = 3000 ms]{\includegraphics[scale=0.27]{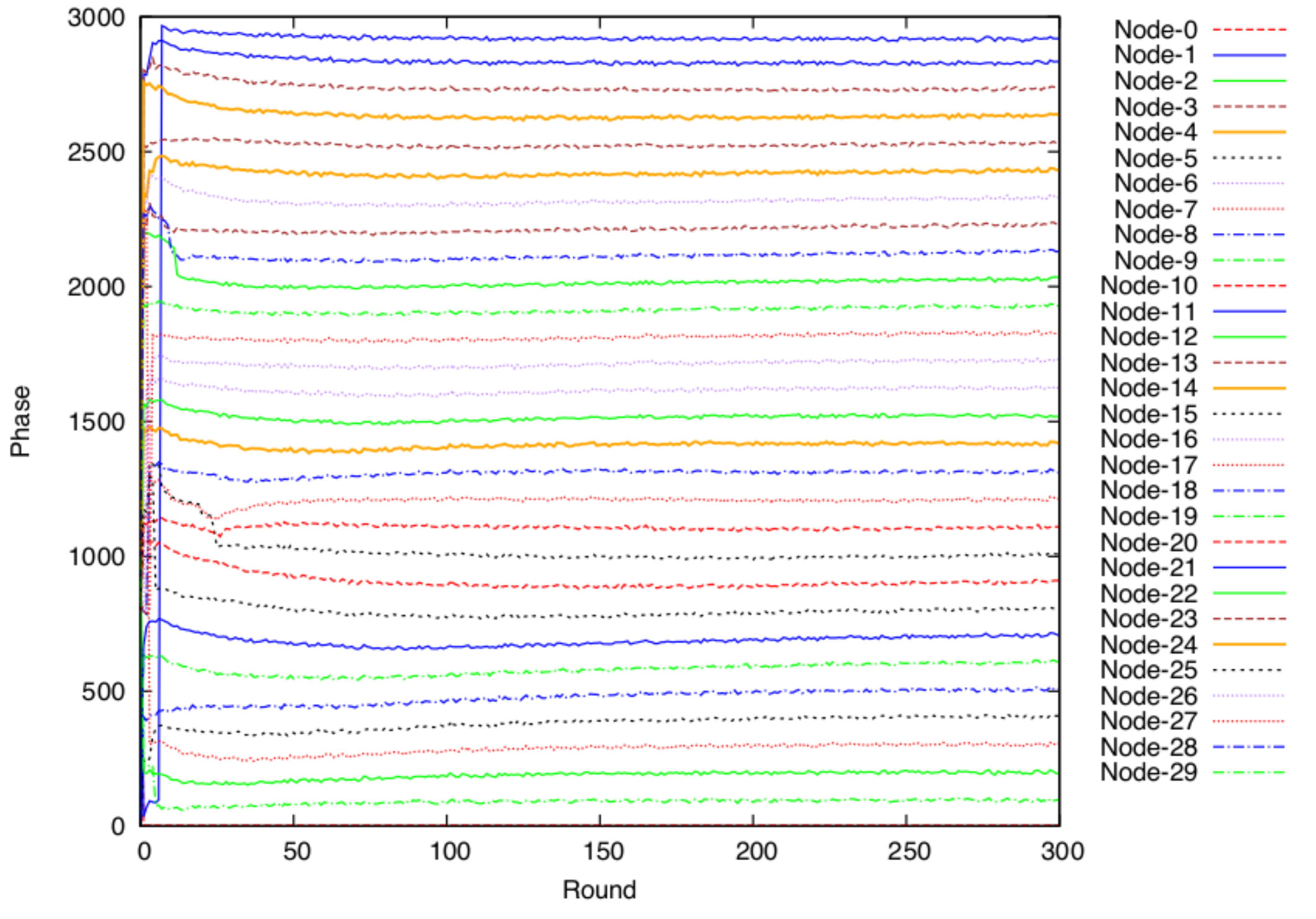}%
	\label{fig:period3000-mdwarf}}
}
\caption{Impact of period length ($T$) in M-DWARF}
\label{fig:period-mdwarf}
\end{figure*}

\begin{figure*}
\centerline{
	\subfloat[$T$ = 1000 ms]{\includegraphics[scale=0.27]{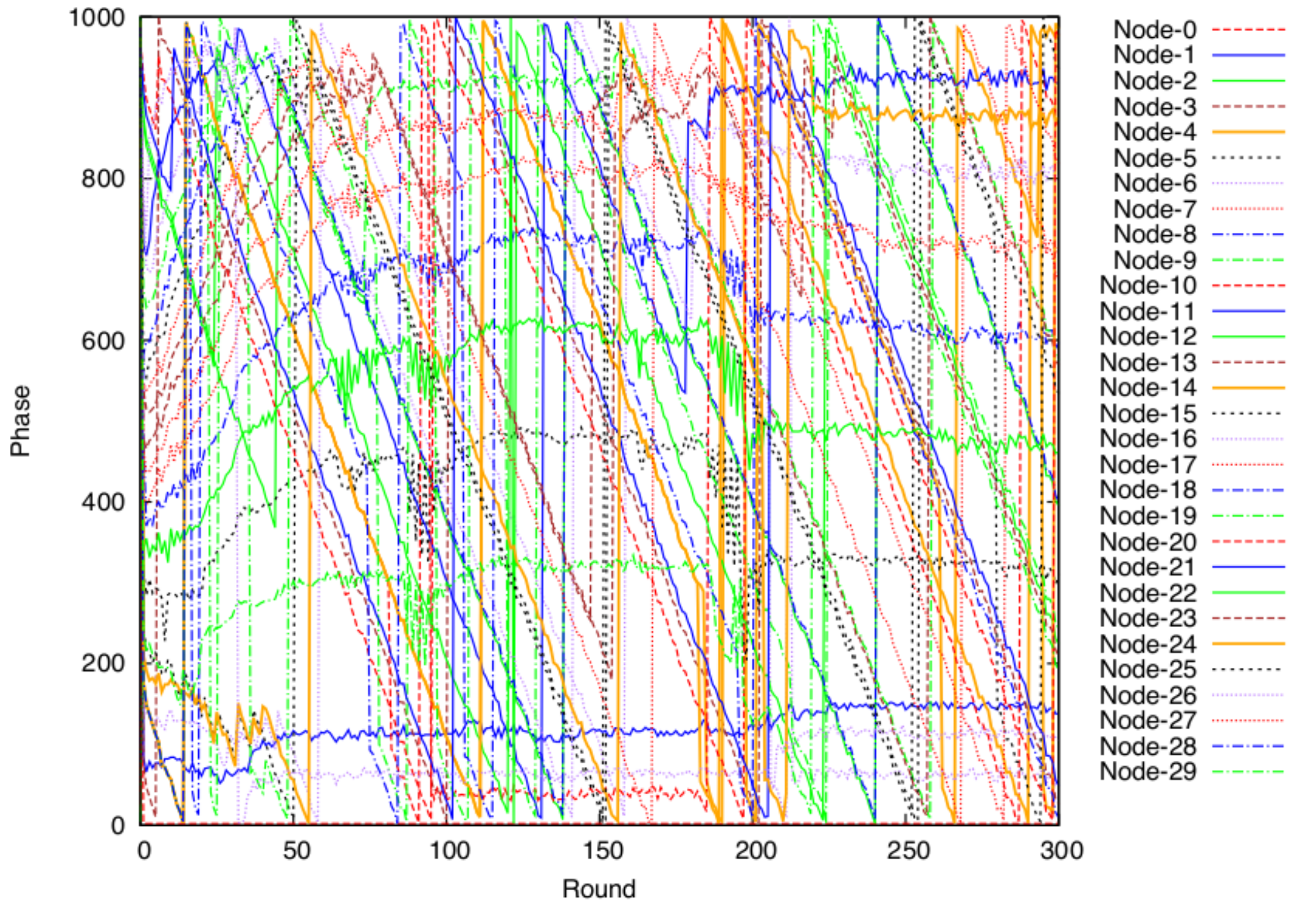}%
	\label{fig:period1000-ext}}
	\subfloat[$T$ = 2000 ms]{\includegraphics[scale=0.27]{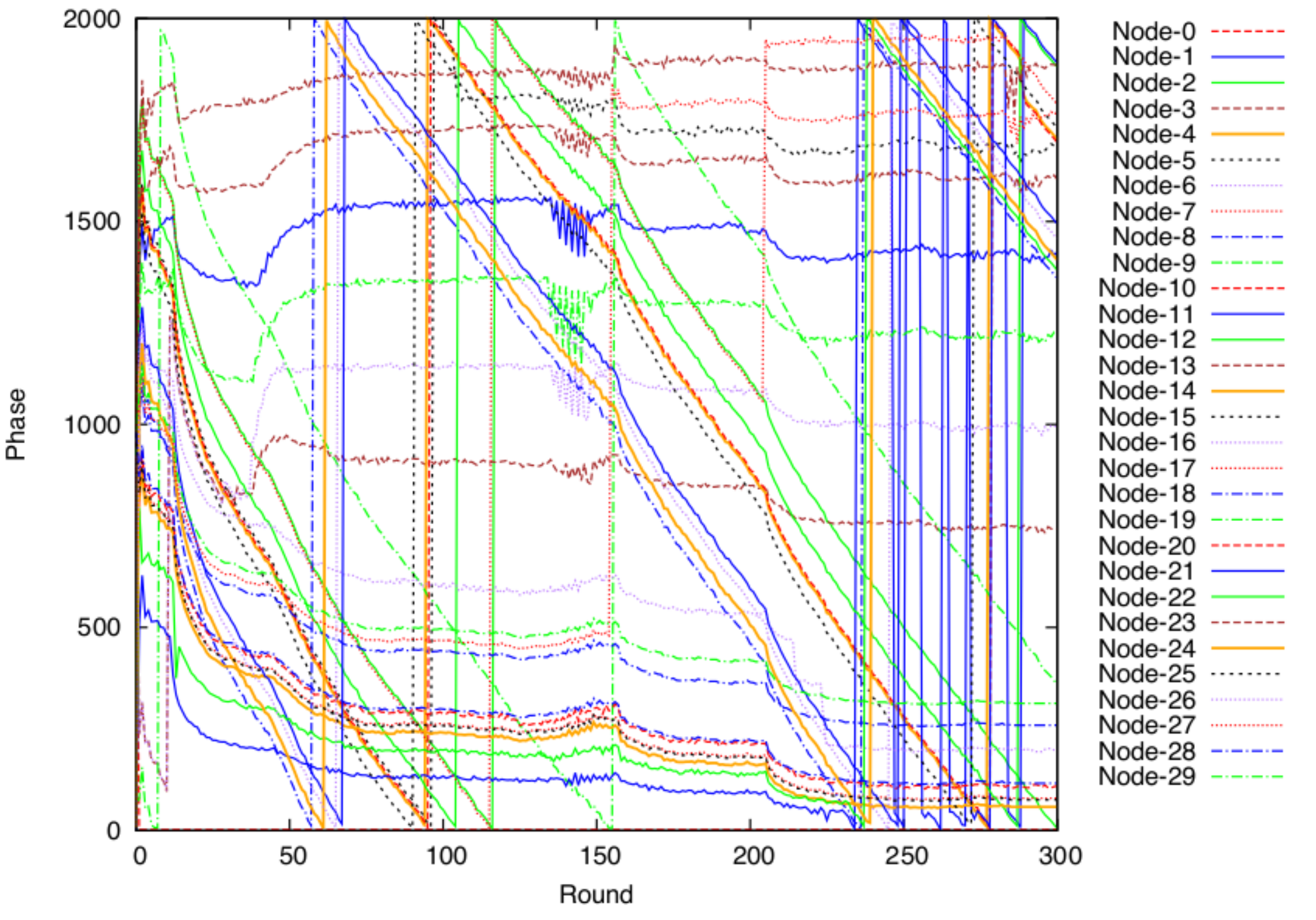}%
	\label{fig:period2000-ext}}
	\subfloat[$T$ = 3000 ms]{\includegraphics[scale=0.27]{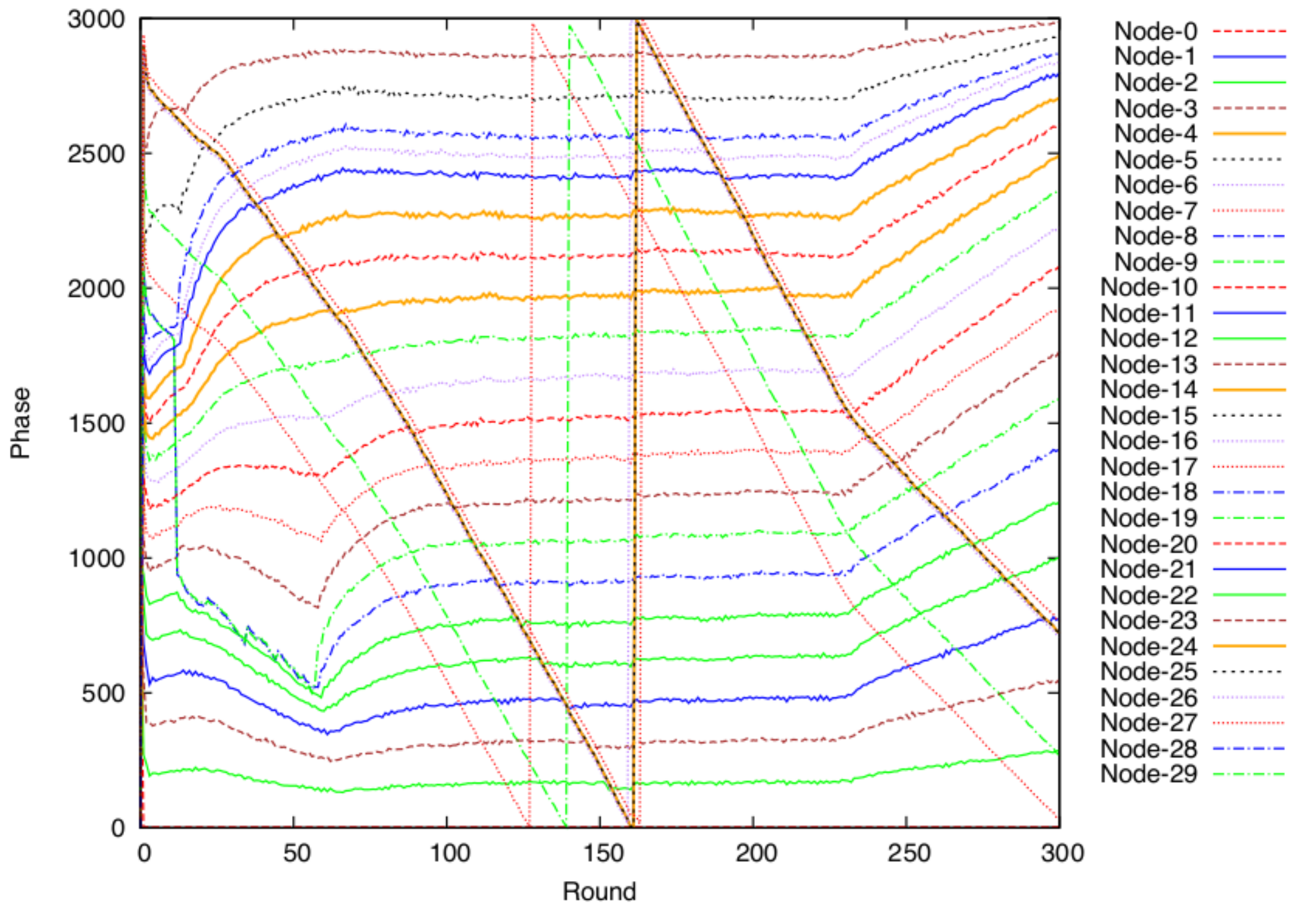}%
	\label{fig:period3000-ext}}
}
\caption{Impact of period length ($T$) in EXT-DESYNC}
\label{fig:period-ext}
\end{figure*}

Figure \ref{fig:period-mdwarf} and \ref{fig:period-ext} show the phase graphs when the algorithms vary the period lengths of M-DWARF and EXT-DESYNC respectively. At the 1,000-ms period, the network is saturated for 30 nodes and both the algorithms seem not to be able to function properly but EXT-DESYNC shows more phase fluctuation. At the 2,000-ms period, M-DWARF still is able to desynchronize the network even if the convergence rate is slow. EXT-DESYNC, on the other hand, fails to desynchronize because the fluctuation rate is too high for the network to converge. When the network is relaxed with the 3,000-ms period, M-DWARF reaches a state of desynchrony within approximately 50 time periods from the total of 300. In contrast, node phases in EXT-DESYNC still fluctuate even at the end of the simulations.

We argue that desynchronization algorithms do not need to stress the network with such a small 1,000-ms period on a single-hop 30-node topology. They can easily adjust the period length according to the network size. For example, if the number of neighbor nodes is 1 to 10, desynchronization algorithms may select $T$ = 1,000 ms. Whereas, the period length of a 30-node network shall be set to be $T$ = 3,000 ms, allowing more gaps between neighbor phases.

\section{Channel Utilization Fairness}
\label{sec:channel_util_fair}

\begin{figure*}
\centerline{
	\subfloat[$T$ = 1000 ms]{\includegraphics[scale=0.27]{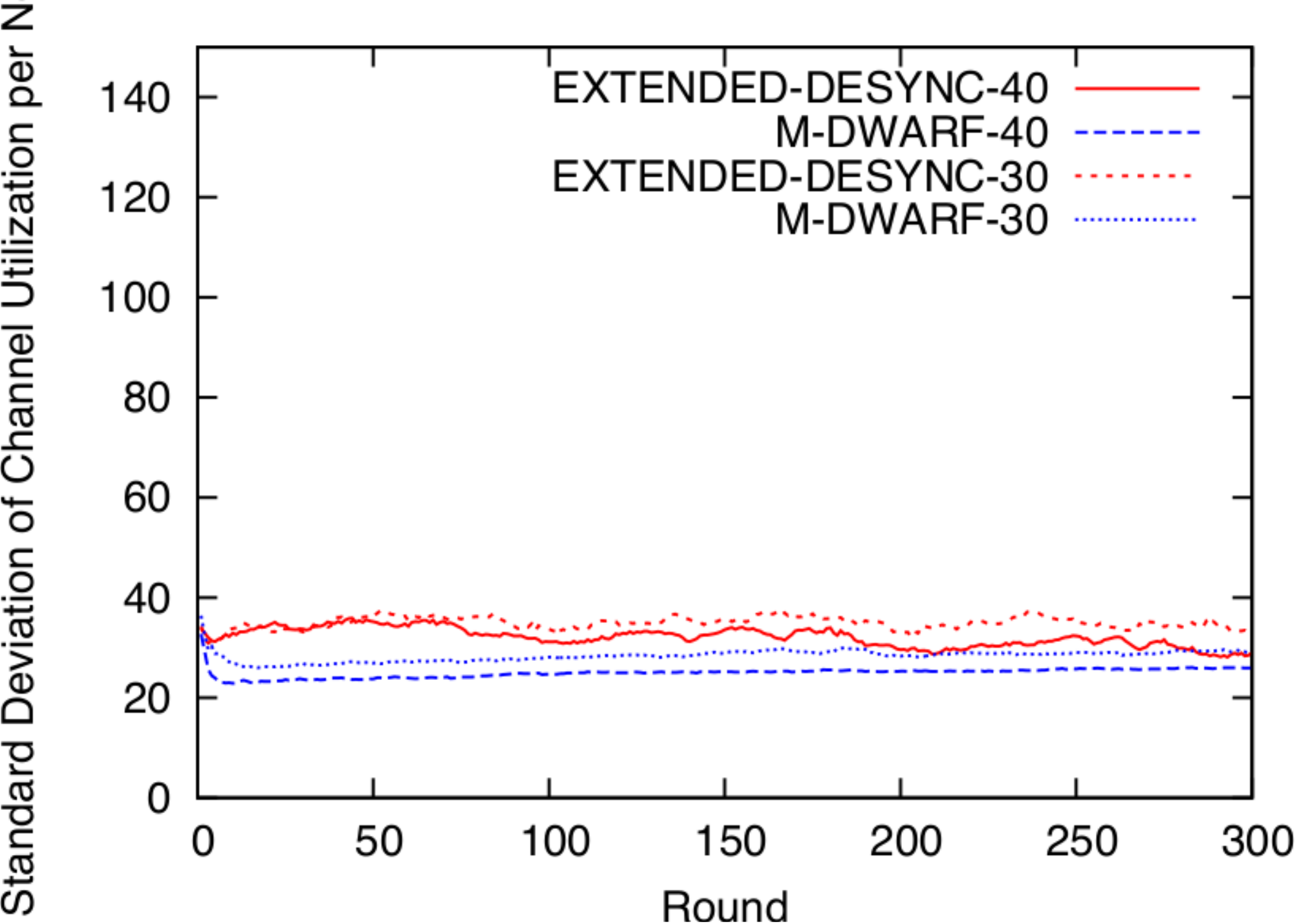}%
	\label{fig:period1000-fairness-30-40}}
	\hfil
	\subfloat[$T$ = 2000 ms]{\includegraphics[scale=0.27]{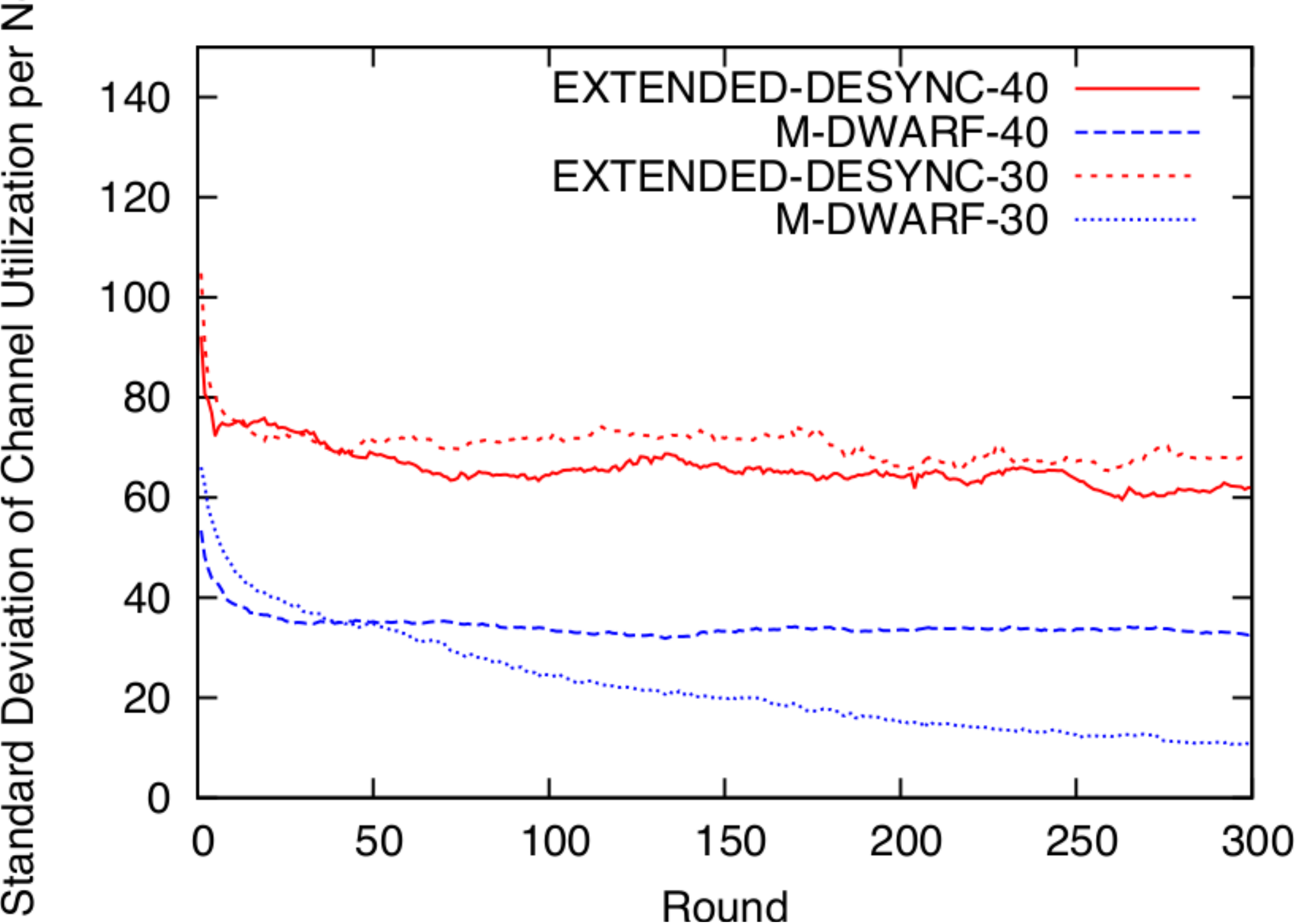}%
	\label{fig:period2000-fairness-30-40}}
	\hfil
	\subfloat[$T$ = 3000 ms]{\includegraphics[scale=0.27]{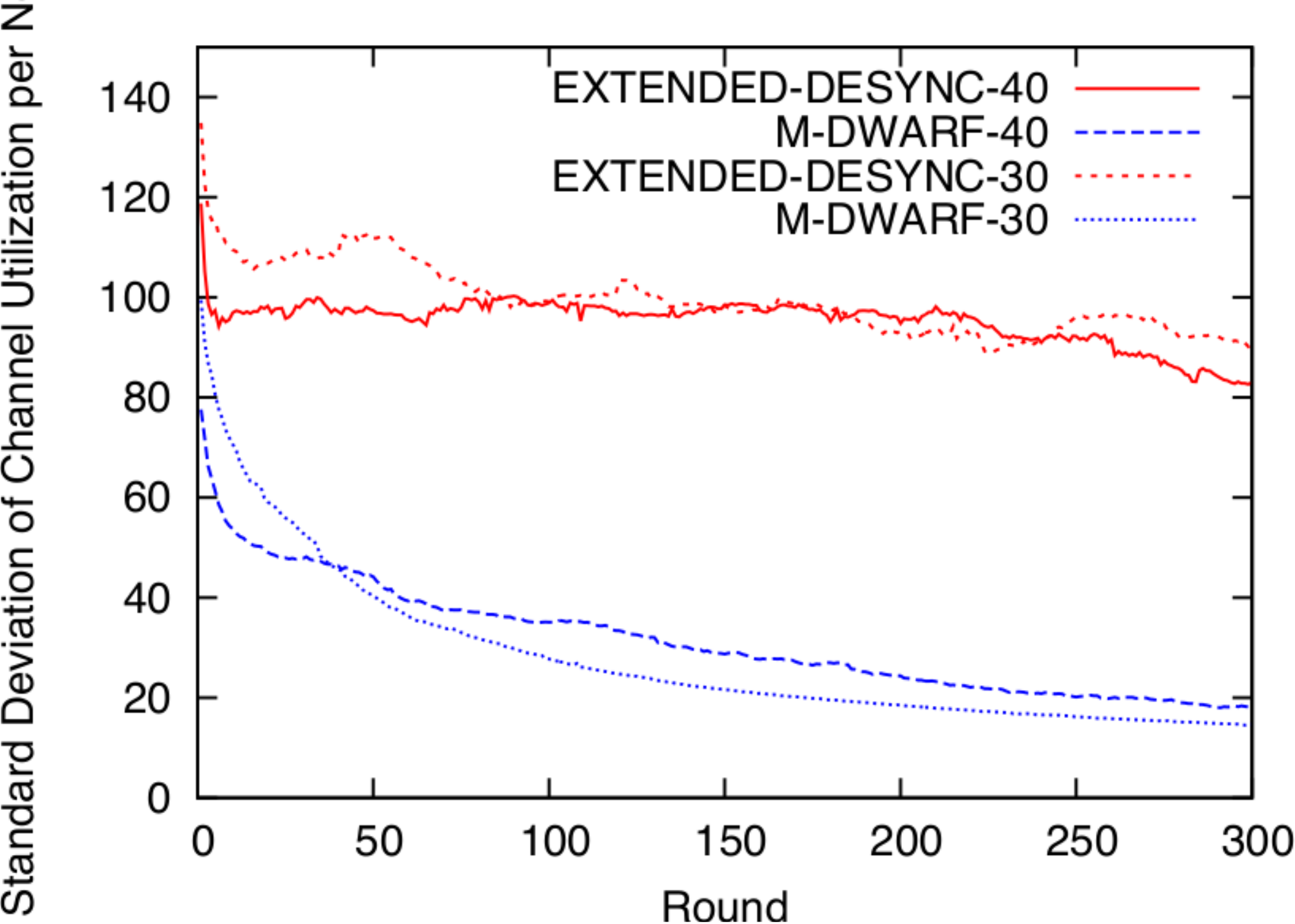}%
	\label{fig:period3000-fairness-30-40}}
}
\caption{Average standard deviation of channel utilization per node of 30 and 40 node networks}
\label{fig:period-fairness-30-40}
\end{figure*}

\begin{figure*}
\centerline{
	\subfloat[$T$ = 1000 ms]{\includegraphics[scale=0.27]{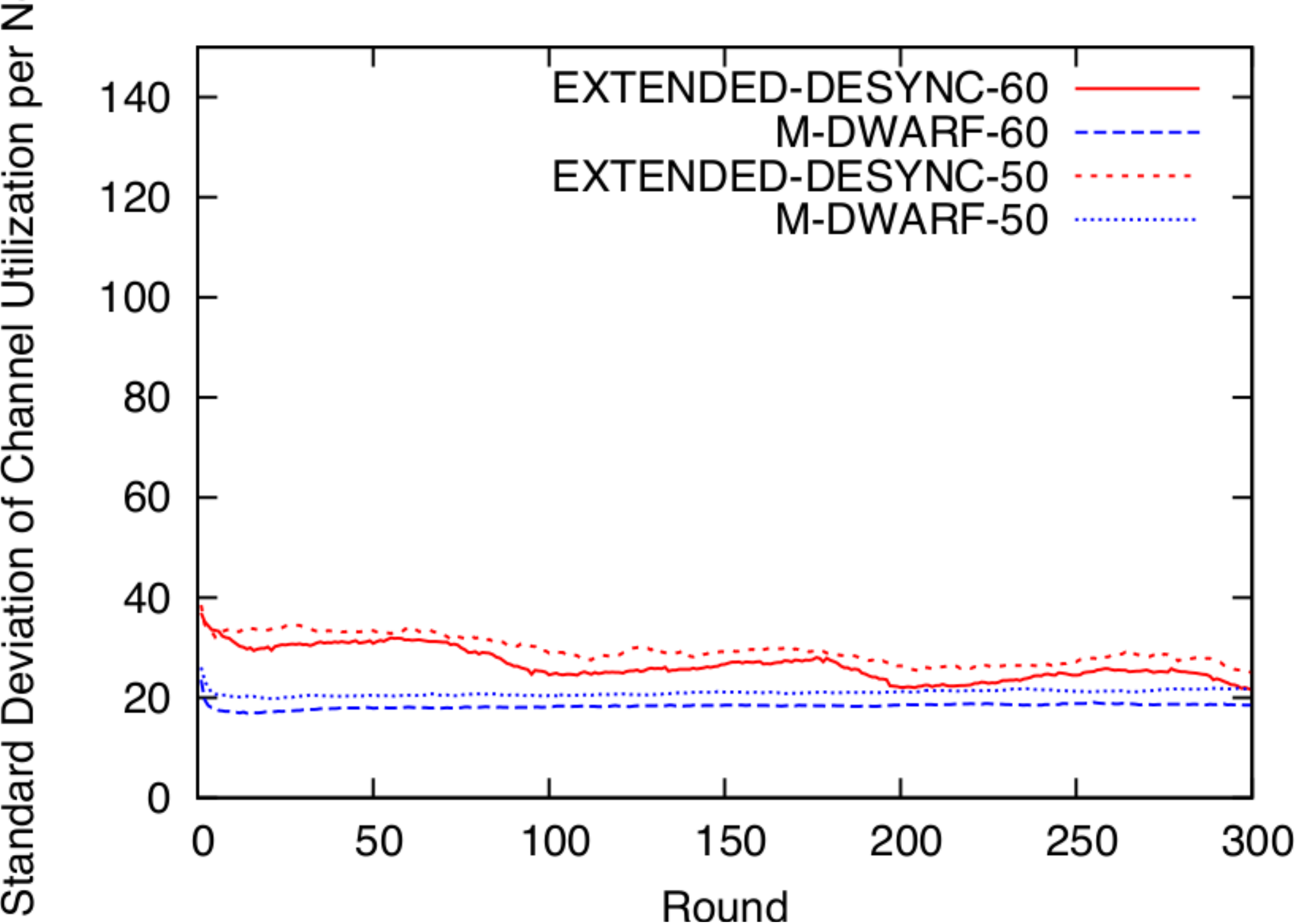}%
	\label{fig:period1000-fairness-50-60}}
	\hfil
	\subfloat[$T$ = 2000 ms]{\includegraphics[scale=0.27]{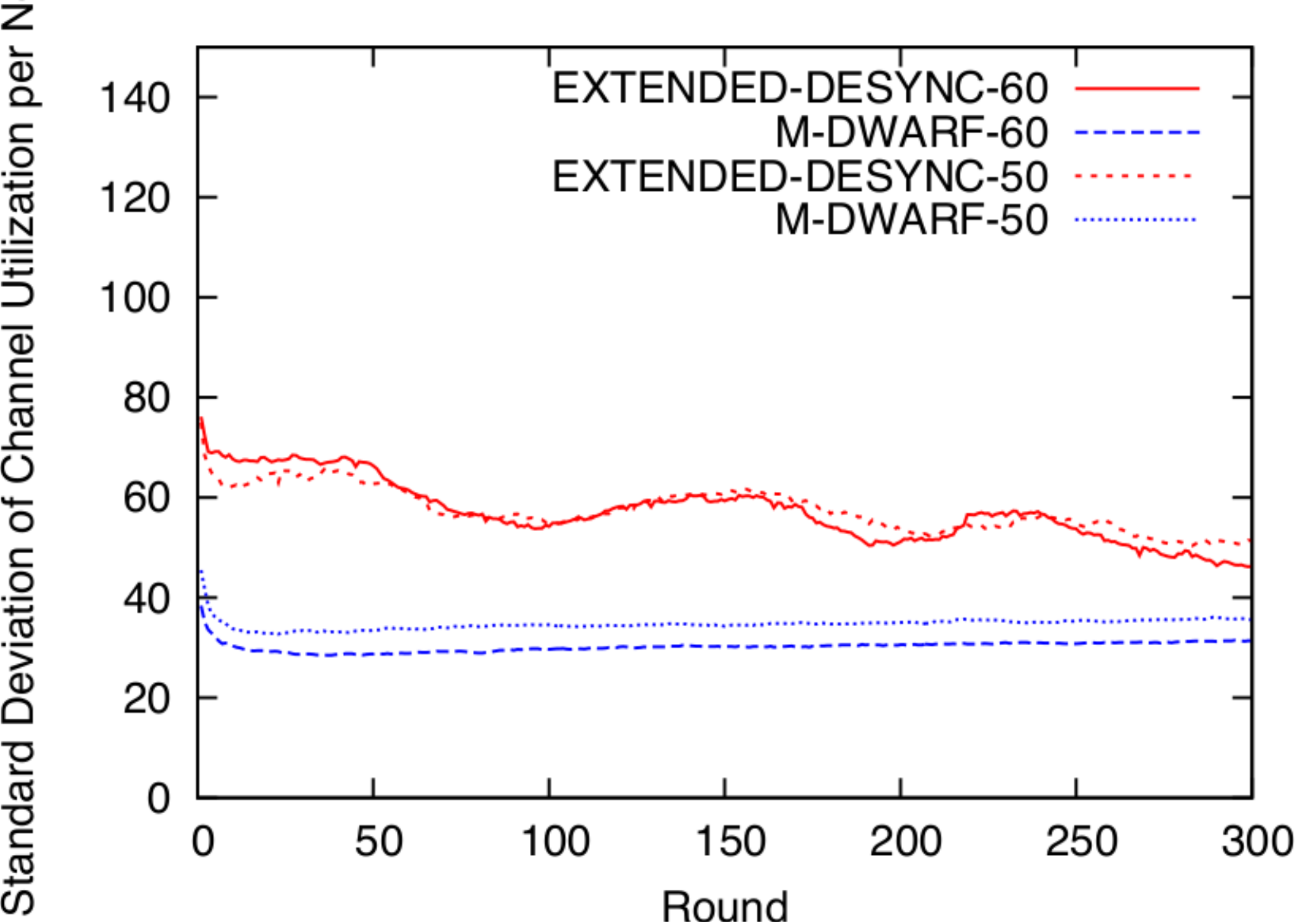}%
	\label{fig:period2000-fairness-50-60}}
	\hfil
	\subfloat[$T$ = 3000 ms]{\includegraphics[scale=0.27]{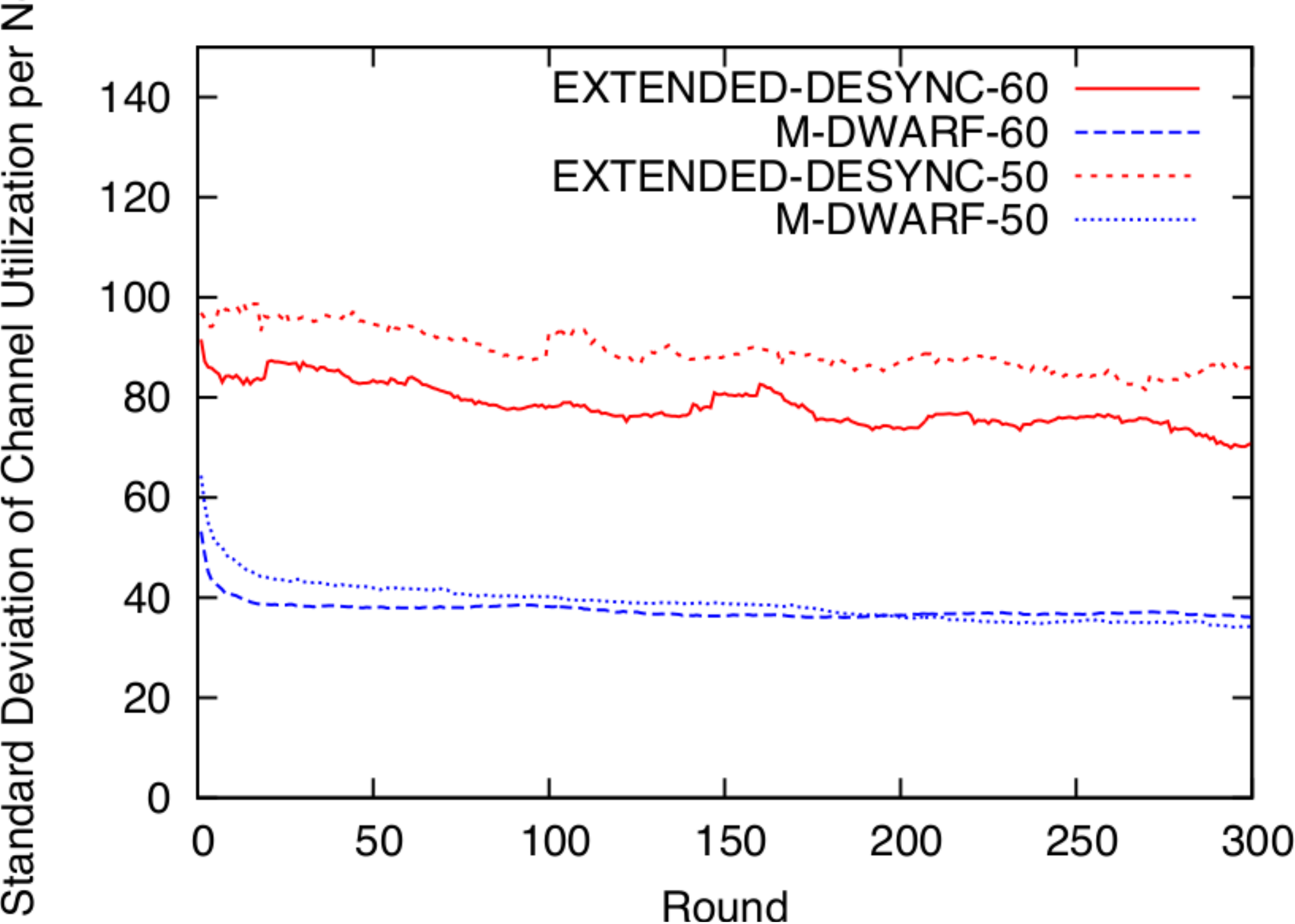}%
	\label{fig:period3000-fairness-50-60}}
}
\caption{Average standard deviation of channel utilization per node of 50 and 60 node networks}
\label{fig:period-fairness-50-60}
\end{figure*}

We measure the channel utilization fairness of nodes. If we average the channel utilization per node, the result is approximately the same for M-DWARF and EXT-DESYNC but M-DWARF is insignificantly slightly better (which is the reason why the average channel utilization of M-DWARF is only slightly better). However, to measure the fairness, we are interested in the value of the average standard deviation of the channel utilization per node. If nodes fairly utilize the network, this deviation value must be low.
The results of 30- and 40-node networks are shown in Figure \ref{fig:period-fairness-30-40} and the results of 50- and 60-node networks are shown in Figure \ref{fig:period-fairness-50-60}.
In saturated networks (\textit{i.e.} 1000-millisecond period), the fairness of both M-DWARF and EXT-DESYNC is approximately the same when time passes by. However, the difference of fairness becomes significantly large when the period length is increased. The average standard deviation of the channel utilization per node of EXT-DESYNC is 30 - 400\% higher than that of M-DWARF. Therefore, M-DWARF achieves better fairness. One of the main reasons that nodes in M-DWARF fairly utilize a network is that the stability of M-DWARF is better than that of EXT-DESYNC which is sensitive to neighbor phase errors leading to high fluctuation.

\section{Overhead Optimization}
\label{sec:overhead_opt}
Because a firing message contains relative phase information of one-hop neighbors, if there are too many neighbors, the algorithm incurs large overhead. Therefore, in this evaluation, we attempt to reduce such overhead. We straightforwardly reduce the number of relative phases broadcastings. We only sporadically attach neighbors' relative phases in a firing message; therefore, in most firing messages, relative phases are not included, minimizing the size of such messages. 
Each node keeps records of phase neighbors within two hops even it does not receive updated neighbor phase information. We set a time-to-live counter for each record to 3 periods. If a node does not receive the phase information from another node within 3 periods, it removes that node from its records.
To investigate how much we can reduce the overhead, we vary a saving gain parameter $\beta$ which is the frequency to exclude relative phases from a firing message; for example, $\beta  = 0$ means including relative phases in every firing message, $\beta = 1$ means including relative phases in every two firing messages, $\beta = 2$ means including relative phases in every three firing messages, and so on. 

\begin{figure*}
\centering 
{
	\subfloat[$\beta$ = 0]{\includegraphics[width=2.7in]{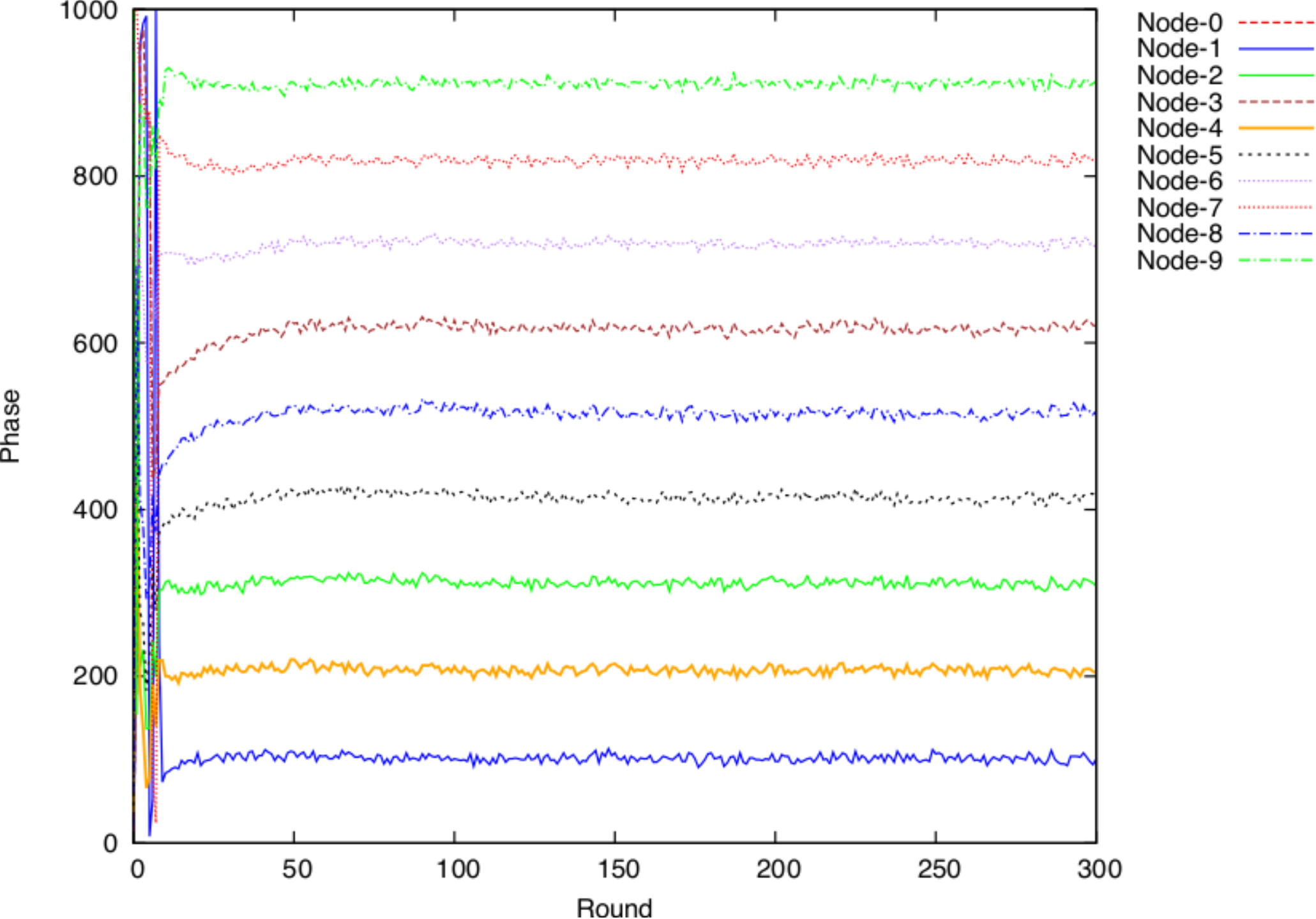}%
	\label{fig:beta-0}}
	\subfloat[$\beta$ = 4]{\includegraphics[width=2.7in]{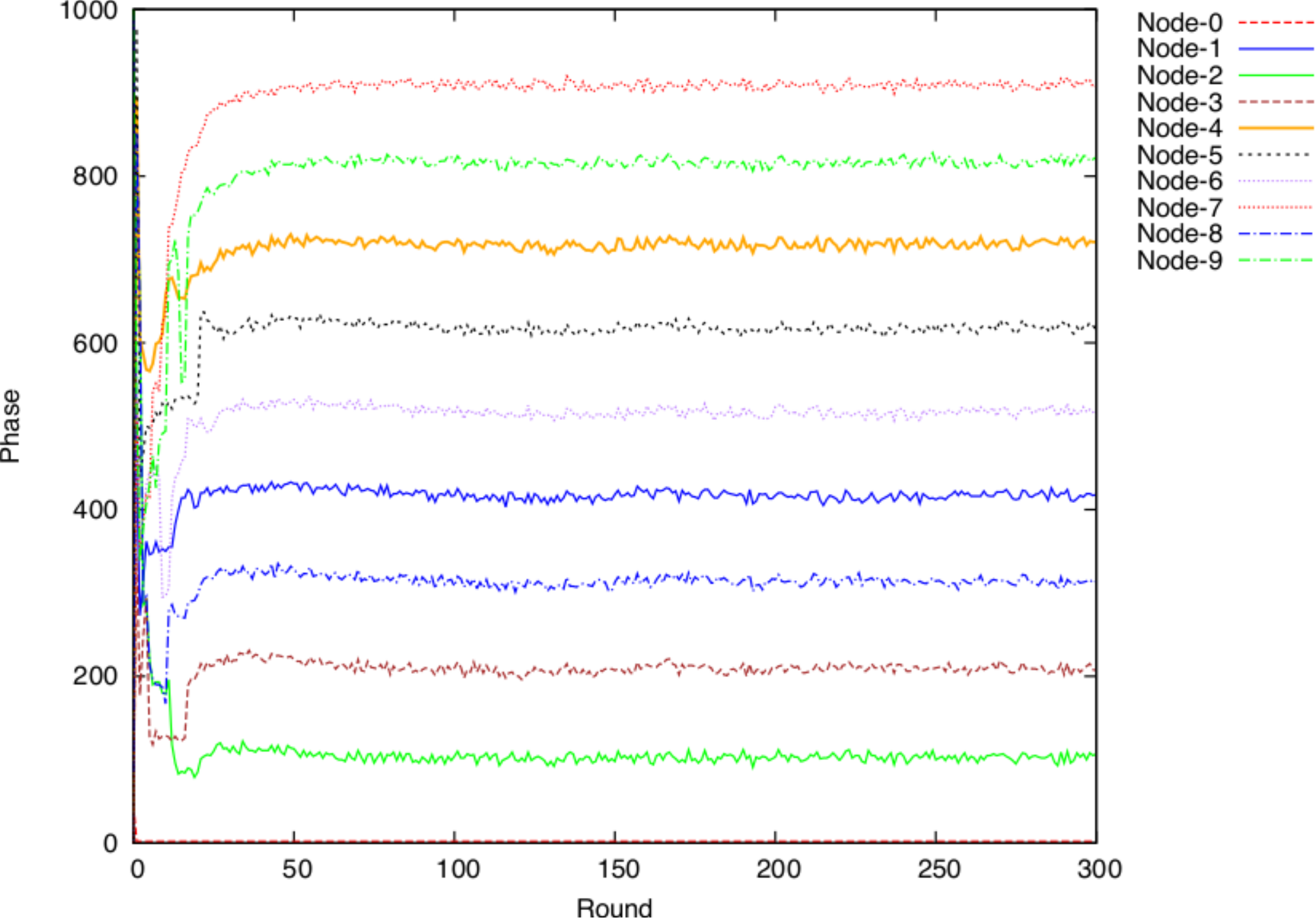}%
	\label{fig:beta-4}}
    \hspace{8pt}%
	\subfloat[$\beta$ = 8]{\includegraphics[width=2.7in]{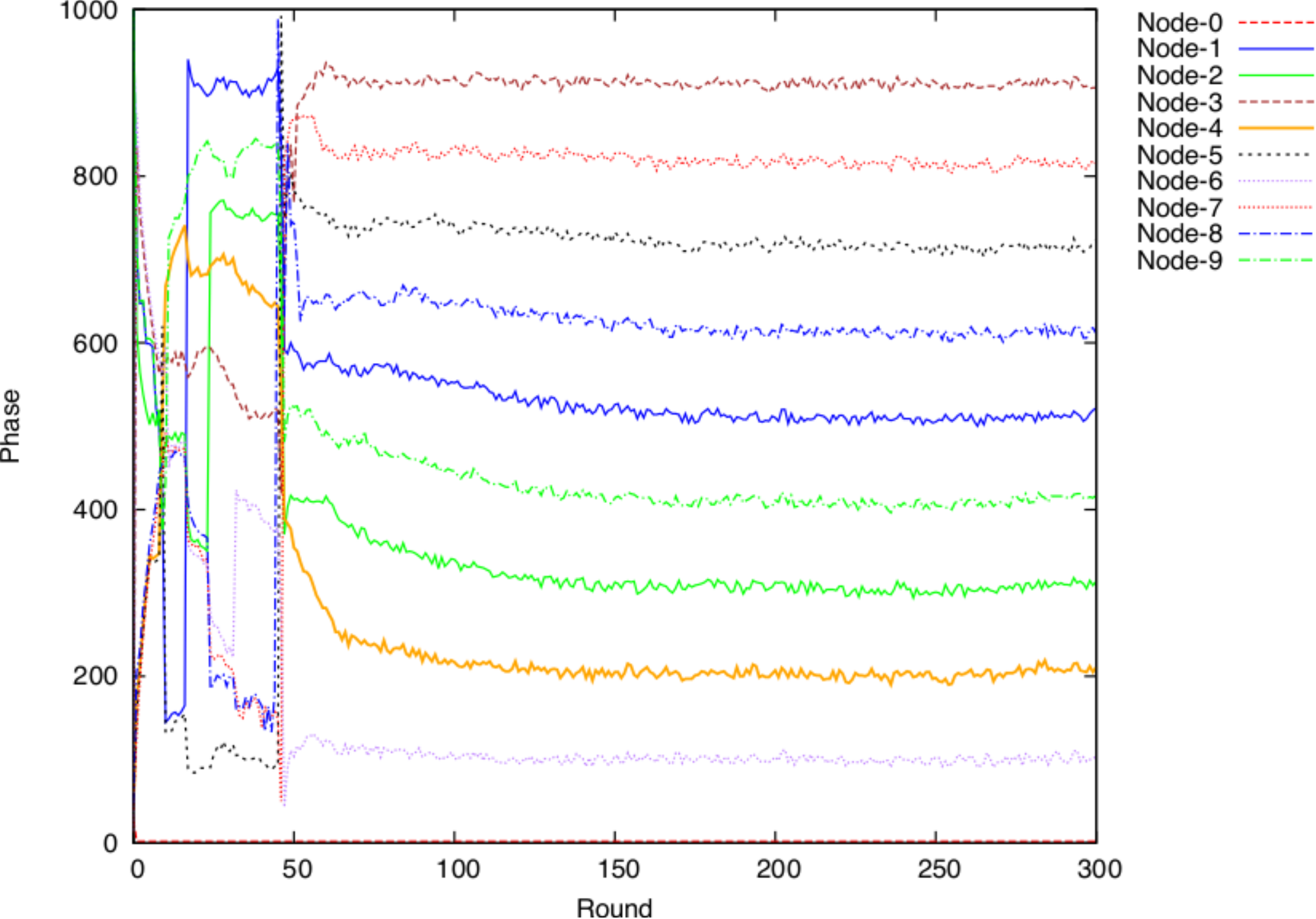}%
	\label{fig:beta-8}}
	\subfloat[$\beta$ = 12]{\includegraphics[width=2.7in]{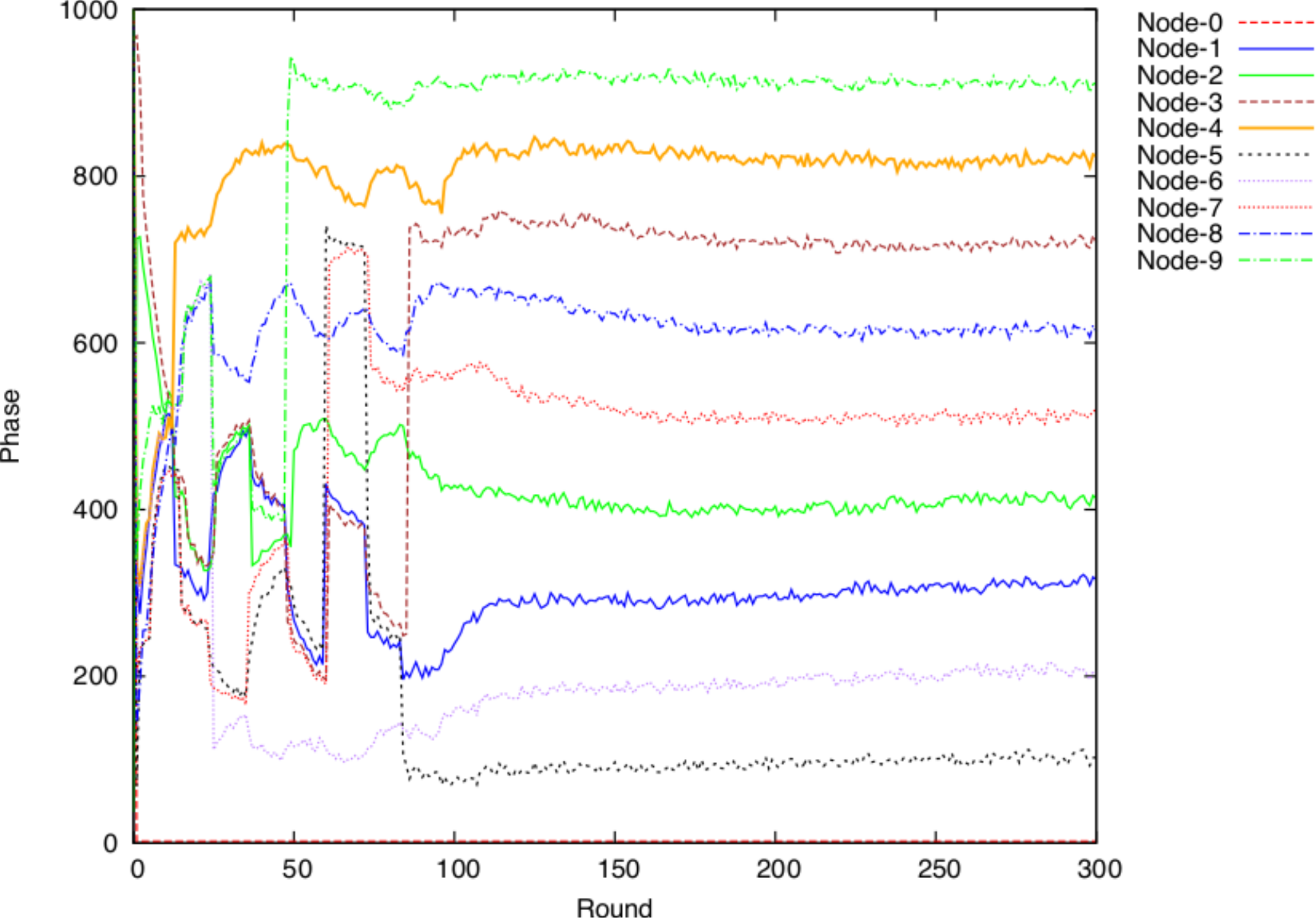}%
	\label{fig:beta-12}}
    \hspace{8pt}%
	\subfloat[$\beta$ = 16]{\includegraphics[width=2.7in]{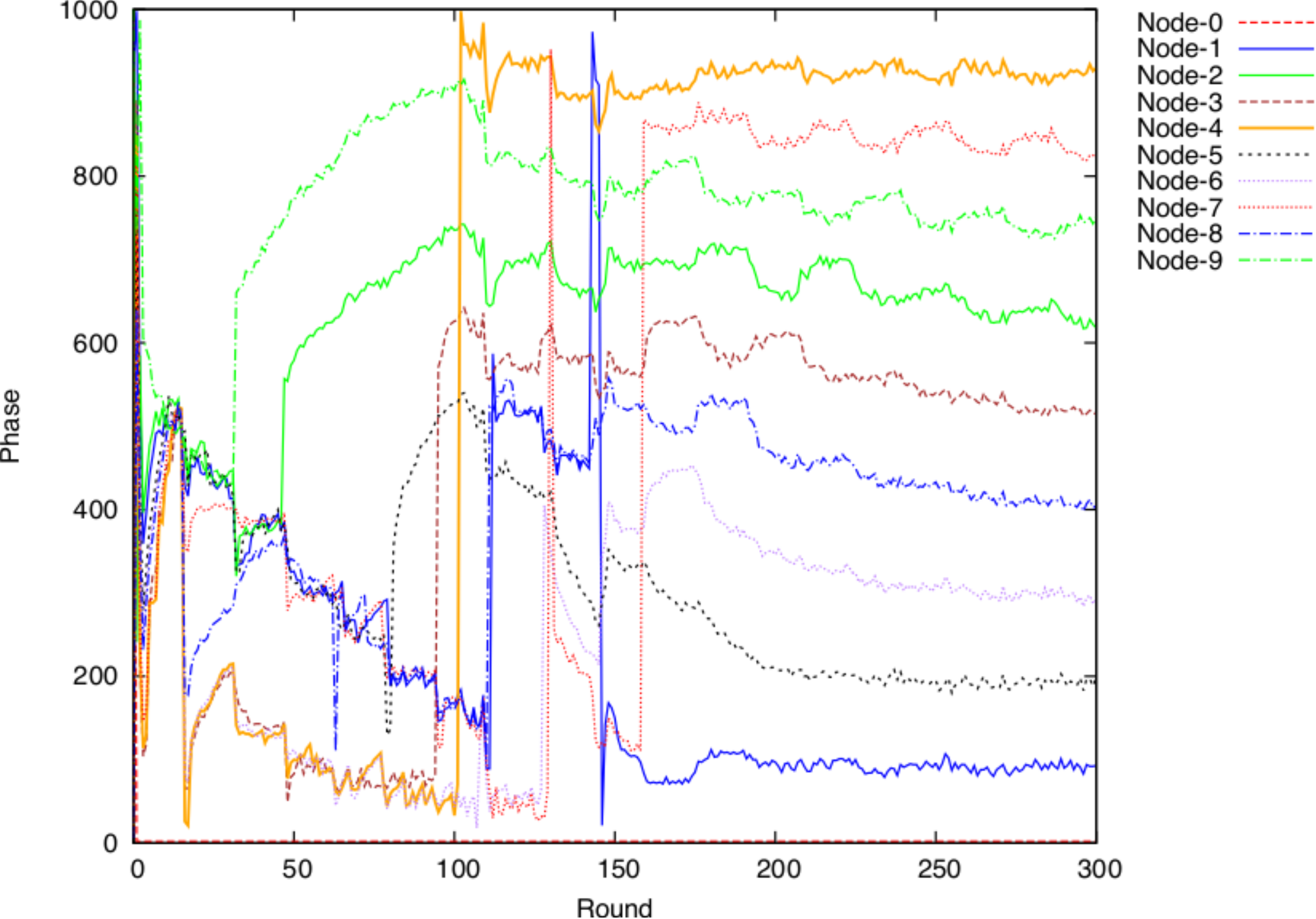}%
	\label{fig:beta-16}}
	\subfloat[$\beta$ = 20]{\includegraphics[width=2.7in]{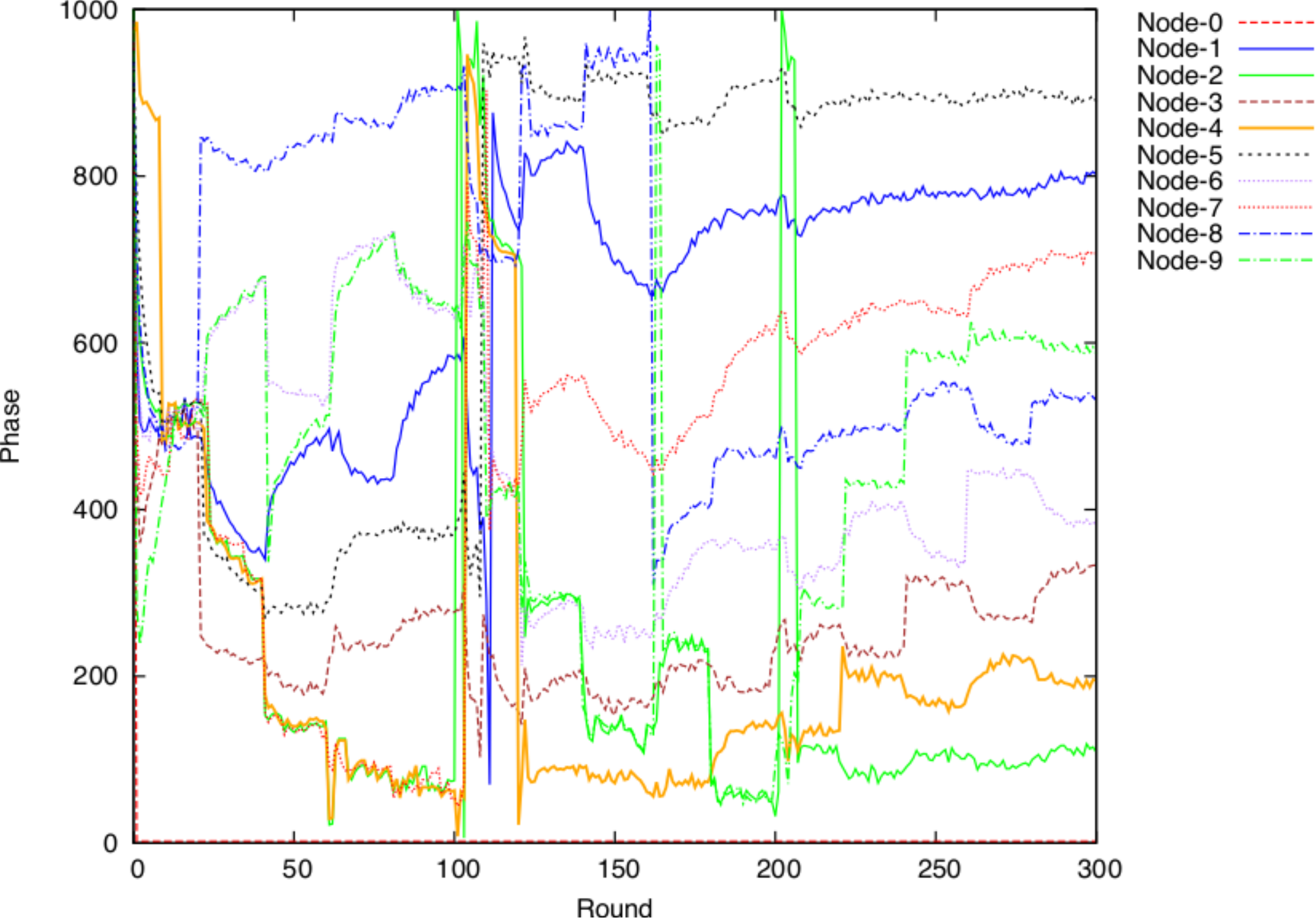}%
	\label{fig:beta-20}}
}
\caption{Optimized M-DWARF: Varying saving gain from 0 to 20.}%
\label{fig:vary-saving}%

\end{figure*}

We evaluate this optimization scheme on the 10-node star topology. We vary the saving gain $\beta$ from 0 to 20. 
The result is shown in Figure \ref{fig:vary-saving}.
From the figure, we find that, even we reduce the number of relative phases broadcastings, the system still converges to the perfect desynchrony state in most cases. However, if the saving gain is greater than 16, the system converges slowly or may not converge at all because the information that a node keeps is removed.
This result indicates two important insights.
First, by trading off with convergence speed, we can considerably reduce the overhead upto a factor of 16 in our investigated scenario. Second, even if each node does not receive relative phases for many consecutive times, the algorithm still operates to desynchronize successfully. Therefore, this implies that our proposed algorithm tolerates packet loss at a certain level. As a result, in a perspective of WSN energy consumption, wireless sensor nodes are allowed to go into a sleep mode for a period of time at the expense of slower convergence rate.

\section{Conclusion and Future Work}
\label{sec:conclusion}
In this paper, we propose a novel desynchronization algorithm for multi-hop networks namely M-DWARF. Our M-DWARF algorithm works without any master node, desynchronizes nodes to possess an equitable slot size, and requires nodes to occupy and de-occupy slots on-demand and on-the-fly. M-DWARF has two additional techniques, relative time relaying and force absorption, that allow it to successfully overcome hidden terminal problems and efficiently support multi-hop networks. We experiment M-DWARF on single-hop networks and various multi-hop topologies including star, chain, cycle, dumbbell, and mesh topologies, with network size variations. Our performance evaluation indicates that M-DWARF has lower desynchronization errors, converges faster, and maintains more stable than existing approaches. Moreover, we propose an overhead optimization that allows M-DWARF to reduce overhead in exchange with slower convergence rates. We also describe limitations of our work that leave room for future research and development. Finally, we provide extensive stability analysis of the single-hop and multi-hop desynchronization algorithms by applying dynamic system analysis.

The future work includes the followings. First, it is interesting to have convergence analysis for our desynchronization algorithms by using a similar model as in \cite{convergence_desync}. This kind of analysis helps quantify how much iterations the algorithms need to reach desynchronization convergence. Second, energy efficiency can be further improved by using techniques like \emph{reduced radio listening}. This can be achieved by cross layer information of data lengths, packet lengths and frame lengths. Our desynchronization algorithms could correctly estimate the \emph{time-to-live} of neighbors' phases without having to turn on the radio and listen to packets all the time. Third, we plan to experiment our desynchronization algorithms on more scalable networks and implement them on other wireless network standards or platforms like Zigbee \cite{zigbee} or WARP \cite{WARP}.

\bibliographystyle{unsrt}      

\bibliography{paper}   

\end{document}